\documentclass[aoas]{imsart}

\RequirePackage{amsthm,amsmath,amsfonts,amssymb}
\RequirePackage[authoryear]{natbib}
\RequirePackage[colorlinks,citecolor=blue,urlcolor=blue]{hyperref}%% uncomment this for coloring bibliography citations and linked URLs
\RequirePackage{graphicx}%% uncomment this for including figures
\RequirePackage{multirow}
\RequirePackage{xcolor}
\RequirePackage{float}

\startlocaldefs
\theoremstyle{plain}
\newtheorem{theorem}{Theorem}[section]

\newtheorem{corollary}[theorem]{Corollary}
\endlocaldefs

\begin{document}

\begin{frontmatter}
\title{Low-rank longitudinal factor regression with application to chemical mixtures}
%\title{A sample article title with some additional note\thanksref{T1}}
\runtitle{Low-rank longitudinal factor regression}
%\thankstext{T1}{A sample of additional note to the title.}

\begin{aug}
%%%%%%%%%%%%%%%%%%%%%%%%%%%%%%%%%%%%%%%%%%%%%%%
%% Only one address is permitted per author. %%
%% Only division, organization and e-mail is %%
%% included in the address.                  %%
%% Additional information can be included in %%
\author[A]{\fnms{Glenn}~\snm{Palmer}\ead[label=e1]{glenn.palmer@duke.edu}},
\author[A]{\fnms{Amy}~\snm{H. Herring}\ead[label=e2]{amy.herring@duke.edu}}
\and
\author[A]{\fnms{David}~\snm{B. Dunson}\ead[label=e3]{dunson@duke.edu}}
%%%%%%%%%%%%%%%%%%%%%%%%%%%%%%%%%%%%%%%%%%%%%%
%% Addresses                                %%
%%%%%%%%%%%%%%%%%%%%%%%%%%%%%%%%%%%%%%%%%%%%%%
\address[A]{Department of Statistical Science,
Duke University\printead[presep={,\ }]{e1,e2,e3}}
\end{aug}

\begin{abstract}

Developmental epidemiology commonly focuses on assessing the association between multiple early life exposures and childhood health. Statistical analyses of data from such studies focus on 
inferring the contributions of individual exposures, while also characterizing time-varying and interacting effects. Such inferences are made more challenging by correlations among exposures, nonlinearity, and the curse of dimensionality.
Motivated by studying the effects of prenatal bisphenol A (BPA) and phthalate exposures on glucose metabolism in adolescence using data from the ELEMENT study, we propose a low-rank longitudinal factor regression (LowFR) model for tractable inference on flexible longitudinal exposure effects. 
LowFR handles highly-correlated exposures using a Bayesian dynamic factor model, which is fit jointly with a health outcome via a novel factor regression approach. The model collapses on simpler and intuitive submodels when appropriate, while expanding to allow considerable flexibility in time-varying and interaction effects when supported by the data. After demonstrating LowFR's effectiveness in simulations, we use it to analyze the ELEMENT data and find that diethyl and dibutyl phthalate metabolite levels in trimesters 1 and 2 are associated with altered glucose metabolism in adolescence.

\end{abstract}

\begin{keyword}
\kwd{Factor analysis}
\kwd{Bayesian statistics}
\kwd{interaction effects}
\kwd{longitudinal data analysis}
\kwd{mixtures problem}
\kwd{maternal and child health}
\end{keyword}

\end{frontmatter}
%%%%%%%%%%%%%%%%%%%%%%%%%%%%%%%%%%%%%%%%%%%%%%
%% Please use \tableofcontents for articles %%
%% with 50 pages and more                   %%
%%%%%%%%%%%%%%%%%%%%%%%%%%%%%%%%%%%%%%%%%%%%%%
%\tableofcontents

%%%%%%%%%%%%%%%%%%%%%%%%%%%%%%%%%%%%%%%%%%%%%%
%%%% Main text entry area:

\section{Introduction}\label{section1}

A key problem in developmental and environmental epidemiology is assessing the association between complex longitudinal, multifactorial early life exposures and subsequent outcomes. Our focus is on exposure to a mixture of chemical exposures, measured over time. Given modern industrial and manufacturing processes, humans are constantly exposed to complex mixtures of environmental and synthetic chemicals, some of which may interfere with biological processes \citep{kahn2020endocrine}. While the health effects of some individual chemicals have been well-studied, relevant groups of chemicals are often related and likely to co-occur, and the potential for interaction effects can lead to biased or misleading inferences when studying only one chemical at a time \citep{dominici2010opinion, joubert2022powering}. Thus, there has been a recent push to develop methods for analyzing the effects of such mixtures jointly \citep{stafoggia2017statistical, hamra2018environmental}. Additional challenges are introduced when studying these exposures during pregnancy and early childhood. In such settings, accounting for the timing of exposures is critical for analysis, given that early-life development is a dynamic process with a number of distinct stages \citep{buckley2019statistical}. For example, a given exposure in the first trimester may have a dramatically larger or smaller effect than in the third trimester. Moreover, one exposure early in pregnancy could plausibly increase vulnerability to a different exposure later on, suggesting the need to evaluate highly-flexible but interpretable interaction effects.

In this article we are directly motivated by studying the effects of prenatal exposure to bisphenol A (BPA) and phthalates on glucose metabolism in adolescence. BPA and phthalates are widely used in the manufacturing of plastics, cosmetics, and other household and industrial products, and their metabolites are found in the overwhelming majority of participants in human observational studies \citep{silva2004urinary, becker2009geres, wenzel2018prevalence, braun2013phthalate}. However, exposure to these chemicals has been linked with a variety of disruptive effects to the endocrine system, leading to such health outcomes as delayed mental and physical development in childhood \citep{kim2011prenatal, whyatt2012maternal}, delayed onset of puberty \citep{frederiksen2012high}, allergic diseases \citep{just2012prenatal, bornehag2004association}, and obesity, insulin resistance, and otherwise disrupted glucose transport and metabolism \citep{stahlhut2007concentrations, eng2013bisphenol, stojanoska2017influence}. Despite the growing literature on these effects, a major limitation is that analyses are typically conducted for one exposure at a time, and for either a single measurement time or averaged over time. To address these weaknesses and better illuminate health effects of phthalates and BPA, we seek to perform an analysis that attains three objectives: (1) to learn interpretable relationships between the measured exposures and the outcome, (2) to understand how these relationships vary over time, and (3) to allow for non-additive effects. In particular, we would like to investigate interactions across both exposures and time.

Despite the need for a method addressing our goals, there are several challenges in developing such an approach. First, measurements are often highly correlated across both different chemicals and time, leading to significant collinearity if one were to try even a simple linear regression approach with all measured values. Second, if exposures are moderate- to high-dimensional and then measured at even a modest number of times, the overall dimensionality can quickly explode, making analysis difficult even if correlations were low. Even in the cross-sectional case, finding interactions in moderate- to high-dimensional linear regression is a challenging problem that has sparked a substantial body of recent work \citep{bien2013lasso, hao2018model, wang2021penalized}. When we also want to consider longitudinal measurements, the problem quickly becomes harder: if we have $p$ exposures each measured at $T$ aligned times, then the number of coefficients in a quadratic regression model becomes $2pT + {pT \choose 2}$. For these reasons, current methods either address only a subset of our objectives, or have some significant limitations, drawbacks, or make simplifying assumptions.

\subsection{Relevant literature}

When modeling the effect of longitudinal exposure, a large body of work has focused on a single exposure measured over time using a distributed lag model \citep{welty2009bayesian}. Extensions account for spatial variation \citep{warren2012spatial},  time-to-event outcomes \citep{chang2015assessment},  multivariate outcomes \citep{warren2016bayesian}, variable selection to zero out some time intervals \citep{warren2020critical}, and allowing nonlinear relationships between the exposure and outcome via splines or regression trees \citep{gasparrini2010distributed, gasparrini2017penalized, mork2022treed}.

In parallel, there has been a recent push to analyze mixtures of exposures jointly. Given the low sample sizes and signal-to-noise ratios typical of studies in this domain, power to detect main and interaction effects within a mixture is quite low using standard regression approaches \citep{nguyen2023power}. Thus, a number of methods have been proposed that attempt to make the problem more tractable by applying some combination of variable selection and dimension reduction. In the cross-sectional case, \cite{bobb2015bayesian} developed Bayesian Kernel Machine Regression (BKMR), which uses Gaussian process regression with variable selection priors to estimate a nonlinear regression function. \cite{liu2022cross} developed a hypothesis testing framework for interaction effects using kernel regression to model nonlinear main effects. Weighted quantile sum regression (WQSR) \citep{carrico2015characterization, czarnota2015assessment} uses a two-stage approach to estimate an overall effect of quantile-binned exposures. In step one, a training set is used to estimate a set of nonnegative weights summing to one. Then in a validation set, outcomes are regressed on the resulting weighted indices. Extensions to this approach include fitting with random subsets of features \citep{curtin2021random}, averaging estimates over cross-validation splits \citep{tanner2019repeated}, and a Bayesian implementation using a Dirichlet prior for the weights \citep{colicino2020per}. See \cite{hamra2018environmental} for an epidemiological perspective comparing BKMR and WQSR with standard high-dimensional regression methods like principal components \citep{massy1965principal}, LASSO \citep{tibshirani1996regression}, and ridge regression \citep{hoerl1970ridge}.

Most relevant to our work is a growing literature building on the above approaches to allow mixtures measured over time.
Leveraging on a distributed lag approach, \cite{wang2022semiparametric} developed quantile regression for time-varying mixtures, but their model does not include interactions. \cite{chen2019distributed} proposed a distributed lag model that allows interactions, but focuses on the two exposure case. \cite{warren2022critical} extended the critical window variable selection approach from \cite{warren2020critical} to the mixture setting, but their method includes constraints so that at a given time, all main effects and interactions must have the same sign. \cite{bello2017extending} and \cite{gennings2020lagged} combined WQSR with distributed lag models, modeling time-varying weights with splines. However, as in cross-sectional WQSR, this approach does not allow for interactions. \cite{antonelli2022multiple} proposed linear regression with all pairwise interactions on a mixture of exposures measured over time, using basis functions to ensure coefficients vary smoothly in time, and assigning spike-and-slab variable selection priors to find critical windows. However, as this approach does not naturally reduce dimensionality, they proposed PCA as an intermediate step, thus underestimating uncertainty and limiting interpretability of the resulting coefficient estimates. \cite{wilson2022kernel} combined the distributed lag approach with BKMR by estimating a smooth lag function for each exposure, using the integral of the product of exposure trajectories and their corresponding lag functions as input to BKMR. However, this approach shares issues with cross-sectional BKMR in terms of the curse of dimensionality implicit in estimation of unconstrained multivariate regression functions. Also building on BKMR, \cite{liu2018lagged} combined group lasso \citep{yuan2006model} and fused lasso \citep{tibshirani2005sparsity} penalties to regularize, shrinking regression functions for nearby times toward each other. However, along with the issues with BKMR noted above, this method only allows interactions between exposures within time points, not interactions across different times \citep{wilson2022kernel}.

\subsection{Our contribution}\label{section1.2}

\begin{figure}[h]
\includegraphics[width=0.95\textwidth]{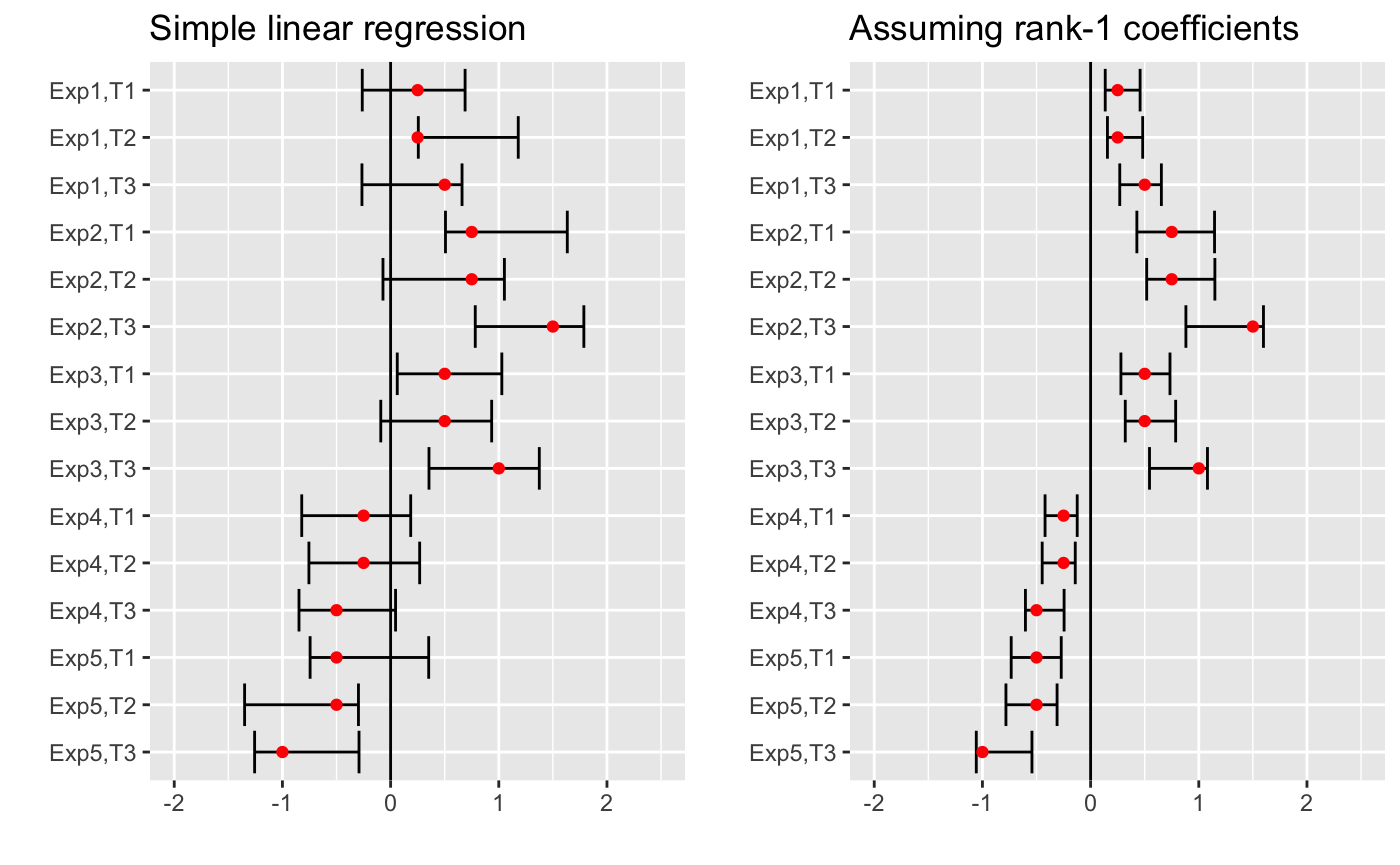}
\caption{Posterior summaries for two linear regression models fit to the same $n=200$ simulated data points. The left panel shows a simple Bayesian linear regression approach using $N(0,10)$ priors for all coefficients. The right panel shows a ``rank-1'' model where each coefficient is modeled as the product of an exposure-specific parameter and a time-specific parameter. Each of these parameters is again given a $N(0,10)$ prior.}
\label{fig_intro_rank1}
\end{figure}

In this work, we propose a novel regression approach for longitudinal data that gains statistical power through dimension reduction similar to weighted quantile sum approaches, but allows for much more flexibility and interpretability, can handle interaction effects, and fully quantifies uncertainty through a single coherent Bayesian model. To get some intuition motivating our approach, consider estimating a linear regression of a real-valued outcome on $p=5$ exposures, each measured at $T=3$ time points. A naive approach would simply be to treat the measurements as 15 separate exposures and estimate a linear regression. A posterior summary for a Bayesian implementation of this approach for $n=200$ simulated data points is shown in the left panel of Figure \ref{fig_intro_rank1}. True regression coefficients are shown as red dots, along with $95\%$ posterior credible intervals using independent $N(0,10)$ priors for the coefficients. Details of the data simulation and model fitting are provided in the Supplemental Information, along with fully reproducible code.

Examining the true coefficients in Figure \ref{fig_intro_rank1}, note that for each exposure, the magnitude is larger for time T3 than for times T1 or T2. This could correspond to a health setting where third-trimester fetal development is strongly linked to the outcome, for example. In a case like this, we would like our model to take advantage of the temporal structure by sharing information across coefficient estimates. The right panel of Figure \ref{fig_intro_rank1} demonstrates the benefit of doing so; for the same data, we estimated linear regression coefficients $\theta \in \mathbb{R}^{15}$ as the outer product of an ``exposure coefficient'' vector $\beta \in \mathbb{R}^5$ and a ``time coefficient'' vector $\omega \in \mathbb{R}^3$, again using $N(0,10)$ priors for the entries of $\beta$ and $\omega$. That is, for each $i$ and $j$ we simply let
$$\theta_{Exp\, i, T j} = \beta_i \cdot \omega_j.$$
Note that in the above, $\beta$ and $\omega$ are not jointly identifiable. Rather than place constraints to address this, we view the setup as an example of parameter expansion \citep{liu1999parameter} and only interpret the posterior of $\theta$. Observe in Figure \ref{fig_intro_rank1} that the credible intervals for this ``rank-1'' model are much narrower, while still containing the true parameter values. If we use posterior means as estimates of the coefficients, the naive model has a mean squared estimation error of $0.055$, while the rank-1 model has mean squared estimation error of $0.012$. This makes sense, since we are able to use fewer parameters in our model, and estimation of each $\beta_i$ and $\omega_j$ shares information across all times and exposures, respectively.

\begin{figure}[h]
\includegraphics[width=\textwidth]{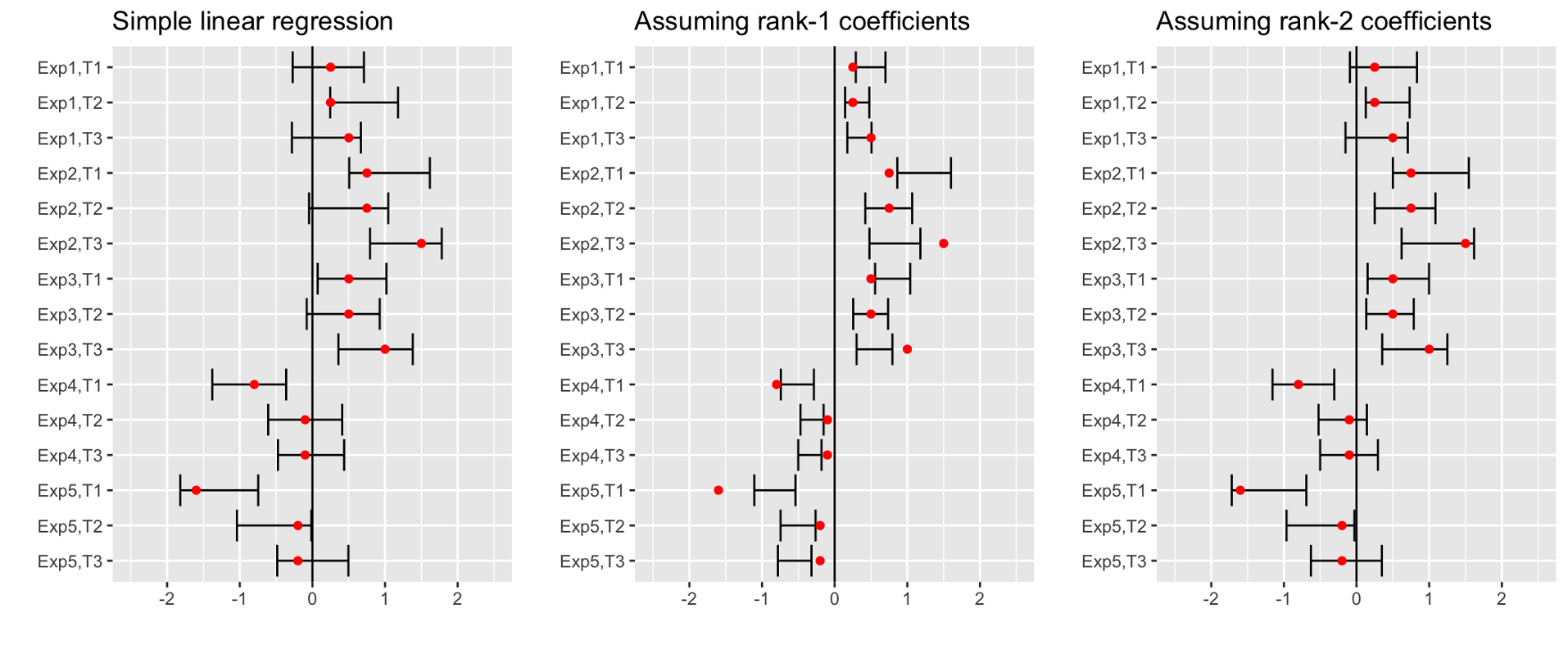}
\caption{Posterior summaries for linear regression models fit to $n=200$ simulated data points with true rank-2 regression coefficients. The left panel shows a simple Bayesian linear regression approach. The center panel shows a rank-1 model as in Figure \ref{fig_intro_rank1}. The right panel shows a rank-2 model. $N(0,10)$ priors are used for the coefficient parameters in all three models.}
\label{fig_intro_rank2}
\end{figure}

While the parsimonious rank-1 model in Figure \ref{fig_intro_rank1} has clear benefits, an immediate criticism of such an approach is that it makes strong a priori assumptions about the model structure that may be unrealistic in practice. As a concrete example, Figure 2 shows data where the temporal dependence is different for exposures 4-5 than for exposures 1-3. Again, this could correspond to a realistic case where some exposures are important early in pregnancy, while others are important later. In this case, the rank-1 model above induces significant bias in estimation. However, if we instead allow two $\beta$ and $\omega$ vectors such that
$$\theta_{Exp\, i, T j} = \beta_{i1} \cdot \omega_{j1} + \beta_{i2} \cdot \omega_{j2},$$
we again outperform a naive linear regression approach, as shown in the right panel of Figure \ref{fig_intro_rank2}, by learning what shared structure there is. The mean squared estimation error of posterior mean coefficients for the naive, rank-1, and rank-2 models are $0.054$, $0.140$, and $0.044$, respectively.

Taken together, the above results suggest a broader approach to estimating longitudinal regression coefficients. For a real-valued outcome regressed on $p$ exposures measured at $T$ aligned times, we propose modeling the $pT$ linear regression coefficients as a matrix $\theta \in \mathbb{R}^{p \times T}$, and we postulate that in many settings, $\theta$ can be well-approximated by a matrix $\Tilde{\theta}$ of rank $l < \min\{p, T\}$. In doing so, we reduce the dimensionality of the regression surface, and can more precisely and accurately estimate associations. We note that this idea relates to approaches in \cite{hung2013matrix} and \cite{zhou2013tensor}, who proposed rank decompositions of regression coefficients when regressing on a matrix or tensor in the context of medical imaging. However, in addition to our different applied motivation, our approach is distinct from these in several key ways. First, to avoid making assumptions about the needed rank of $\Tilde{\theta}$, we learn $l$ from data using an increasing shrinkage prior distribution \citep{bhattacharya2011sparse} for the exposure and time vectors. Building upon this insight, we extend our model to handle interactions with an analogous decomposition, and combine our regression approach with a latent factor model \citep{bartholomew2011latent} for the exposures to handle strong correlations and further reduce dimensionality. We refer to our full model as low-rank longitudinal factor regression (LowFR). To our knowledge, LowFR is the first approach using this type of multifaceted dimension reduction for longitudinal regression, all within a coherent Bayesian model that fully quantifies uncertainty in all parameter estimates.

In Section \ref{section2}, we describe the motivating application and data set. In Section \ref{section3}, we describe our full LowFR model and expand on its flexibility and interpretability. In Section \ref{section4}, we compare LowFR to competitors in a simulation study. In Section \ref{section5}, we analyze data from the Early Life Exposure in Mexico to ENvironmental Toxicants (ELEMENT) study \citep{goodrich2022trimester}. Section \ref{section6} presents  discussion, limitations, and future directions to expand upon our approach.

\section{Glucose metabolism data analysis}\label{section2}

\subsection{Motivation} 
Although BPA and phthalates are widely used in consumer products, there is growing evidence that they may have disruptive health effects. Both BPA and a number of phthalates appear to interfere with functioning of nuclear receptor proteins \citep{hurst2003activation, grimaldi2015reporter}, which are involved in a wide range of physiological processes including glucose metabolism and homeostasis, and whose disruption has been linked with diabetes, obesity, and insulin resistance among other poor outcomes \citep{stojanoska2017influence, desvergne2009ppar}. In a laboratory study, \cite{alonso2006estrogenic} found that injecting rats with small amounts of BPA led to an increase in plasma insulin along with a decrease in blood glucose. \cite{guven2016low} treated human pancreatic beta cells with monoethyl phthalate (MEP) in an in vitro study and also observed an increase in insulin levels, along with an increase in activity levels of GLUT1, a glucose transporter responsible for moving glucose molecules from plasma into cells. Although this type of glucose metabolism hyperactivity would initially lead to lower blood glucose levels, when sustained long-term (e.g., over the course of childhood and adolescence) it has been proposed as a cause of insulin resistance leading to type 2 diabetes \citep{shanik2008insulin}. Correspondingly, a number of observational studies have linked high urinary BPA and phthalate levels with prevalence of obesity, insulin resistance, and type 2 diabetes in adults \citep{stahlhut2007concentrations, james2012urinary, sun2014association, lang2008association}.

Since disorders like insulin resistance and type 2 diabetes typically develop slowly over many years, there is concern about whether exposures during critical windows of fetal development could cause small but persistent changes in glucose homeostasis leading to long-term disease \citep{grun2007perturbed}. To investigate such long-term effects, the Early Life Exposure in Mexico to ENvironmental Toxicants (ELEMENT) study \citep{goodrich2022data} measured maternal urinary levels of nine phthalates and BPA in each trimester of pregnancy, and then followed up with each child in early adolescence, recording their physical development and collecting a plasma sample. Analyzing these samples, \cite{goodrich2022trimester} found negative associations between first-trimester MEP, first-trimester monobutyl phthalate (MBP), and first-trimester mono-(3-carboxypropyl) phthalate (MCPP) with followup plasma levels of 2-deoxy-D-glucose (2-DG). As 2-DG is a glucose molecule that is transported out of plasma by GLUT1, these negative associations are consistent with the MEP-induced GLUT1 upregulation found in \cite{guven2016low}, but in human subjects and measured over a vastly longer window of time. If these early prenatal exposures lead to a years-long shift in glucose metabolism, it would be a significant medical finding with immediate implications for policy makers, medical providers, and pregnant individuals. MEP, MBP, and MCPP are metabolites of phthalates commonly used in cosmetics, personal care products, pharmaceuticals and supplements \citep{braun2013phthalate}, so understanding their safety directly affects recommendations for patient behaviors and product usage.

These results should be replicated in separate studies to better understand their validity. However, given that such studies take decades to conduct, we also see an opportunity to glean more information from the already-collected and published ELEMENT data. To analyze the associations between each of the 30 prenatal exposures (10 chemicals measured 3 times) and childhood 2-DG, \cite{goodrich2022trimester} simply performed 30 separate linear regressions of 2-DG on one exposure at a time, substantially limiting the interpretability of their results. In this work, we apply our longitudinal regression approach to analyze all these associations jointly in a single model, allowing us to study mixture effects and give more nuance to the conclusions about exposure timing. As it is difficult to detect individual significant effects for the reasons listed in Section \ref{section1}, we summarize our model fit in stages. First, we note that much of the correlation among the measured phthalates arises because they are actually metabolites of a smaller set of parent phthalates, which we summarize in Table \ref{table_parentlist}, compiled from \cite{braun2013phthalate} and \cite{koch2012di}. Based on this list, we consider the expected change in 2-DG if all metabolites of one parent phthalate at a time were to increase together across all three trimesters. By pooling effects in this way, we may have  power to detect associations for these parent compounds even when posterior credible intervals for all of the corresponding individual main and interaction terms include zero. Then, based on our results for these pooled effects, we systematically evaluate effects of smaller sets of both metabolites and measurement times. In particular, we summarize regression surfaces for both single chemicals and sets of chemicals based on parent phthalates at individual trimesters, pairs of trimesters, and cumulatively over all three trimesters. In doing so, we are able to draw conclusions unavailable from separate analyses, and that competitor models are not well-equipped to find.

\begin{table*}[]
\caption{Parent compound phthalates corresponding to the 9 measured phthalate metabolites and some examples of where they are found. For a more detailed summary and additional references, see \cite{braun2013phthalate}.}
\begin{tabular}{lll}
\hline
Parent phthalate                                                              & Examples of places found                                                                                                                                         & Measured metabolites                                                                                                                                                                                                                      \\ \hline
\begin{tabular}[c]{@{}l@{}}Dibutyl \\ phthalate (DBP)\end{tabular}            & \begin{tabular}[c]{@{}l@{}}Cosmetics,\\ personal care products,\\ pharmaceuticals/supplements\end{tabular}                                                       & \begin{tabular}[c]{@{}l@{}}Mono-butyl \\ phthalate (MBP),\\ Mono-3-carboxypropyl\\ phthalate (MCPP)\end{tabular}                                                                                                                          \\ \hline
\begin{tabular}[c]{@{}l@{}}Diethyl \\ phthalate (DEP)\end{tabular}            & \begin{tabular}[c]{@{}l@{}}Cosmetics,\\ personal care products,\\ pharmaceuticals/supplements\end{tabular}                                                       & \begin{tabular}[c]{@{}l@{}}Mono-ethyl \\ phthalate (MEP)\end{tabular}                                                                                                                                                                     \\ \hline
\begin{tabular}[c]{@{}l@{}}Butylbenzyl \\ phthalate (BBzP)\end{tabular}       & \begin{tabular}[c]{@{}l@{}}Flooring,\\ adhesives,\\ food packaging\end{tabular}                                                                                  & \begin{tabular}[c]{@{}l@{}}Mono-benzyl \\ phthalate (MBzP),\\ Mono-butyl \\ phthalate (MBP)\end{tabular}                                                                                                                                  \\ \hline
\begin{tabular}[c]{@{}l@{}}Di-(2-ethylhexyl) \\ phthalate (DEHP)\end{tabular} & \begin{tabular}[c]{@{}l@{}}Polyvinyl chloride plastics\\ (used e.g. in medical tubing,\\ food packaging, plastic toys,\\ shower curtains, rainwear)\end{tabular} & \begin{tabular}[c]{@{}l@{}}Mono-ethyl-hexyl \\ phthalate (MEHP),\\ Mono-2-ethyl-5-oxo-hexyl \\ phthalate (MEOHP),\\ Mono-2-ethyl-5-hydroxyl-hexyl \\ phthalate (MEHHP),\\ Mono-2-ethyl-5-carboxy-pentyl \\ phthalate (MECPP)\end{tabular} \\ \hline
\begin{tabular}[c]{@{}l@{}}Di-n-octyl \\ phthalate (DnOP)\end{tabular}        & \begin{tabular}[c]{@{}l@{}}Polyvinyl chloride plastics,\\ paints,\\ adhesives,\\ flooring\end{tabular}                                                           & \begin{tabular}[c]{@{}l@{}}Mono-3-carboxypropyl \\ phthalate (MCPP)\end{tabular}                                                                                                                                                          \\ \hline
\begin{tabular}[c]{@{}l@{}}Di-isobutyl \\ phthalate (DIBP)\end{tabular}       & \begin{tabular}[c]{@{}l@{}}Cosmetics,\\ personal care products,\\ pharmaceuticals/supplements\end{tabular}                                                       & \begin{tabular}[c]{@{}l@{}}Mono-isobutyl \\ phthalate (MIBP)\end{tabular}                                                                                                                                                                 \\ \hline
\end{tabular}
%\caption{Parent compound phthalates corresponding to the 9 measured phthalate metabolites and some examples of where they are found. For a more detailed summary and additional references, see \cite{braun2013phthalate}.}
\label{table_parentlist}
\end{table*}

Finally, we fit a second LowFR model with interaction terms added between all exposures and sex. In addition to their analysis of the entire data set, \cite{goodrich2022trimester} also split the data into male and female child subgroups and again fit linear regressions for each chemical and trimester. In these models, they did not detect any significant associations with 2-DG for either sex subgroup. However, as BPA and phthalates are endocrine disruptors, it is reasonable to think that any effects on 2-DG could vary by sex. We use this second model to evaluate associations between exposures and 2-DG for male and female children separately, again gaining power through our pooled chemical and time effects, and to evaluate any direct evidence that the associations differ by sex. More broadly, our overall analysis suggests a general strategy in this domain to fit a single model that balances structure with flexibility, and then make inferences at increasing levels of granularity, all while accounting for the mechanism by which the data arose.

\subsection{Data description and preprocessing}

The ELEMENT study recruited 234 pregnant mothers and measured levels of 9 phthalates and BPA in spot urine samples during each trimester of pregnancy. When children (110 boys and 124 girls) were between 8 and 14 years old, serum samples were collected and analyzed using an untargeted metabolomics approach. Investigators also recorded the BMI, sex, and age of each child, as well as whether they had begun puberty. The resulting data set includes measurements of 572 metabolites from the serum samples (of which we focus on 2-DG as our outcome) for each child, as well as the linked pregnancy exposures, demographic data, and followup data for each child. The data set is available for download from the University of Michigan \citep{goodrich2022data}.

As part of data cleaning before publishing the released data set, the metabolomics data were log-transformed and adjusted for batch effects, and any missing values were imputed via $K$-nearest-neighbors with $K=5$. The BPA and phthalate measurements were adjusted for specific gravity and log-transformed, and any measurements below the limit of detection (LOD) were imputed as $\text{LOD} / \sqrt{2}$. See \cite{goodrich2022trimester} for more information about data collection and processing. In the released data set, 7 mothers do not have any measured values of BPA or phthalate metabolites, so we removed these before analysis. Additionally, 55 individuals are missing BPA and phthalate measurements at either one or two trimesters. We imputed these values at each step of MCMC when fitting our model. We also standardized all measurements of 2-DG, as well as the BPA and phthalates within each time point, to have mean 0 and variance 1. After standardizing, inspection of the 2-DG data revealed one extreme outlier with a standardized value of $-8.4$. We removed this observation and re-standardized the remaining 2-DG values prior to model fitting.

\section{Model}\label{section3}
Building on the intuition developed in the introduction, in Section \ref{section3.1} we specify a low-rank linear regression model and extend it to a quadratic regression to include pairwise interactions. After a brief review of Bayesian factor analysis in Section \ref{section3.2}, Section \ref{section3.3} demonstrates that this quadratic regression approach can be paired with a dynamic factor model for the exposures in the case that they are high-dimensional and/or highly-correlated. In doing so, we reduce the dimensionality of both the exposures and the subsequent regression surface, while still learning quite general linear and interaction coefficients for all exposures.

\subsection{Low-rank quadratic regression}\label{section3.1}
To formalize the linear regression approach motivated in Section \ref{section1.2}, let $y_i \in \mathbb{R}$ be an outcome of interest for $i = 1,...,n$, and let $x_i = (x_{i11},...,x_{i1T}, ..., x_{ip1},..., x_{ipT})^T \in \mathbb{R}^{pT}$ be a vector of $p$ exposures, each measured at $T$ aligned measurement times. To take advantage of shared structure in the regression surface over both times and factors, we let
\begin{equation}\label{y_linreg}
   y_i = \mu + \theta^T x_i + \varepsilon_i^{(y)}, \,\,\,\,\,\,
\theta = vec\left(\sum_{l=1}^{H_1} \omega_l \beta_l^T\right),
\end{equation}
where $\mu \in \mathbb{R}$ is an intercept, $\omega_1,...,\omega_{H_1} \in \mathbb{R}^T$, and $\beta_1,...,\beta_{H_1} \in \mathbb{R}^p$. Observe that if $H_1 \geq \min\{p, T\}$, then $\theta$ can take on completely general values in $\mathbb{R}^{pT}$. However, we propose that $\theta$ may be well approximated using some smaller value of $H_1$. To allow our model to take advantage of such a situation, we set $H_1 = \min\{p, T\}$, but place a multiplicative gamma process (MGP) prior distribution \citep{bhattacharya2011sparse} on the $\omega_l$s and $\beta_l$s. The MGP shrinks parameter values more and more strongly toward zero for increasing $l$ at a rate learned from the data. Such a setup allows us to effectively reduce $H_1$, while allowing for fully-general $\theta$ when supported by the data.

To extend the above to include interaction effects, we can simply add a quadratic term and perform an analogous decomposition. Specifically, let
\begin{equation}\label{y_quadreg}
   y_i = \mu + \theta^T x_i + x_i^T \Omega x_i + \varepsilon_i^{(y)}, \,\,\,\,\,\,
\theta = vec\left(\sum_{l=1}^{H_1} \omega_l \beta_l^T\right), \,\,\,\,\,\, \Omega = \sum_{l = 1}^{H_2} B_{l} \otimes W_{l},
\end{equation}
where $B_1,...,B_{H_2} \in \mathbb{R}^{p \times p}$, $W_1,...,W_{H_2} \in \mathbb{R}^{T \times T}$, and $\otimes$ denotes a Kronecker product. To learn a fully-general matrix $\Omega \in \mathbb{R}^{pT \times pT}$, we would need $H_2 \geq \min\{p^2, T^2\}$. However, note that since $\Omega$ only appears in a quadratic form, its elements are not uniquely identified in the above model (e.g., letting $\Tilde{\Omega} = (\Omega + \Omega^T)/2$, we have that $x_i^T \Tilde{\Omega} x_i = x_i^T \Omega x_i$ for all $x_i \in \mathbb{R}^{pT}$). Thus, we can learn fully general interaction and quadratic coefficients using $H_2 = \min\{p(p+1)/2, T(T+1)/2\}$. To facilitate computation however, we suggest fixing $H_2 = 1$ in practice. In this case, if we organize the interaction terms between exposures $j_1$ and $j_2$ in a $T \times T$ matrix, that matrix can take on values given by any linear combination of $W$ and $W^T$. So in particular, if we let $\gamma_{j_1,t_1,j_2,t_2}$ be the interaction coefficient for $x_{i j_1 t_1} x_{i j_2 t_2}$, we have
$$\gamma_{j_1, t_1, j_2, t_2} = B_{j_1,j_2} W_{t_1,t_2} + B_{j_2,j_1} W_{t_2,t_1}.$$
While this restriction does not allow for completely general interaction terms, we demonstrate in Section \ref{section4} that our approach is still flexible enough to outperform competitors in simulations, even under misspecification. For $H_2 >1$, the right-hand side of the above equation simply becomes a sum of such expressions over $l$.

If $p$, $T$, and the correlations between the exposures are moderate, model (\ref{y_quadreg}) can be applied directly to learn an interpretable quadratic regression surface. However, in many data sets of interest, including the ELEMENT data we analyze here, some exposures may arise from a common source (for example, groups of exposures measured in urine may actually be metabolites of the same environmental toxin), leading to within-time correlations close to 1. In this case, regressing directly on $x_i$ is unstable due to collinearity. To handle this situation, we propose combining our regression approach with a dynamic factor model that reduces the dimensionality of the exposures.

\subsection{Bayesian factor analysis}\label{section3.2}
Factor analysis has a long history with roots in psychometrics \citep{spearman1904general}. For a comprehensive introduction to both frequentist and Bayesian factor modeling, see \cite{bartholomew2011latent}. Briefly, a factor model seeks to summarize a centered data matrix $X \in \mathbb{R}^{n \times p}$ as a matrix of $k-$dimensional latent vectors $\eta_1,...,\eta_n$ (where $k < p$) through the setup
$$x_i \sim N_p(\Lambda \eta_i, \Sigma),\quad \eta_i \sim N_k(0,I),\quad \text{ for } i = 1,...,n,$$
where $\Lambda \in \mathbb{R}^{p \times k}$ is referred to as the \textit{factor loadings matrix}, $\eta_1,...,\eta_n$ are called \textit{latent factors}, and $\Sigma = \text{diag}(\sigma_1^2, ..., \sigma_p^2)$. Additional constraints are required to make $\Lambda$ identifiable, but often we may be interested in inferences on the induced covariance for $X$, 
$$x_1,...,x_n \overset{iid}{\sim} N_p(0, \Lambda \Lambda^T + \Sigma)$$
in which case we do not need to impose any such constraints. In this covariance modeling setting, modern Bayesian methods have been developed to estimate $k$ along with the parameters of the model \citep{bhattacharya2011sparse, legramanti2020bayesian}. A key use of such covariance models is that if we also model an outcome of interest $y_i$ as a linear function of the latent factors, so that
$$y_i = \mu + \theta^T \eta_i + \epsilon_i, \,\,\,\, \epsilon_i \sim N(0, \sigma^2),$$
then we have induced a joint normal distribution for $(y_i, x_i)^T$ and thus can compute coefficients of a linear regression of $y_i$ on $x_i$ via standard multivariate normal theory \citep{west2003bayesian}. This is particularly useful as an alternative to regularization methods such as lasso regression \citep{tibshirani1996regression} when $X$ is high-dimensional and we do not wish to assume sparsity.

Recently, \cite{ferrari2021bayesian} showed that if one uses a quadratic regression model for $y_i$ on $\eta_i$ rather than a linear model, i.e.,
$$y_i = \mu + \theta^T \eta_i + \eta_i^T \Omega \eta_i + \epsilon_i, \,\,\,\, \epsilon_i \sim N(0, \sigma^2),$$
a corresponding quadratic regression is induced for $y_i$ on $x_i$, thus allowing for inferences about interaction effects in high-dimensional regression problems that would otherwise be intractable.

\subsection{Low-rank longitudinal factor regression}\label{section3.3}

Based on EDA for the ELEMENT exposure data, we propose that a reasonable covariance model for centered exposures $x_i$ is
\begin{equation}\label{xi_cov}
    x_i \sim N_{pT}(0, (\Lambda \Lambda^T + \Sigma) \otimes \Phi),
\end{equation}
where $\Lambda \in \mathbb{R}^{p \times k}$, $\Sigma = diag(\sigma_1^2,...,\sigma_p^2)$, and $\Phi \in \mathbb{R}^{T \times T}$ is a correlation matrix. Note that (\ref{xi_cov}) is the vectorized form of a matrix-normal model \citep{de1985matrix} with covariance matrices $(\Lambda \Lambda^T + \Sigma)$ and $\Phi$, and is related to the separable factor model of \cite{fosdick2014separable}. To induce such a model, letting $x_{it} \in \mathbb{R}^p$ be the exposures for individual $i$ and time $t$, we assign $x_{it}$ a factor model
\begin{eqnarray}\label{xit_model}
 x_{it} & = & \Lambda \eta_{it} + \varepsilon_{it}\\
 \eta_{it} \sim N_k(0, I), & & \varepsilon_{it} \sim N_p(0, \Sigma)\nonumber
\end{eqnarray}
for $t=1,...,T$. To induce the temporal correlation $\Phi$, we simply let each entry of $\eta_{it}$ and $\varepsilon_{it}$ have correlation $\Phi$ among times $t=1,...,T$. Defining $\eta_i = (\eta_{i11},...,\eta_{i1T},...,\eta_{ik1},...,\eta_{ikT})^T$ and $\varepsilon_i = (\varepsilon_{i11},...,\varepsilon_{i1T},...\varepsilon_{ip1},...,\varepsilon_{ipT})^T$, we can express all of this concisely in a single factor model for $x_i$:
\begin{eqnarray}\label{xi_model}
 x_{i} & = & (\Lambda \otimes I_T) \eta_{i} + \varepsilon_{i}\\
 \eta_{i} \sim N_{kT}(0, I_k \otimes \Phi), & & \varepsilon_{i} \sim N_{pT}(0, \Sigma \otimes \Phi).\nonumber
\end{eqnarray}
It is easily verified that model (\ref{xi_model}) implies the covariance model (\ref{xi_cov}) for $x_i$.

Now, given the latent factors from model (\ref{xi_model}), we can model the relationship between $y_i$ and $x_i$ by regressing $y_i$ on $\eta_i$. That is, we let
\begin{equation}\label{y_factor_reg}
    y_i \sim N(\mu + \theta^T \eta_i + \eta_i^T \Omega \eta_i, \sigma^2).
\end{equation}
Doing so, Theorem \ref{quad_reg_1} demonstrates that we have induced a quadratic regression of $y_i$ on $x_i$.

\begin{theorem}\label{quad_reg_1}
    Under the model in (\ref{xi_model}) - (\ref{y_factor_reg}),
    $$\mathbb{E}[y_i | x_i] = \mu + tr(\Omega V) + (\theta^T A) x_i + x_i^T (A^T \Omega A) x_i$$
    where
    $$V = (\Lambda^T \Sigma^{-1} \Lambda + I_k)^{-1} \otimes \Phi$$ and
    \begin{align*}
        A &= V (\Lambda^T \Sigma^{-1} \otimes \Phi^{-1})
        = (\Lambda^T \Sigma^{-1} \Lambda + I_k)^{-1}\Lambda^T \Sigma^{-1} \otimes I_T.
    \end{align*}
\end{theorem}

The proof of Theorem \ref{quad_reg_1} is given in Appendix \ref{app_th1_proof}, and closely mirrors the proof of Proposition 1 from \cite{ferrari2021bayesian}.

We note here that if the covariance structure in (\ref{xi_cov}) is too restrictive for a particular application, it can be generalized by giving $\eta_i$ and $\varepsilon_i$ more general correlation structures in (\ref{xi_model}) that could allow, for example, a temporal correlation structure that varies by exposure. We include a more general form of Theorem \ref{quad_reg_1} in the online supplement for completely general $\text{Cov}(\eta_i)$ and $\text{Cov}(\varepsilon_i)$. However, when applicable, the Kronecker covariance in (\ref{xi_cov}) yields some appealing properties of the induced regression coefficients for $y_i$ on $x_i$. In particular, if we again let
\begin{equation}\label{theta_omega}
    \theta = vec\left(\sum_{l=1}^{H_1} \omega_l \beta_l^T\right), \,\,\,\,\,\, \Omega = \sum_{l = 1}^{H_2} B_{l} \otimes W_{l}
\end{equation}
where now $\beta_l \in \mathbb{R}^k$ and $B_l \in \mathbb{R}^{k \times k}$, Corollary \ref{quad_reg_2} shows a correspondence between the structure of $\theta$ and $\Omega$ and that of the induced coefficients on $x_i$.

\begin{corollary}\label{quad_reg_2}
    Under the model in (\ref{xi_model}) - (\ref{theta_omega}),
    $$\mathbb{E}[y_i | x_i] = \alpha_0 + \alpha^T x_i + x_i^T \Gamma x_i$$
    where
    $$\alpha_0 = \mu + \sum_{l=1}^{H_2} tr(B_l \Tilde{V}) tr(W_l \Phi),\,\,\,\,\,\,\,\,\,\, \alpha = vec(\sum_{l=1}^{H_1} \omega_l \Tilde{\beta_l}^T), \,\,\,\,\,\,\,\,\,\, \Gamma = \sum_{l=1}^{H_2} \Tilde{B_l} \otimes W_l,$$
    and
    $$\Tilde{V} = (\Lambda^T \Sigma^{-1} \Lambda + I_k)^{-1}, \,\,\,\,\,\,\,\,\,\, \Tilde{A} = \Tilde{V} \Lambda^T \Sigma^{-1},$$
    $$\Tilde{\beta_l} = \Tilde{A}^T \beta_l, \,\,\,\,\,\,\,\,\,\, \Tilde{B_l} = \Tilde{A}^T B_l \Tilde{A}.$$
\end{corollary}

In the above expressions, note the form of the linear and quadratic coefficients $\alpha$ and $\Gamma$. In both cases, their structure mirrors the structure developed for $\theta$ and $\Omega$ with the decomposition along exposures and times. The time parameters $\omega_l$ and $W_l$ go through exactly, while the factor parameters $\beta_l$ and $B_l$ are modified from $\mathbb{R}^k$ and $\mathbb{R}^{k \times k}$ to $\mathbb{R}^p$ and $\mathbb{R}^{p \times p}$ based on the covariance structure of $x_{it}$. An appealing implication of this form is that modifying the $t$th index of $\omega_l$ for some $l$ only alters the induced coefficients corresponding to exposures at time $t$, and similarly, $W_{l,t_1, t_2}$ only affects the interaction parameters corresponding to times $t_1$ and $t_2$. Thus, the intuitive motivation for our low-rank model structure goes through exactly from the latent factor regression parameters to the induced coefficients. Note that this property does not hold for general $\text{Cov}(\eta_i)$ and $\text{Cov}(\varepsilon_i)$. Similarly, approaches for reducing the dimensionality of $x_i$ across both the exposure and time dimensions so that $\eta_i \in \mathbb{R}^{kT_0}$ for some $T_0 < T$ also do not share this appealing property. We present an additional alternate version of Theorem 3.1 for such an approach in the online supplement for the $x_i$ model of \cite{hung2012multilinear} and \cite{jiang2020bayesian}.

\subsection{Prior distributions and implementation details}\label{section3.4}

To finalize our Bayesian model specification, we assign prior distributions to all parameters. We place normal priors on the elements of $\Lambda$, the intercept term, and coefficients for any additional covariates we include in the model. To select the rank $k$ of $\Lambda$, we adapt the suggestion of \cite{ferrari2021bayesian} to our longitudinal setting and select $k$ such that $\sum_{j=1}^k \nu_j / \sum_{j=1}^p \nu_j > 0.9$, where the $\nu_j$s are singular values of the matrix of all $x_{it}$s. We place inverse-gamma priors on $\sigma^2$ and the diagonal elements of $\Sigma$. For our application of interest, we choose to model the correlation matrix $\Phi$ as compound symmetric, and place a $\text{Uniform}(0,1)$ prior on the correlation parameter $\phi$, thus assuming positive correlation for a given exposure over time. This compound symmetric structure makes sense for this setting, since exposure to a given compound within individuals is likely to arise from a constant source (e.g., due to daily product use or work environment) and thus is well-modeled as having constant correlation between any two measurement times. This structure is also supported by EDA in Section \ref{section5}. Note, however, that for other applications, this structure for $\Phi$ could easily be modified, for example to allow $\phi$ to range from $-1$ to $1$ rather than assuming non-negativity, or to allow different or more general correlation structures.

As mentioned above, we place a multiplicative gamma process prior \citep{bhattacharya2011sparse} on the $\omega_l$s and $\beta_l$s. Under this setup, each $\beta_{lj}$ and $\omega_{lt}$ is given a normal prior with the scale modeled as the product of a local term $\xi_{lj}$ and global term $\tau_l$, i.e., 
$$\beta_{lj} \sim N(0,\xi_{lj}^{-1} \tau_l^{-1}) \,\,\,\, \text{ for } l = 1,...,H_1, \,\,\,\, j=1,...,k$$
$$\omega_{lt} \sim N(0,\xi_{l(k+t)}^{-1} \tau_l^{-1}) \,\,\,\, \text{ for } l = 1,...,H_1, \,\,\,\, t=1,...,T.$$
The $\xi_{lj}$s are given independent gamma priors, while the $\tau_l$s are modeled as a running product of gamma random variables for increasing $l$, set up so that $\tau_l$ increases stochastically (and thus the global variance of the prior decreases) with $l$ at a rate learned from the data. Thus, if $\theta$ is well-modeled as rank-1, we expect $\tau_l^{-1}$ to shrink quickly for $l > 1$, while if $\theta$ requires a higher-rank approximation, $\tau_l^{-1}$ should shrink more slowly. We also propose modeling the $W_l$s and $B_l$s with a multiplicative gamma process prior. In the case we set $H_2 = 1$, as we do in our simulations, we still use the same setup but can drop the index $l$. Our full LowFR model setup as implemented in the next section is described in the online supplement, including all hyperparameter choices.

We implement LowFR using the probabilistic programming language Stan \citep{carpenter2017stan}. In general, we found that running four chains in parallel, each for 1000 burn-in iterations and 1000 subsequent posterior draws provided quite good mixing, both within and across chains. In simulations and real-data fits, we found R-hat values very close to 1 and effective sample sizes for the induced main and interaction effects almost always above 2000.

\section{Simulations}\label{section4}

To evaluate our LowFR approach, we compare it to five competitors: FIN \citep{ferrari2021bayesian} as implemented in the ``infinitefactor'' R package \citep{poworoznek2020package}, quadratic regression with a sparsity-inducing horseshoe prior on all coefficients \citep{carvalho2009handling} as implemented in the ``horseshoe'' R package \citep{van2016horseshoe}, BKMR \citep{bobb2015bayesian} with variable selection on all exposure measurements individually and BKMR\_group with hierarchical variable selection with one group for each exposure over all times, both as implemented with the ``bkmr'' R package \citep{bobb2018statistical}, and a Bayesian quadratic regression model with learned correlation parameters $\psi_{main}$ and $\psi_{int}$ for the correlations among main and interaction effects within exposures or pairs of exposures across times. This final approach, which we refer to as ``CorrQuadReg'', is well-suited to a data-generating process where the effects of given exposures are similar across times, since the correlation among regression parameters allows for information-sharing. The full model description for CorrQuadReg including all prior distributions is presented in the Supplemental Information. We did not consider approaches based on WQSR, as these do not provide full coefficient estimates with uncertainty quantification and are not designed to find interaction effects. We also did not consider approaches based on a distributed lag model, as they are designed for a much larger number of measurement times than we consider here.

We simulated data with $n=200$, $p=10$, and $T=3$ to match the structure of the motivating ELEMENT data set, and considered three data-generation scenarios. In scenarios 1 and 2, we generated the $x_i$s from our specified longitudinal factor model with $k=5$ factors at each time, temporal correlation $\phi = 0.5$, all elements of the loading matrix $\Lambda$ drawn from $N(0,1)$, and the variances $\sigma_1^2,...,\sigma_p^2$ all set to $0.25$. In scenario 1, we generated the outcomes $y_i$ as variance 1 Gaussian noise plus a quadratic function of the latent factors with rank-1 coefficients and interactions. We drew these coefficients by drawing $\omega_1$ and $W_1$ from flat Dirichlet distributions, and then randomly choosing two elements of $\beta_1$ and three elements of $B_1$ to make nonzero. The nonzero elements of $\beta_1$ and $B_1$ were drawn from $\text{Uniform}[(-2,-1) \cup (1,2)]$. In scenario 2, we repeated the above but also generated $\omega_2$ and $\beta_2$ in the same manner to yield rank-2 main effects.

In scenario 3, we generated $x_1,...,x_n \sim N_{pT}(0, \Phi^{(Exp)} \otimes \Phi^{(Time)})$, where $\Phi^{(Exp)} \in \mathbb{R}^{p \times p}$ and $\Phi^{(Time)} \in \mathbb{R}^{T \times T}$ are both compound symmetric correlation matrices with correlation parameter $0.7$. We generated outcomes $y_i$ as variance 1 Gaussian noise plus a quadratic function of $x_i$. To generate coefficients, we randomly selected 4 exposures to have nonzero main effects (at all times) and 10 pairs of exposures to have nonzero interactions (at all times). Then, for each main effect exposure, we drew a mean main effect from $\text{Uniform}[(-0.4,-0.2) \cup (0.2,0.4)]$ and then drew the coefficients at each time as that mean plus Gaussian noise with standard deviation $0.05$. We generated interactions the same way, but with means drawn from $\text{Uniform}[(-0.15,-0.05) \cup (0.05,0.15)]$ and noise standard deviation $0.01$. This data-generation scheme simulates a setting in which the effects are not low-rank, but where main and interaction coefficients within a given exposure or pair of exposures are related. Note that CorrQuadReg is very well-suited for Scenario 3, while LowFR is misspecified both in the factor model for $x_i$ and the assumption of low-rank coefficients.

To compare the accuracy of the models, we calculated the mean squared estimation error across the main effects and interactions, using posterior means as coefficient estimates. We computed coverage as the fraction of coefficients for which $95\%$ posterior credible intervals included the true value. For the main effects, we also computed true positive (TP) rates as the fraction of $95\%$ credible intervals excluding zero and having the correct sign among coefficients with nonzero true values, and true negative (TN) rates as the fraction of $95\%$ credible intervals including zero among coefficients with a true value of zero. Since BKMR estimates a general regression surface rather than coefficients, we computed BKMR ``main effects'' as the change in expected outcome when a given $x_{jt}$ is increased from $-0.5$ to $0.5$ with all other exposures held constant at $0$; this calculation for all of the quadratic regression approaches simply yields the appropriate linear coefficient. Similarly, we computed BKMR ``interaction effects'' as the difference between the change in expected outcome when a given $x_{jt}$ is increased from $-0.5$ to $0.5$ when all other exposures are $0$ vs. when a single interacting exposure has a value of $1$. Finally, given that in these settings, part of the goal is to identify chemicals that have an association with the outcome, regardless of the particular timing, we also calculated MSE, coverage, and TP and TN rates for estimated effects of cumulative exposure over all times. We defined the cumulative exposure effect value as the expected increase in the outcome when a given exposure was increased from $-1$ to $1$ at all three times, with all other exposures held constant at $0$ for all times. Note that the particular choice of change from $-1$ to $1$ is somewhat arbitrary, and other values could be used to summarize bigger or smaller shifts in exposure profiles in cases where those are of interest.

For all scenarios, we fit LowFR in Stan using 4 chains, each for 1000 burn-in and 1000 subsequent samples. Mixing was good, with all effective sample sizes for main and interaction effects at least 600 across all simulations, and most of them much higher.

The results of these simulations averaged over 100 replications are shown in Table \ref{table1}. Observe that LowFR has the lowest MSE for both main and interaction effects in all three scenarios, and the coverage of LowFR's $95\%$ posterior intervals is quite good. In Scenarios 1 and 2, LowFR also has the highest main effect TP rate, and substantially outperforms competitors in all cumulative effect evaluations. In Scenario 3 where LowFR is misspecified, the main effect TP rate is lower than that of CorrQuadReg, and the misspecification appears to affect cumulative effect accuracy, for which LowFR has the highest MSE among all models. However, note that the cumulative effect TN rate for LowFR is still almost 1, suggesting that the misspecification does not lead to high rates of false positive effects. FIN has almost perfect coverage across all three scenarios, but much higher MSE than LowFR. BKMR, BKMR\_group, and Horseshoe also have higher MSE, and their main effect coverage in Scenarios 1 and 2 is quite poor compared to LowFR, FIN, and CorrQuadReg. 
This is consistent with over-aggressive variable selection or shrinkage by BKMR, BKMR\_group, and Horseshoe with some nonzero coefficients  effectively zeroed out in the analysis.

\begin{table*}
\caption{Comparison of six models for simulated data with $n=200$, $p=10$, and $T=3$. Longitudinal exposures in Scenarios 1 and 2 were generated from a factor model as in the LowFR setup. Scenario 1 outcomes were generated with rank-1 main effects and interactions coefficients, while Scenario 2 outcomes had rank-2 main effects. Scenario 3 exposures were generated as multivariate normal with a Kronecker covariance, with main and interaction coefficients for outcomes generated as constant plus noise within exposures and pairs of exposures. CE indicates cumulative effects, defined as the expected change in outcome if a single exposure increases from $-1$ to $1$ at all measurement times, with all other exposures held constant at zero.}
\resizebox{\textwidth}{!}{%
\begin{tabular}{llllllll}
\hline
                     &               & LowFR                        & BKMR                         & BKMR\_group                  & FIN                          & Horseshoe                    & CorrQuadReg                  \\ \hline
\multirow{10}{*}{S1} & Main MSE      & 1.1 * 10\textasciicircum{}-3 & 2.5 * 10\textasciicircum{}-2 & 2.8 * 10\textasciicircum{}-2 & 3.8 * 10\textasciicircum{}-2 & 9.0 * 10\textasciicircum{}-3 & 5.6 * 10\textasciicircum{}-3 \\
                     & Int MSE       & 3.4 * 10\textasciicircum{}-5 & 2.8 * 10\textasciicircum{}-1 & 2.5 * 10\textasciicircum{}-1 & 1.6 * 10\textasciicircum{}-3 & 1.2 * 10\textasciicircum{}-4 & 1.3 * 10\textasciicircum{}-3 \\
                     & Main Coverage & 1.00                         & 0.58                         & 0.44                         & 0.98                         & 0.64                         & 0.98                         \\
                     & Int Coverage  & 1.00                         & 0.46                         & 0.34                         & 1.00                         & 0.97                         & 0.97                         \\
                     & Main TP Rate  & 0.30                         & 0.071                        & 0.074                        & 0.078                        & 0.055                        & 0.17                         \\
                     & Main TN Rate  & N/A                          & N/A                          & N/A                          & N/A                          & N/A                          & N/A                          \\ \cline{2-8} 
                     & CE MSE        & 0.020                        & 0.24                         & 0.18                         & 0.89                         & 0.12                         & 0.14                         \\
                     & CE Coverage   & 1.00                         & 0.80                         & 0.88                         & 0.94                         & 0.63                         & 0.96                         \\
                     & CE TP Rate    & 0.53                         & 0.24                         & 0.27                         & 0.16                         & 0.22                         & 0.29                         \\
                     & CE TN Rate    & N/A                          & N/A                          & N/A                          & N/A                          & N/A                          & N/A                          \\ \hline
\multirow{10}{*}{S2} & Main MSE      & 2.2 * 10\textasciicircum{}-3 & 4.3 * 10\textasciicircum{}-2 & 6.4 * 10\textasciicircum{}-2 & 5.4 * 10\textasciicircum{}-2 & 2.0 * 10\textasciicircum{}-2 & 8.1 * 10\textasciicircum{}-3 \\
                     & Int MSE       & 4.0 * 10\textasciicircum{}-5 & 4.0 * 10\textasciicircum{}-1 & 4.1 * 10\textasciicircum{}-1 & 1.7 * 10\textasciicircum{}-3 & 1.2 * 10\textasciicircum{}-4 & 1.4 * 10\textasciicircum{}-3 \\
                     & Main Coverage & 0.99                         & 0.58                         & 0.36                         & 0.96                         & 0.58                         & 0.98                         \\
                     & Int Coverage  & 1.00                         & 0.45                         & 0.29                         & 1.00                         & 0.97                         & 0.96                         \\
                     & Main TP Rate  & 0.38                         & 0.10                         & 0.11                         & 0.14                         & 0.092                        & 0.23                         \\
                     & Main TN Rate  & N/A                          & N/A                          & N/A                          & N/A                          & N/A                          & N/A                          \\ \cline{2-8} 
                     & CE MSE        & 0.027                        & 0.41                         & 0.33                         & 1.3                          & 0.19                         & 0.17                         \\
                     & CE Coverage   & 0.99                         & 0.78                         & 0.82                         & 0.90                         & 0.64                         & 0.96                         \\
                     & CE TP Rate    & 0.60                         & 0.31                         & 0.39                         & 0.26                         & 0.31                         & 0.40                         \\
                     & CE TN Rate    & N/A                          & N/A                          & N/A                          & N/A                          & N/A                          & N/A                          \\ \hline
\multirow{10}{*}{S3} & Main MSE      & 9.9 * 10\textasciicircum{}-3 & 4.2 * 10\textasciicircum{}-2 & 6.3 * 10\textasciicircum{}-2 & 2.2 * 10\textasciicircum{}-2 & 2.9 * 10\textasciicircum{}-2 & 1.0 * 10\textasciicircum{}-2 \\
                     & Int MSE       & 1.4 * 10\textasciicircum{}-3 & 3.3 * 10\textasciicircum{}-1 & 3.2 * 10\textasciicircum{}-1 & 7.6 * 10\textasciicircum{}-3 & 2.2 * 10\textasciicircum{}-3 & 3.4 * 10\textasciicircum{}-3 \\
                     & Main Coverage & 0.97                         & 0.97                         & 0.77                         & 1.00                         & 0.90                         & 1.00                         \\
                     & Int Coverage  & 0.95                         & 0.63                         & 0.51                         & 1.00                         & 0.98                         & 1.00                         \\
                     & Main TP Rate  & 0.27                         & 0.12                         & 0.17                         & 0.00                         & 0.10                         & 0.51                         \\
                     & Main TN Rate  & 1.00                         & 1.00                         & 1.00                         & 1.00                         & 1.00                         & 1.00                         \\ \cline{2-8} 
                     & CE MSE        & 0.27                         & 0.19                         & 0.18                         & 0.27                         & 0.14                         & 0.14                         \\
                     & CE Coverage   & 0.84                         & 0.98                         & 0.97                         & 1.00                         & 0.90                         & 0.96                         \\
                     & CE TP Rate    & 0.83                         & 0.88                         & 0.87                         & 0.00                         & 0.84                         & 0.97                         \\
                     & CE TN Rate    & 0.98                         & 0.98                         & 0.98                         & 1.00                         & 1.00                         & 0.96                         \\ \hline
\end{tabular}}
\label{table1}
\end{table*}

\begin{figure}
\includegraphics[width=0.85\textwidth]{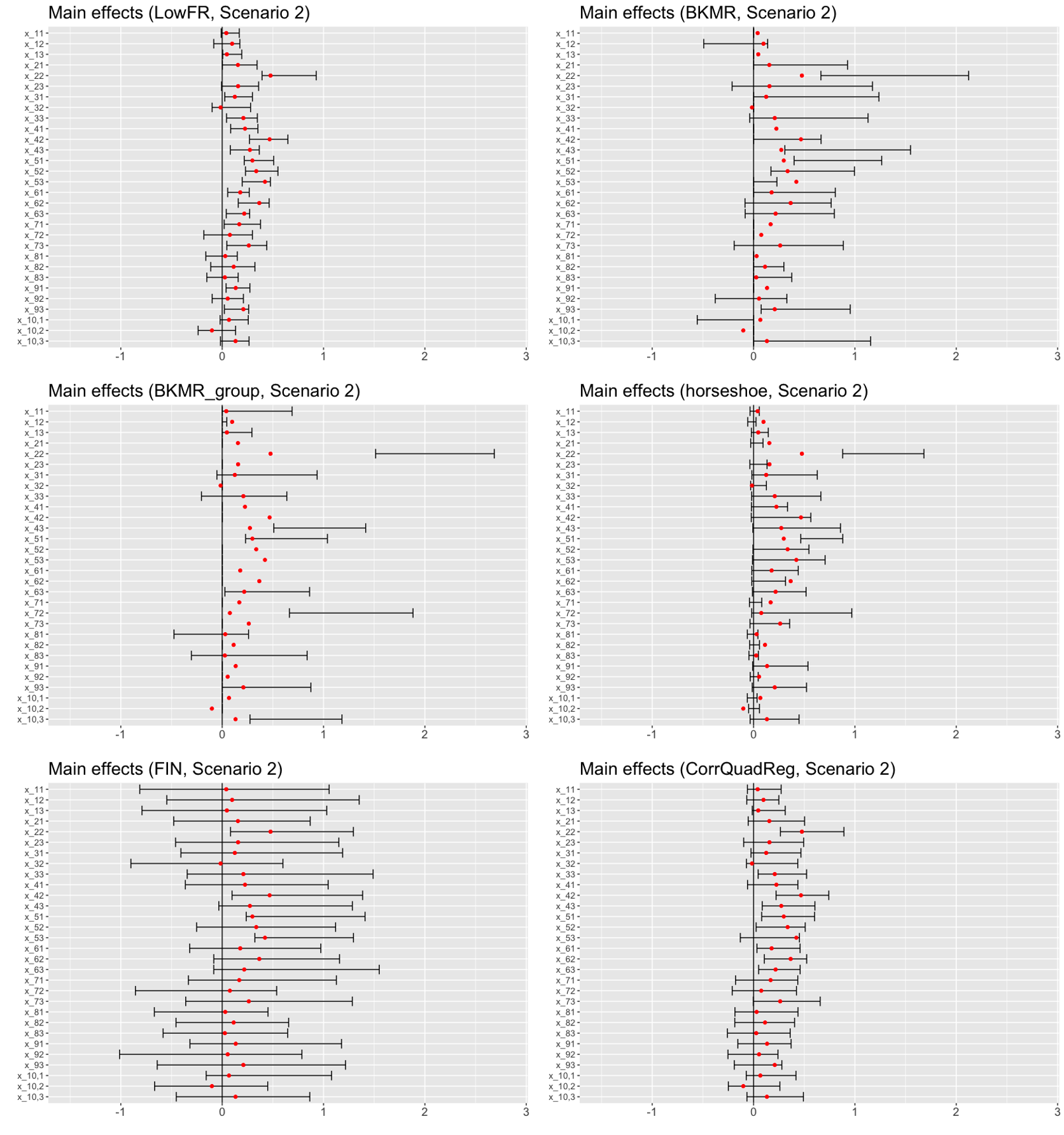}
\caption{Estimated main effect 95\% posterior credible intervals for Scenario 2 simulated data. Red dots indicate true values of coefficients.}
\label{fig_s2_main}
\end{figure}

To better understand these results, Figure \ref{fig_s2_main} shows true coefficients and estimated $95\%$ posterior credible intervals for the main effects in one simulation under Scenario 2. The intervals for FIN look reasonable, but are quite wide. The shrinkage or variable selection approaches of Horseshoe and variants of BKMR overestimate some coefficients (in particular, they dramatically overestimate the effect of $x_{22}$), while zeroing out or underestimating others. In contrast, LowFR successfully shrinks intervals toward the true values, while handling the rank-2 coefficients by not overshrinking toward a rank-1 setup. CorrQuadReg does a reasonable job of estimating coefficients, with narrower intervals than BKMR, FIN, or Horseshoe due to the information sharing induced by the correlated coefficients, but it still has wider intervals than LowFR since it fails to take advantage of the effects being less than full rank.

\begin{figure}[b]
\includegraphics[width=0.85\textwidth]{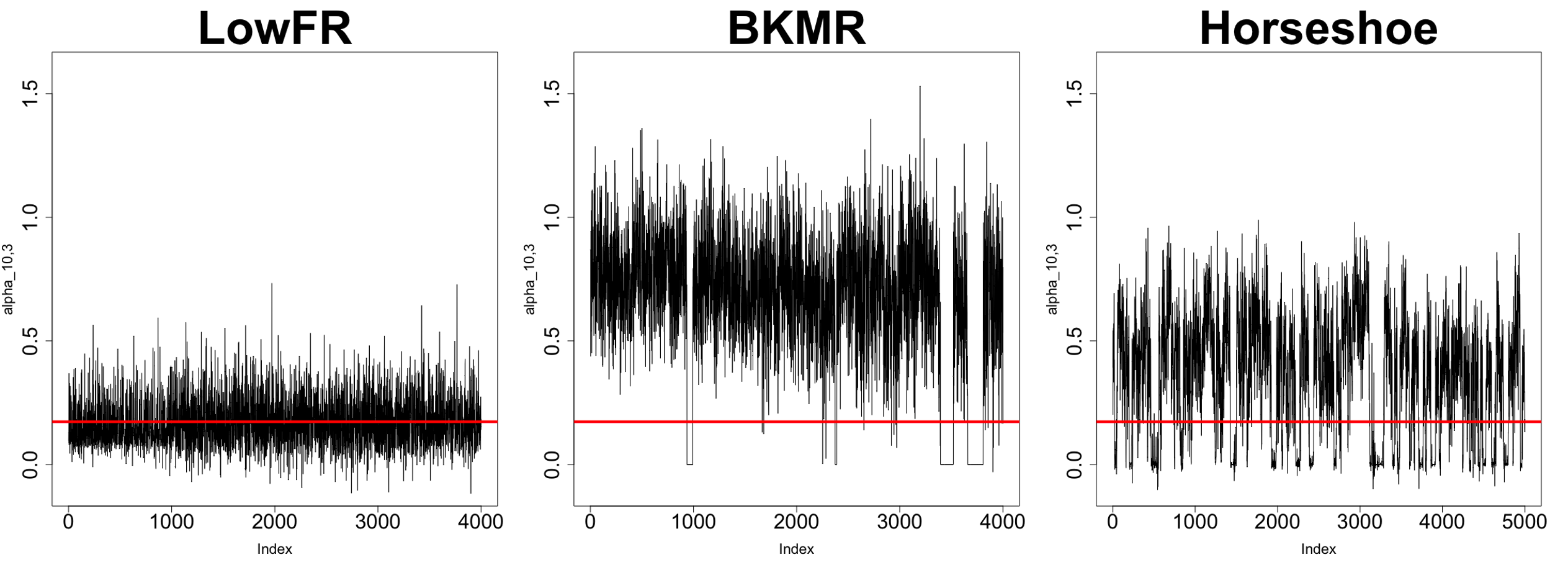}
\caption{Trace plots for the main effect of $x_{10,3}$ for LowFR, BKMR, and Horseshoe for Scenario 3 data.}
\label{fig_s3_trace}
\end{figure}

When viewing the intervals in Figure \ref{fig_s2_main}, it is worth noting that for Horseshoe and both versions of BKMR, many of the wider intervals reflect a mode-switching in posterior samples between overestimation and underestimation, with little posterior mass assigned near the true value. This partially explains the higher coverage but worse MSE for Horseshoe compared to LowFR for interaction effects in Scenario 3. To illustrate this, Figure \ref{fig_s3_trace} shows posterior trace plots for LowFR, BKMR, and Horseshoe for the main effect of $x_{10,3}$ in Scenario 3. All three resulting equal-tailed intervals include the true coefficient, but the posterior density for LowFR is much more concentrated around the true value.

\begin{figure}[t]
\includegraphics[width=\textwidth]{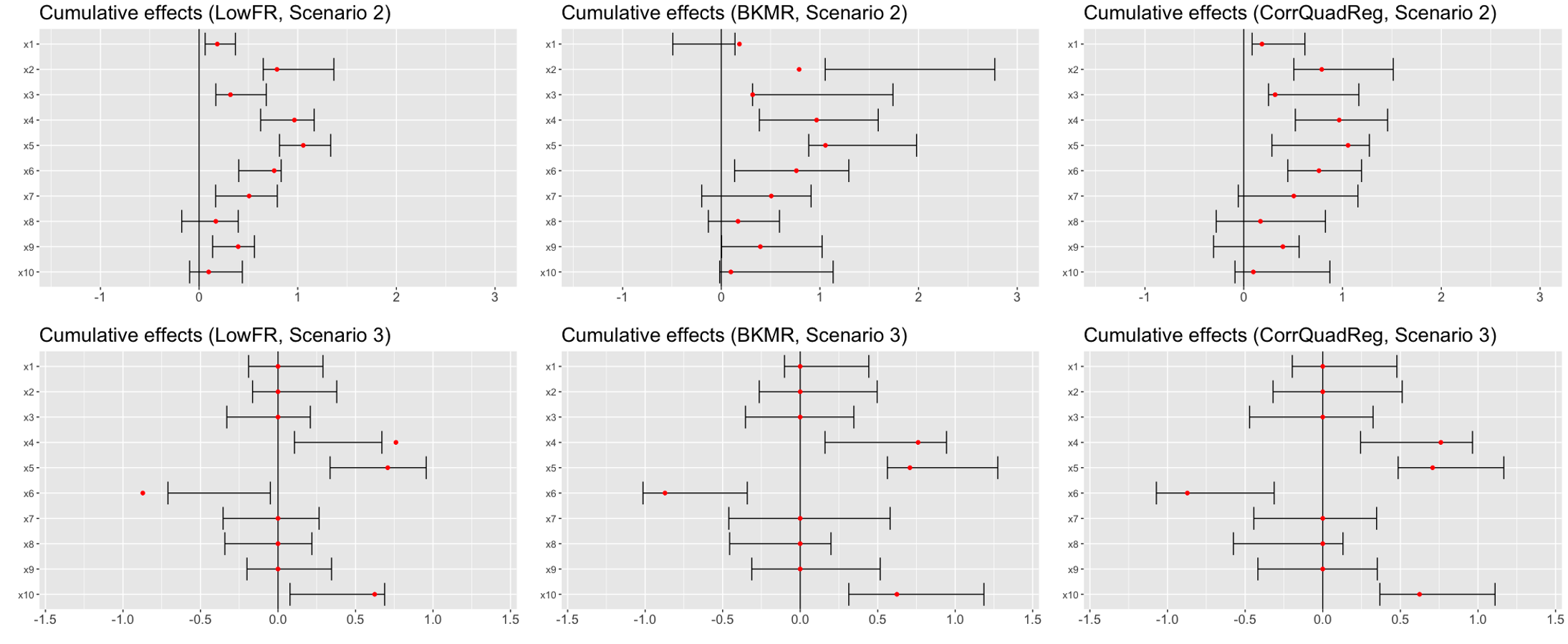}
\caption{$95\%$ posterior credible intervals for cumulative effects of the 10 exposures in a single simulation for LowFR, BKMR, and CorrQuadReg fit to Scenario 2 and Scenario 3 simulated data. Red dots indicate true values.}
\label{cum_exp_s2s3}
\end{figure}

Finally, to illustrate how LowFR fares in estimating cumulative effects of chemicals, Figure \ref{cum_exp_s2s3} shows true values and $95\%$ posterior intervals for LowFR, BKMR, and CorrQuadReg for one run each of Scenarios 2 and 3. In Scenario 2, LowFR both detects the most true effects as significant, and estimates them the most accurately with the narrowest intervals. In Scenario 3, LowFR badly underestimates the magnitude of the effects for $x_4$ and $x_6$, and generally has intervals that are closer to zero than those of BKMR or CorrQuadReg. However, all intervals that exclude zero correspond with truly nonzero effects of the correct signs. This trend is maintained across the simulation replicates, and as shown in Table \ref{table1}, these results suggest that when overall effects of chemicals are of interest, if LowFR is well-suited for the data, it can detect more effects more accurately than competitors. Meanwhile, when LowFR is misspecified as in Scenario 3, it will possibly miss or underestimate some effects, but the effects it does find are likely to be true associations, as illustrated by its TN rate of 0.98.

Before proceeding, we note that while these simulations were designed for our particular application in which $T=3$, many similar studies involve a higher, but still moderate, number of measurement times. To evaluate LowFR's ability to handle such situations, we have included results in the Supplemental Information for the same data generation mechanism as Scenario 2 (i.e., rank-2 linear terms and Kronecker interactions), but for $T=p=10$. In this higher-$T$ scenario, LowFR still substantially outperforms all other models with respect to main, interaction, and cumulative effect accuracy and coverage, suggesting that it continues to capture low-rank structure even as dimensionality grows.

\section{Application}\label{section5}

\subsection{Exploratory data analysis}\label{EDA}

\begin{figure}[t]
\includegraphics[width=0.82\textwidth]{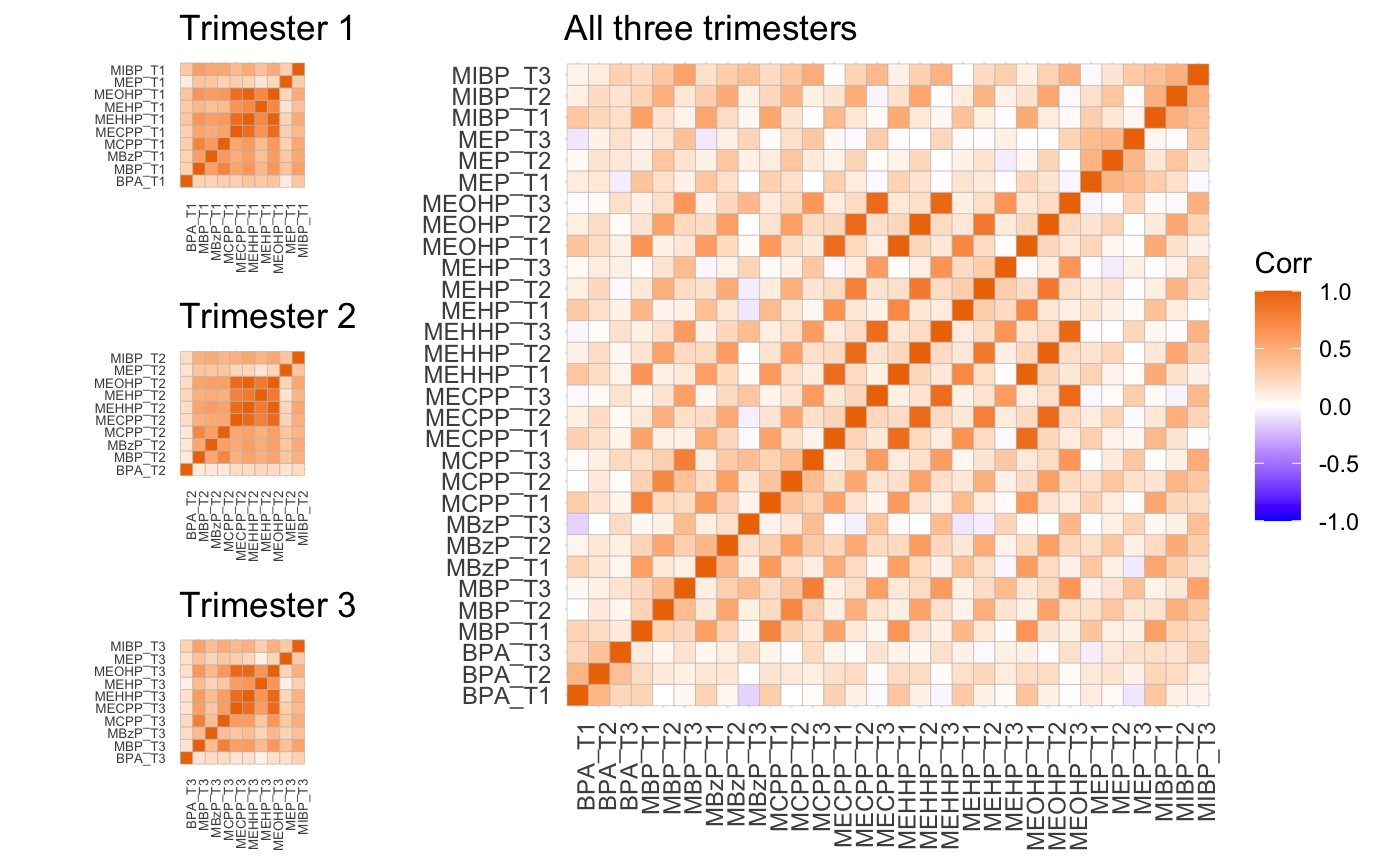}
\caption{Pearson correlation matrices for the exposure levels of BPA and 9 phthalates in the ELEMENT data across trimesters 1, 2, and 3.}
\label{fig_corr}
\end{figure}

Before fitting our model to the ELEMENT data, we did some data visualization to evaluate the assumptions of the model. In Figure \ref{fig_corr}, we show the Pearson correlations for the exposure data, both across all times and separately for each trimester. We note two things from this visualization. First, the within-trimester correlation structure appears consistent across the three trimesters, supporting our approach of using a single factor loadings matrix $\Lambda$ across time. Second, all of the highest correlations are across exposures within times, with some values as high as 0.99. In contrast, the across-time within-chemical correlations are all positive but moderate, with none exceeding 0.5. Taken together, these observations lend support to the Kronecker covariance structure in (\ref{xi_cov}) being reasonable for these data.

\subsection{Model fitting}

We fit LowFR to the ELEMENT data with standardized 2-DG as the outcome and the exposure data processed as described in Section \ref{section2}. We 
specified prior distributions as described in the supplemental information, letting $k=7$ based on $\sum_{j=1}^7 \nu_j / \sum_{j=1}^{10} \nu_j \approx 0.9$. We also added linear terms for sex, age at followup, BMI z-score, and pubertal onset following the analysis of \cite{goodrich2022trimester}. We assigned these additional coefficients $N(0,10)$ prior distributions. As above, we fit the model with Stan \citep{carpenter2017stan}. To handle missing data, we simply imputed values for all missing exposure measurements at each MCMC step based on our multivariate normal model for $x_i$. In Stan, this is accomplished by treating the entries of each $x_i$ as a combination of known data (for observed exposures) and parameters to be learned (representing any missing values). Fully reproducible code for the analysis is provided in the online supplement. We ran four chains, each for 1000 burn-in iterations and 1000 subsequent posterior samples. Convergence and mixing was good, with all Rhat values below $1.02$ and all bulk and tail effective sample sizes above $2100$ across all induced main and interaction coefficients.

For comparison, we also fit BKMR and CorrQuadReg to the same data. We implemented BKMR using the ``bkmr'' R package \citep{bobb2018statistical} for 4000 burn-in iterations and 4000 subsequent samples, and with the variable selection option selected. We fit CorrQuadReg using Stan, running four chains for 1000 burn-in and 1000 subsequent samples. The ``bkmr'' package has a built-in option to include additional linear covariate effects; for CorrQuadReg, we added these terms to the Stan model with $N(0,10)$ priors as in our LowFR implementation. Since neither of these approaches includes a model for $x_i$, we used MICE \citep{van2011mice} to impute all missing exposure values before analysis.

\begin{figure}[h]
\includegraphics[width=\textwidth]{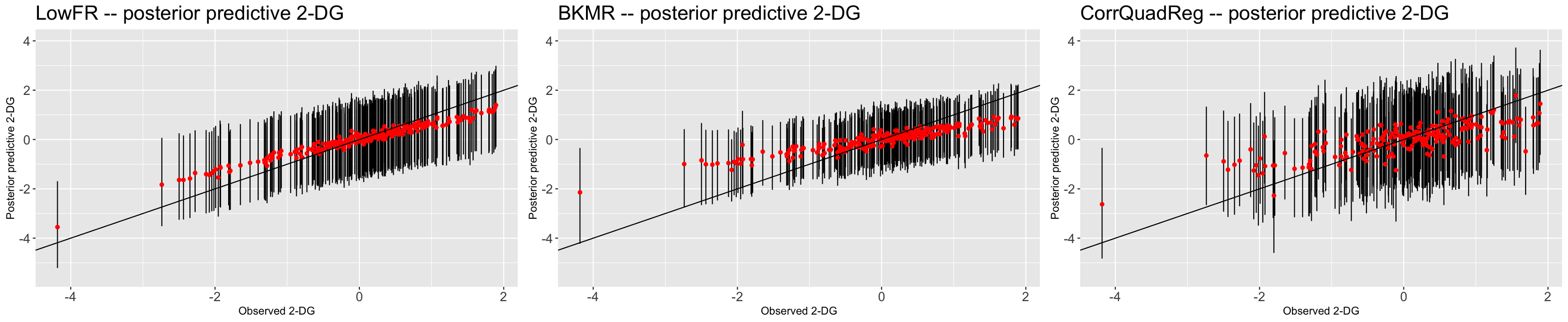}
\caption{Posterior predictive mean 2-DG values and associated $95\%$ predictive intervals for each child, plotted against observed 2-DG values.}
\label{fig_ppcheck_perpatient}
\end{figure}

To evaluate how well the models fit the data, we ran posterior predictive checks. Posterior predictive distributions for the marginal distribution of 2-DG are shown in the online supplement, and indicate that LowFR fits the data reasonably well, while BKMR tends to assign far too much density close to zero, and CorrQuadReg has a left-skewed peak and overestimates the variance. The posterior predictive distributions for each child's 2-DG level are shown in Figure \ref{fig_ppcheck_perpatient}, plotted against the true values. 
The estimates for LowFR tend to be closest to the observed values, indicating that the dimension reduction of our low-rank factor regression approach does not keep LowFR from being sufficiently expressive to capture the variation in the data. We also ran 10-fold cross-validation for each model to check for overfitting. Treating the out-of-sample MSE of LowFR as the reference, BKMR had 5\% higher MSE and CorrQuadReg had 45\% higher MSE.
In combination with the in-sample fit shown in Figure \ref{fig_ppcheck_perpatient}, these results suggest that LowFR does the best job of parsimoniously fitting the data.

\subsection{Posterior summaries}

We first computed posterior summaries of the linear covariate effects, shown in Table \ref{table_covariates}. For LowFR, the coefficient for female sex has a posterior mean of $0.27$ with a strictly positive $95\%$ credible interval, suggesting that female children have 
higher 2-DG levels on average accounting for the exposures and other covariate effects. The posterior means for female sex are also positive for BKMR and CorrQuadReg, but both of their $95\%$ CIs include zero. All $95\%$ intervals for age, BMI z-score, and pubertal onset include zero across all models.

\begin{figure}[b]
\includegraphics[width=\textwidth]{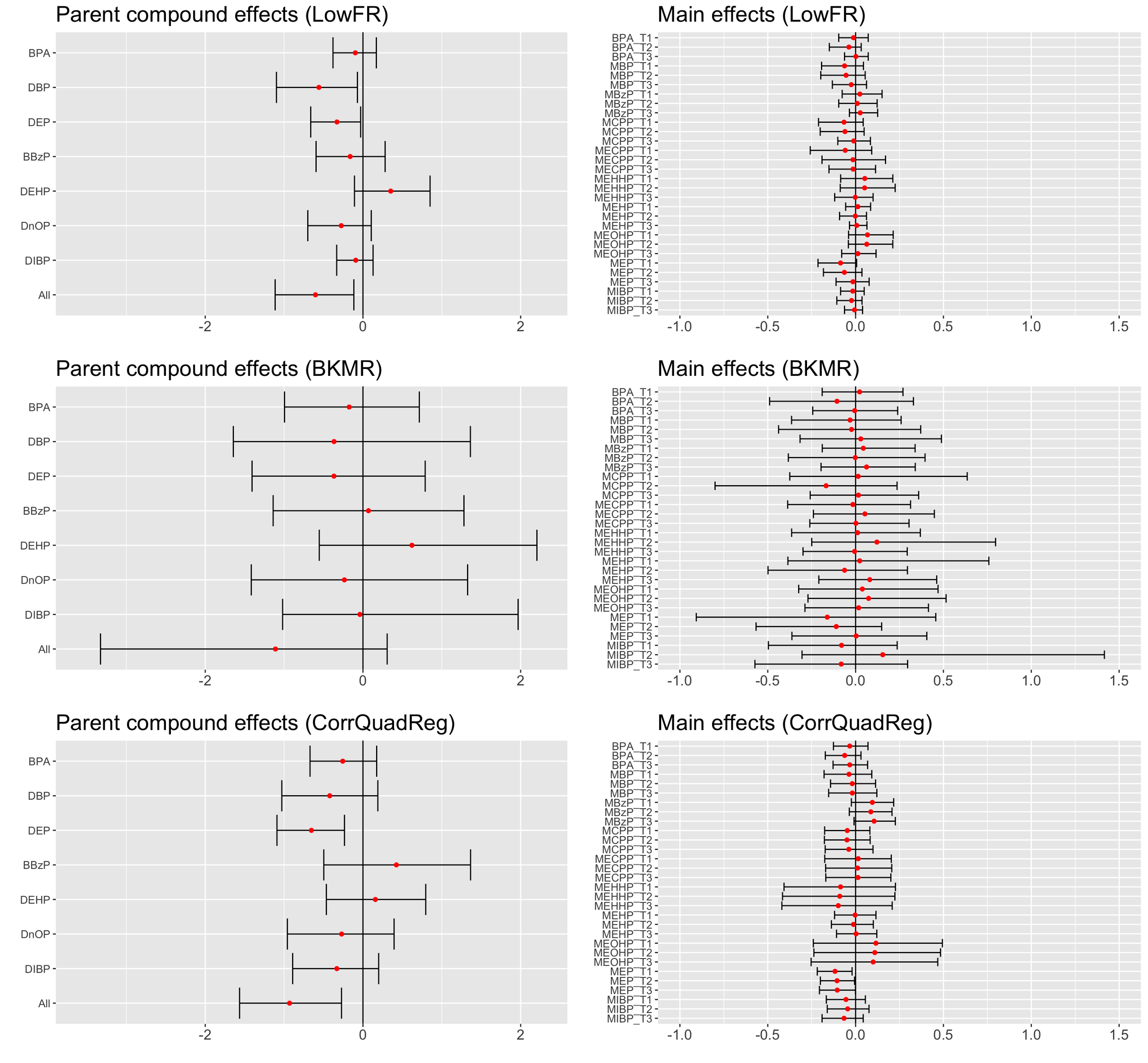}
\caption{Left: The expected change in 2-DG level when all metabolites of the given phthalates (listed in Table 1) are are increased from $-1$ to $1$ at all three trimesters, with all other exposures fixed to a value of $0$. Right: Main effect coefficients for all measured exposures. For BKMR, these main effects are calculated as the expected change in 2-DG when the given exposure increases from $-0.5$ to $0.5$ with all other exposures held constant at zero. In all plots, red dots indicate posterior means, and intervals represent $95\%$ posterior credible intervals.}
\label{fig_parent_main}
\end{figure}

To evaluate the chemical exposure effects, we began by examining the cumulative effects of BPA and the combined metabolites of each parent compound listed in Table \ref{table_parentlist}. The left column of Figure \ref{fig_parent_main} shows posterior means and $95\%$ CIs for the expected change in 2-DG when all metabolites of each given compound increase from $-1$ to $1$ (i.e., from 1 standard deviation below their mean to 1 standard deviation above their mean) at all three trimesters, with all other exposures fixed to 0 (i.e., to their mean values). The BPA intervals represent BPA increasing from $-1$ to $1$ at all times, and the ``All'' intervals represent all measured exposures increasing from $-1$ to $1$ at all times. Given the low signal-to-noise ratio and sample size in this and similar studies, these cumulative effects give a way to gain power and generate actionable insights about potentially harmful compounds by pooling information across measurements, while also flagging a subset of exposures to examine more closely.

\begin{table*}[b]
\caption{Posterior means and $95\%$ credible intervals for covariate effects in LowFR, BKMR, and CorrQuadReg models fit to the ELEMENT data with 2-DG as the outcome.}
\begin{tabular}{llll}
\hline
      & LowFR              & BKMR                & CorrQuadReg        \\ \hline
Sex.F   & 0.27 (0.01, 0.53)  & 0.17 (-0.07, 0.43)  & 0.15 (-0.15, 0.44) \\
Age   & 0.11 (-0.06, 0.27) & 0.10 (-0.06, 0.27)  & 0.07 (-0.12, 0.26) \\
BMI   & 0.09 (-0.03, 0.22) & 0.07 (-0.06, 0.20)  & 0.00 (-0.14, 0.14) \\
Onset & 0.03 (-0.31, 0.38) & -0.01 (-0.32, 0.31) & 0.14 (-0.24, 0.53) \\ \hline
\end{tabular}
\label{table_covariates}
\end{table*}

For both LowFR and CorrQuadReg, the All and DEP intervals are strictly negative, suggesting that overall higher exposure to the measured chemicals is associated with lower followup 2-DG levels, and that in particular, MEP (the only measured metabolite of DEP) is negatively associated with 2-DG. This aligns with the analysis of \cite{goodrich2022trimester} for first-trimester MEP, and with the laboratory study of \cite{guven2016low}, which suggested that MEP can induce higher glucose transport activity out of the blood. Interestingly, the interval for DBP (metabolites MBP and MCPP) is also strictly negative for LowFR, but not for CorrQuadReg. This result for LowFR also aligns with \cite{goodrich2022trimester}, who found that first-trimester MBP and MCPP were each associated with lower 2-DG. All intervals for BKMR include zero, even for the overall effect.

To get a sense of the individual effects driving these cumulative inferences, the right column of Figure \ref{fig_parent_main} shows main effects for all measured exposures at all times. The intervals for BKMR are by far the widest, corresponding to BKMR's wide cumulative effect intervals. The differences between the main effects for LowFR and CorrQuadReg give some insights that can inform interpretation of their differing cumulative effect estimates. Observe that for CorrQuadReg, the posterior means for coefficients within chemicals across trimesters are extremely similar. This makes sense given the structure of the model, which allows for information sharing across times via a learned correlation parameter $\psi_{main}$, but not across chemicals. Aligning with this observation, $\psi_{main}$ has a posterior mean of $0.98$. Similarly $\psi_{int}$, the correlation for within-exposure-pair interaction terms, has a posterior mean of $0.99$. By allowing information sharing among times but not among chemicals, CorrQuadReg shrinks the effects of chemicals over time so strongly toward each other that it can smooth out important temporal effects. Even if overall cumulative effects are of interest, this smoothing can still be harmful. In particular, if two chemicals both have an effect that is strongest at a particular time, CorrQuadReg may handle this by estimating too-high consistent effects for one chemical across all times and too-low consistent effects for the other chemical. This is particularly a concern when the chemicals are correlated. This phenomenon may contribute to the wide performance gap between LowFR and CorrQuadReg for cumulative effects in Scenarios 1 and 2 in Section \ref{section4}. Also note that this concern is not specific to CorrQuadReg, but applies for any model that handles data like these by shrinking time-specific regression surfaces toward each other.

In contrast to CorrQuadReg, since LowFR flexibly shares information across both times and chemicals through its rank decomposition of coefficients, it can easily handle a situation where a group of exposures have shared temporal structure in their effects. Observe in Figure \ref{fig_parent_main} that although all main effect intervals for LowFR include zero, all of MEP, MBP, and MCPP have posterior mean effects with the highest magnitude at trimester 1, then trimester 2, and then a much smaller magnitude for trimester 3. This is a useful insight on its own, but having this shared structure also allows LowFR to find moderate but significant cumulative effects for both DEP and DBP, rather than only a single large effect for one of them.

\begin{figure}[b]
\includegraphics[width=0.8\textwidth]{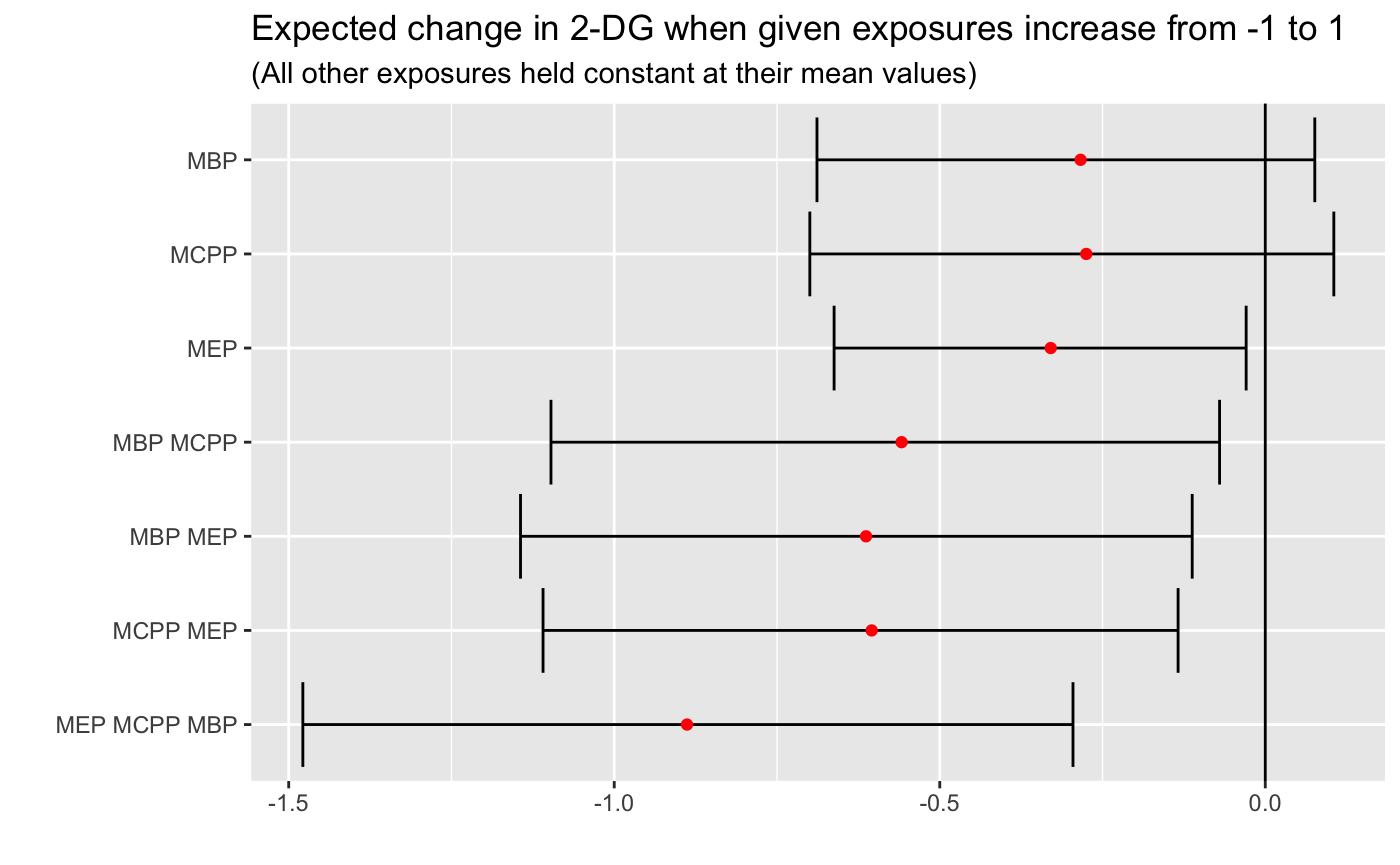}
\caption{Expected change in 2-DG level when the noted combinations of exposures are increased from -1 to 1 at all three trimesters, with all other exposures fixed to a value of 0. The red dots indicate posterior means, and the intervals represent $95\%$ posterior credible intervals.}
\label{fig_time_sums}
\end{figure}

Given these initial inferences, we can use the fitted LowFR model to more closely examine the joint effects of MEP, MBP, and MCPP. In Figure \ref{fig_time_sums}, we show the expected change in 2-DG when each possible combination of these metabolites increases from $-1$ to $1$ jointly at all times. Although MEP has a significantly negative cumulative effect corresponding to the DEP interval in Figure \ref{fig_parent_main}, intervals for the individual effects of MBP and MCPP both include zero. However, intervals for all combinations of two or more of these chemicals do exclude zero, and have increasing magnitude as more are added.

\begin{figure}[t]
\includegraphics[width=\textwidth]{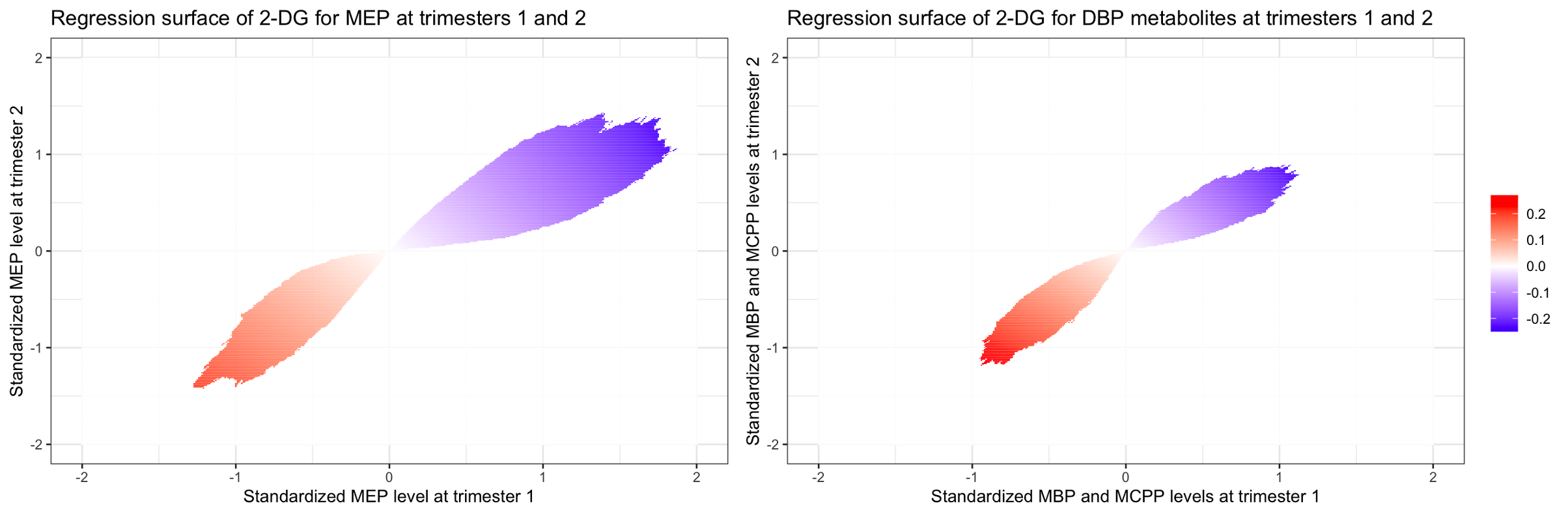}
\caption{Regression surface of 2-DG on trimester 1 and 2 MEP, and on trimester 1 and 2 MCPP and MBP varying together. Each surface is after linear correction for covariates and with all other chemical exposure values set to their mean. White regions indicate regions where the $95\%$ interval for the regression surface includes $0$.}
\label{fig_mep_dbp_surface}
\end{figure}

Based on these results, we can also examine the role of time more closely for MEP, MBP, and MCPP exposure. Since all intervals for individual trimester effects included zero, we focus on pairs of trimesters. Similarly, we focus on joint exposure of MBP and MCPP rather than either individually based on the intervals in Figure \ref{fig_time_sums}. In addition to their being the two metabolites of DBP, MBP and MCPP have an average within-trimester correlation of $0.78$ and only moderate correlations with any other chemicals, so treating them as varying together is relevant for understanding realistic exposure profiles. For each of MEP and DBP (MBP and MCPP), we visualized regression surfaces for their measured levels at pairs of trimesters after accounting for covariate effects and with all other exposures held constant at $0$. For trimester pairs $(1,3)$ and $(2,3)$, both regression surfaces had $95\%$ credible intervals that contained zero everywhere. For trimesters $1$ and $2$, the surfaces are shown in Figure \ref{fig_mep_dbp_surface}, with white regions indicating $95\%$ intervals that include zero. Observe that for both MEP and DBP, when exposure levels for both trimester $1$ and $2$ increase away from $0$, expected 2-DG decreases. This suggests there is not sufficient evidence to claim that a single trimester is particularly important, as \cite{goodrich2022trimester} suggested for trimester $1$, but there does appear to be a strong association for some combination of trimester $1$ and $2$ MEP and DBP levels. Thus, we are able to make some conclusions about the role of time even in low sample-size, high noise settings such as this, but without necessarily needing to conclude that a particular trimester is significant while others are not. This both strengthens the results of \cite{goodrich2022trimester} and adds additional nuance in interpretation.

\subsection{Interactions between exposures and sex}

Finally, given that phthalates and BPA act by disrupting the endocrine system, their effects on 2-DG could reasonably differ by biological sex. To evaluate this possibility, we adapted our LowFR model to include rank-1 interaction terms between all exposure measurements and sex. 
We accomplished this using the same rank-1 MGP prior setup as for the entries of $W$ and $B$, but for additional vectors $\beta^{(int)} \in \mathbb{R}^k$ and $\omega^{(int)} \in \mathbb{R}^T$.
\begin{figure}[h]
\includegraphics[width=\textwidth]{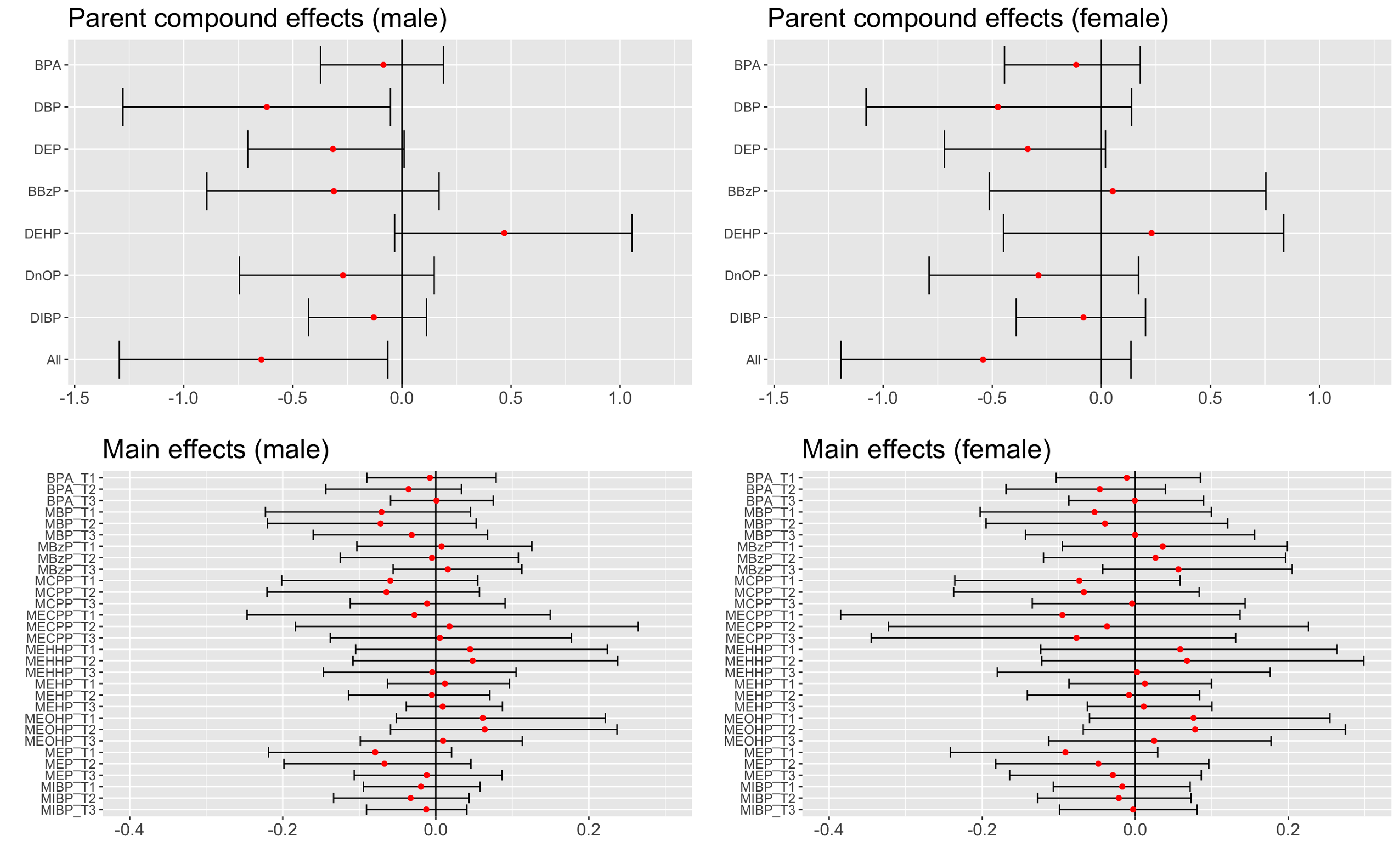}
\caption{Top row: Sex-specific expected changes in 2-DG level when all metabolites of the given phthalates (listed in Table 1) are are increased from $-1$ to $1$ at all three trimesters, with all other exposures fixed to a value of $0$. Bottom row: Sex-specific main effect coefficients for all measured exposures. Results were generated from a LowFR model with added rank-1 interaction terms between all exposures and the indicator for female sex. In all plots, red dots indicate posterior means, and intervals represent $95\%$ posterior credible intervals.}
\label{fig_sex_specific}
\end{figure}
\begin{figure}[b]
\includegraphics[width=\textwidth]{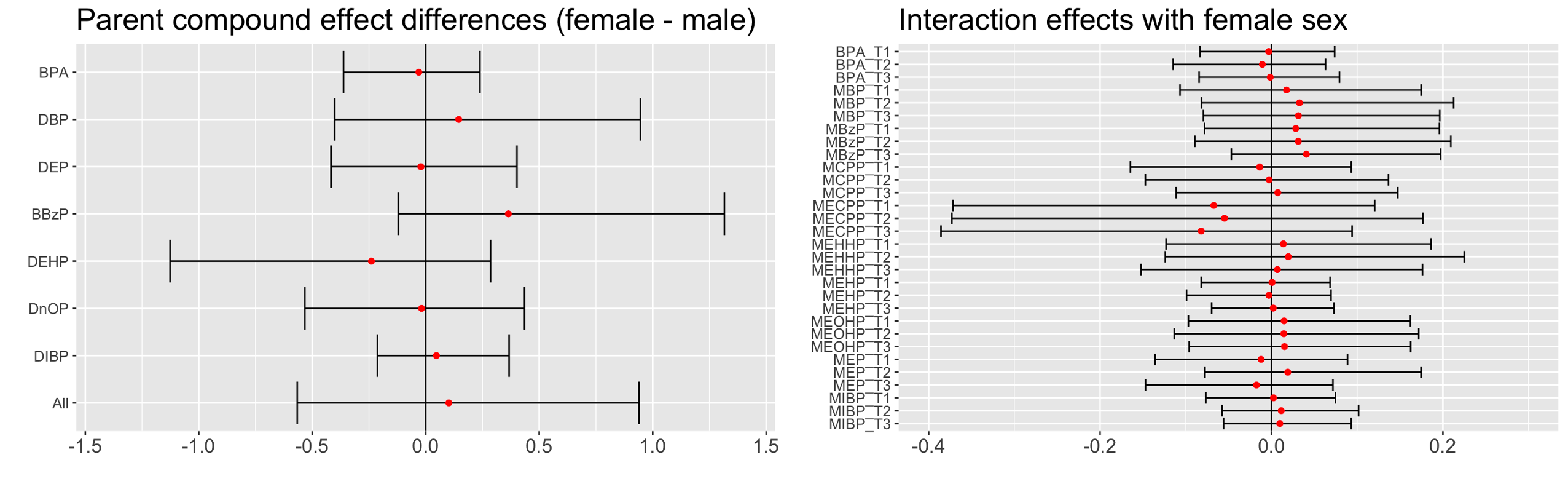}
\caption{Left: Posterior means and $95\%$ credible intervals of differences between the female and male parent compound effects shown in Figure \ref{fig_sex_specific}. Right: Posterior means and $95\%$ credible intervals of the interaction effects between the $30$ exposures and sex, corresponding to the difference between the female and male main effects shown in Figure \ref{fig_sex_specific}.}
\label{fig_sex_interactions}
\end{figure}
Using our fitted model, Figure \ref{fig_sex_specific} shows posterior summaries for sex-specific main and cumulative parent compound effects. In the parent compound plots, observe that all intervals in the female plot contain zero, even the ``All'' interval for all measured exposures. In the male plot, the ``All'' and DBP intervals are still strictly negative, but the DEP interval now includes zero. After adding these interactions, the $95\%$ credible interval for female sex now includes zero, although just barely (mean $0.26$, $95\%$ CI $(-0.001, 0.52)$). Note that this mean and CI are very close to their values for the LowFR model without the sex interactions, shown in Table \ref{table_covariates}. Taken together, these results suggest that in addition to some evidence of female children having higher 2-DG after accounting for prenatal phthalate and BPA exposure, there is also less evidence of a negative association between these exposures and 2-DG for females than males. Additionally, after including the interactions with sex, it is less clear whether cumulative MEP exposure is associated with lower followup 2-DG in either male or female children.

To directly evaluate the evidence for differences in exposure effects by sex, Figure \ref{fig_sex_interactions} shows posterior summaries for the interaction terms between the exposures and sex, and for the differences in cumulative parent compound effects for girls compared to boys. All $95\%$ credible intervals in both plots contain zero, so we are not able to conclude that there are differences by sex.

\section{Discussion}\label{section6}

In this work, we investigated the associations between prenatal BPA and phthalate exposure and glucose metabolism in adolescence as measured by serum 2-DG levels. By first focusing on inferences for pooled effects motivated by parent compounds of the measured phthalates, we identified overall negative associations for both DEP and DBP with 2-DG. Then, using the same fitted model, we found evidence that these associations are stronger for trimester 1 and 2 exposures compared to trimester 3. Finally, we considered interactions between these exposures and sex, and found stronger evidence of the above effects for boys than girls. To make these insights possible, we developed LowFR, a flexible Bayesian model motivated by our application that estimates all linear, quadratic, and two-way interaction terms for all exposures both within and across time. LowFR accomplishes this estimation by reducing dimensionality of both the exposures, via a latent factor model, and the regression coefficients, by modeling them as low-rank. Unlike variable selection approaches, LowFR can easily capture relationships where a number of highly-correlated exposures all have small but significant effects, while also fully quantifying uncertainty, unlike WQSR or methods using PCA. LowFR outperformed competitors for estimating both individual coefficients and cumulative effects over time in simulations, and was the only model that found the association of DBP with lower 2-DG among the competitors we tried.

There are several promising directions for future work. From a public health and clinical perspective, these early-pregnancy associations of DEP and DBP metabolites with 2-DG have implications for advising individual patients, crafting national guidelines, and advocating for regulation of particular chemicals. As DEP and DBP are both commonly found in cosmetics, personal care products, and pharmaceuticals and supplements \citep{braun2013phthalate}, informing pregnant mothers of the potential risks associated with certain products could be a key step to preventing complications. However, additional followup studies should be done to validate these results. In particular, a larger-scale study with more patients, more covariates collected for each patient, and more repeated measurements of both exposures and followup serum levels could allow for a stronger level of confidence in our conclusions. Additionally, in such a study, observational causal inference methods such as matching \citep{stuart2010matching} could be applied to yield a more complete understanding of the direct causal effects of these chemicals.

From a modeling perspective, one key assumption of our approach is that the measured exposures follow a Gaussian distribution. While we found this assumption reasonable for the ELEMENT exposures based on EDA, this will not always be the case, for example due to right-skewedenss or multimodality. In such cases, there are several approaches to address this issue while still taking advantage of our low-rank model structure. One approach would be to use the copula factor modeling approach of \cite{murray2013bayesian}, which combines the copula model of \cite{hoff2007extending} with a Gaussian factor model, thereby allowing arbitrary marginal distributions for the exposures. Alternately, a simpler approach would be to transform the exposures prior to analysis, for example with an empirical inverse CDF or Box-Cox transformation. Although we do not use these approaches in our primary analysis since the ELEMENT exposures appear reasonably Gaussian, and such transformations affect the interpretability of coefficient magnitudes, we have included a version of our ELEMENT data analysis after applying an inverse CDF transformation in the online supplement. The results are quite similar to those we present in Section \ref{section5}.

There are also several other directions for improving the generalizability of LowFR for analysis of related data sets. First, while our current Stan implementation has very good mixing and runs in under two hours on a laptop for the ELEMENT data, it may not scale sufficiently well for use in a significantly larger study. A Gibbs sampler implementation could help to some extent; however, the conditional posterior for the $\eta_i$s, which make up the vast majority of the parameters to be sampled, is not available in closed form, and thus would still require a Metropolis-Hastings step at each iteration. Alternately, if only the induced coefficients for $\mathbb{E}[y_i | x_i]$ are of interest, one could try marginalizing out the $\eta_i$s so they do not need to be sampled, and instead simply modeling the induced distribution of $(x_i, y_i)$ directly. To do so, one could note that the marginal distribution of $x_i$ is given in equation \ref{xi_cov}, while $y_i | x_i$ has a generalized chi-squared distribution with parameters depending on $x_i$, $A$, $V$, $\mu$, $\theta$, $\Omega$, and $\sigma^2$. There is no closed-form expression for the density of a generalized chi-squared random variable, but numerical approaches have been developed to estimate the distribution function (see e.g. \cite{de2017package}), which could potentially be leveraged within a posterior sampler. We note that while such an approach may yield computational savings by eliminating the need to sample the $\eta_i$s, it would generate an identical posterior distribution for the regression coefficients to the one we currently estimate, as our approach samples all parameters jointly in a single step. Second, while LowFR works well for a relatively small number of aligned measurement times (e.g., the three trimesters of pregnancy), it may become unstable for many fine-grained measurements, and is not equipped to handle misaligned times between patients. To address a large number of fine-grained measurements, the individual $\omega_{lt}$s and $W_{lt_1t_2}$s could be assumed to vary smoothly along 1D and 2D surfaces, modeled for example with splines. This mirrors many of the distributed lag approaches reviewed in Section \ref{section1}. Addressing misaligned measurement times would require more care, given that such data are often sparse with potentially different numbers of measurements for each patient. However, it is an important use case and an interesting avenue for future inquiry.

Third, while our current approach learns the dependence structure among the exposures without assuming any group structure a priori (for example, due to families of related chemicals), it would be interesting to investigate methods of adding this group information to aid in our modeling goals. One approach for this could be to allow the factor loadings matrix $\Lambda$ to depend on this information, inducing a covariate-dependent correlation structure among the exposure levels, related to the covariance regression model of \cite{hoff2012covariance}. Alternately, one could allow such group structure to inform the relationships between the exposures and outcome, perhaps through an analogous approach to that of \cite{ovaskainen2017make}, which includes a learned amount of phylogenetic information in the prior covariance of regression coefficients in an ecological species distribution model. It would be interesting to explore how such a structured covariance for the $\beta_l$s and $B_l$s would impact the overall model when combined with the low-rank structure of the coefficients.
Finally, since latent factor models learn the latent factors using the joint distribution of $(x_i, y_i)$ rather than modeling $y_i | x_i$ directly, the learned regression of $y_i$ on $x_i$ sometimes yields suboptimal out-of-sample predictive performance. Thus, another interesting avenue for future work could be to combine the structure of our approach with a more ``supervised'' approach to latent factor regression, for example the method of \cite{hahn2013partial} which allows additional dependence directly between $x_i$ and $y_i$ in addition to the dependence induced by $\eta_i$.

%%%%%%%%%%%%%%%%%%%%%%%%%%%%%%%%%%%%%%%%%%%%%%
%% Single Appendix:                         %%
%%%%%%%%%%%%%%%%%%%%%%%%%%%%%%%%%%%%%%%%%%%%%%
%\begin{appendix}
%\section*{???}%% if no title is needed, leave empty \section*{}.
%\end{appendix}
%%%%%%%%%%%%%%%%%%%%%%%%%%%%%%%%%%%%%%%%%%%%%%
%% Multiple Appendixes:                     %%
%%%%%%%%%%%%%%%%%%%%%%%%%%%%%%%%%%%%%%%%%%%%%%
\begin{appendix}

\section{Proof of Theorem \ref{quad_reg_1}}\label{app_th1_proof}
Recall the model setup:
$$y_i = \mu + \theta^T \eta_i + \eta_i^T \Omega \eta_i + \varepsilon_i^{(y)}, \,\,\,\,\, \varepsilon_i^{(y)} \sim N(0, \sigma^2),$$
$$x_i = (\Lambda \otimes I_T) \eta_i + \varepsilon_i, \,\,\,\,\, \varepsilon_i \sim N(0, (\Sigma \otimes \Phi)),$$
$$\eta_i \sim N(0, I_k \otimes \Phi).$$
Now, first observe that
\begin{align*}
    \mathbb{E}[y_i | x_i] &= \mathbb{E}[\mathbb{E}[y_i | \eta_i] | x_i] \\
    &= \mathbb{E}[\mu + \theta^T \eta_i + \eta_i^T \Omega \eta_i | x_i] \\
    &= \mu + \theta^T \mathbb{E}[\eta_i | x_i] + \mathbb{E}[\eta_i^T \Omega \eta_i | x_i].
\end{align*}
Consider the second term of the above. By Bayes' rule and some algebra, we get that
\begin{align*}
    p(\eta_i | x_i) &\propto p(x_i | \eta_i) p(\eta_i) \\
    & \propto \exp \left(-\frac{1}{2}(x_i - (\Lambda \otimes I_T)\eta_i)^T (\Sigma \otimes \Phi)^{-1} (x_i - (\Lambda \otimes I_T) \eta_i) \right) \exp\left(-\frac{1}{2} \eta_i^T (I_k \otimes \Phi)^{-1} \eta_i \right) \\
    &\propto \exp \left(-\frac{1}{2} (\eta_i^T ((\Lambda^T \Sigma^{-1} \Lambda + I_k) \otimes \Phi^{-1}) \eta_i - 2 \eta_i^T (\Lambda^T \Sigma^{-1} \otimes \Phi^{-1}) x_i\right)
\end{align*}
and thus, letting
\begin{align*}
    V &= ((\Lambda^T \Sigma^{-1} \Lambda + I_k) \otimes \Phi^{-1})^{-1}\\
    &= (\Lambda^T \Sigma^{-1} \Lambda + I_k)^{-1} \otimes \Phi
\end{align*}
and
\begin{align*}
    A &= V(\Lambda^T \Sigma^{-1} \otimes \Phi^{-1})\\
    &= (\Lambda^T \Sigma^{-1} \Lambda + I_k)^{-1} \Lambda^T \Sigma^{-1} \otimes I_T
\end{align*}
we get that
$$\eta_i | x_i \sim N(Ax_i, V)$$
so that $\mathbb{E}[\eta_i | x_i] = A x_i$. For the third term, note that the expectation of a quadratic form $\eta^T \Omega \eta$ of a random vector $\eta$ with mean $m$ and covariance $K$ is equal to $tr(\Omega K) + m^T \Omega m$. Using this, we get that
\begin{align*}
    \mathbb{E}[y_i | x_i] &= \mu + \theta^T (A x_i) + tr(\Omega V) + (Ax_i)^T \Omega (Ax_i) \\
    &= \mu + tr(\Omega V) + (\theta^T A) x_i + x_i^T (A^T \Omega A) x_i.
\end{align*}
$\square$

%
%\section{???}
%
\end{appendix}

%%%%%%%%%%%%%%%%%%%%%%%%%%%%%%%%%%%%%%%%%%%%%%
%% Support information, if any,             %%
%% should be provided in the                %%
%% Acknowledgements section.                %%
%%%%%%%%%%%%%%%%%%%%%%%%%%%%%%%%%%%%%%%%%%%%%%
\begin{acks}[Acknowledgments]
This work was supported by grants R01ES035625 and P42-ES010356 from the NIH National Institute of Environmental Health Sciences. We acknowledge the Duke Compute Cluster for computational time. Amy Herring is also affiliated with the Duke Global Health Institute and the Department of Biostatistics and Bioinformatics at Duke University. David Dunson is also affiliated with the Department of Mathematics at Duke University.
\end{acks}
%%%%%%%%%%%%%%%%%%%%%%%%%%%%%%%%%%%%%%%%%%%%%%
%% Funding information, if any,             %%
%% should be provided in the                %%
%% funding section.                         %%
%%%%%%%%%%%%%%%%%%%%%%%%%%%%%%%%%%%%%%%%%%%%%%
%\begin{funding}
%This work was supported by NIH grant R01ES035625.
%\end{funding}

%%%%%%%%%%%%%%%%%%%%%%%%%%%%%%%%%%%%%%%%%%%%%%
%% Supplementary Material, including data   %%
%% sets and code, should be provided in     %%
%% {supplement} environment with title      %%
%% and short description. It cannot be      %%
%% available exclusively as external link.  %%
%% All Supplementary Material must be       %%
%% available to the reader on Project       %%
%% Euclid with the published article.       %%
%%%%%%%%%%%%%%%%%%%%%%%%%%%%%%%%%%%%%%%%%%%%%%
\begin{supplement}
\stitle{Online supplement contents:}
\sdescription{(1) Data generation details for Section \ref{section1} motivating simulations. (2) Full LowFR model description with prior distributions. (3) Alternative Theorem \ref{quad_reg_1} for general $Cov(\eta_i)$ and $Cov(\varepsilon_i)$. (4) Full CorrQuadReg model description with prior distributions. (5) Marginal 2-DG posterior predictive plots. (6) Code implementing LowFR and reproducing the results in the paper can be found at https://github.com/glennpalmer/LowFR, and is also included as a .zip file.}
\end{supplement}

%%  imsart-nameyear.bst  will be used to                   %%
%%  MR numbers will be added by VTeX.                      %%
%%                                                         %%
%%  Use \cite{...} to cite references in text.             %%
%%                                                         %%
%%%%%%%%%%%%%%%%%%%%%%%%%%%%%%%%%%%%%%%%%%%%%%%%%%%%%%%%%%%%%

%% if your bibliography is in bibtex format, uncomment commands:
\bibliographystyle{imsart-nameyear} % Style BST file
\bibliography{references}       % Bibliography file (usually '*.bib')

\begin{thebibliography}{89}
% BibTex style file: imsart-nameyear.bst, 2017-11-03
% Default style options (sort=1,type=nameyear).
% Used options (sort=1,type=nameyear).

\bibitem[\protect\citeauthoryear{Alonso-Magdalena et~al.}{2006}]{alonso2006estrogenic}
\begin{barticle}[author]
\bauthor{\bsnm{Alonso-Magdalena},~\bfnm{Paloma}\binits{P.}}, \bauthor{\bsnm{Morimoto},~\bfnm{Sumiko}\binits{S.}}, \bauthor{\bsnm{Ripoll},~\bfnm{Cristina}\binits{C.}}, \bauthor{\bsnm{Fuentes},~\bfnm{Esther}\binits{E.}} \AND \bauthor{\bsnm{Nadal},~\bfnm{Angel}\binits{A.}}
(\byear{2006}).
\btitle{The estrogenic effect of bisphenol A disrupts pancreatic $\beta$-cell function in vivo and induces insulin resistance}.
\bjournal{Environmental Health Perspectives}
\bvolume{114}
\bpages{106--112}.
\end{barticle}
\endbibitem

\bibitem[\protect\citeauthoryear{Antonelli, Wilson and Coull}{2022}]{antonelli2022multiple}
\begin{barticle}[author]
\bauthor{\bsnm{Antonelli},~\bfnm{J}\binits{J.}}, \bauthor{\bsnm{Wilson},~\bfnm{A}\binits{A.}} \AND \bauthor{\bsnm{Coull},~\bfnm{BA}\binits{B.}}
(\byear{2022}).
\btitle{Multiple exposure distributed lag models with variable selection.}
\bjournal{Biostatistics}
\bpages{kxac038}.
\end{barticle}
\endbibitem

\bibitem[\protect\citeauthoryear{Bartholomew, Knott and Moustaki}{2011}]{bartholomew2011latent}
\begin{bbook}[author]
\bauthor{\bsnm{Bartholomew},~\bfnm{David~J}\binits{D.~J.}}, \bauthor{\bsnm{Knott},~\bfnm{Martin}\binits{M.}} \AND \bauthor{\bsnm{Moustaki},~\bfnm{Irini}\binits{I.}}
(\byear{2011}).
\btitle{Latent Variable Models and Factor Analysis: A Unified Approach}.
\bpublisher{John Wiley \& Sons}.
\end{bbook}
\endbibitem

\bibitem[\protect\citeauthoryear{Becker et~al.}{2009}]{becker2009geres}
\begin{barticle}[author]
\bauthor{\bsnm{Becker},~\bfnm{Kerstin}\binits{K.}}, \bauthor{\bsnm{G{\"u}en},~\bfnm{Thomas}\binits{T.}}, \bauthor{\bsnm{Seiwert},~\bfnm{Margarete}\binits{M.}}, \bauthor{\bsnm{Conrad},~\bfnm{Andre}\binits{A.}}, \bauthor{\bsnm{Pick-Fu{\ss}},~\bfnm{Helga}\binits{H.}}, \bauthor{\bsnm{M{\"u}ller},~\bfnm{Johannes}\binits{J.}}, \bauthor{\bsnm{Wittassek},~\bfnm{Matthias}\binits{M.}}, \bauthor{\bsnm{Schulz},~\bfnm{Christine}\binits{C.}} \AND \bauthor{\bsnm{Kolossa-Gehring},~\bfnm{Marike}\binits{M.}}
(\byear{2009}).
\btitle{GerES IV: phthalate metabolites and bisphenol A in urine of German children}.
\bjournal{International Journal of Hygiene and Environmental Health}
\bvolume{212}
\bpages{685--692}.
\end{barticle}
\endbibitem

\bibitem[\protect\citeauthoryear{Bello et~al.}{2017}]{bello2017extending}
\begin{barticle}[author]
\bauthor{\bsnm{Bello},~\bfnm{Ghalib~A}\binits{G.~A.}}, \bauthor{\bsnm{Arora},~\bfnm{Manish}\binits{M.}}, \bauthor{\bsnm{Austin},~\bfnm{Christine}\binits{C.}}, \bauthor{\bsnm{Horton},~\bfnm{Megan~K}\binits{M.~K.}}, \bauthor{\bsnm{Wright},~\bfnm{Robert~O}\binits{R.~O.}} \AND \bauthor{\bsnm{Gennings},~\bfnm{Chris}\binits{C.}}
(\byear{2017}).
\btitle{Extending the distributed lag model framework to handle chemical mixtures}.
\bjournal{Environmental Research}
\bvolume{156}
\bpages{253--264}.
\end{barticle}
\endbibitem

\bibitem[\protect\citeauthoryear{Bhattacharya and Dunson}{2011}]{bhattacharya2011sparse}
\begin{barticle}[author]
\bauthor{\bsnm{Bhattacharya},~\bfnm{Anirban}\binits{A.}} \AND \bauthor{\bsnm{Dunson},~\bfnm{David~B}\binits{D.~B.}}
(\byear{2011}).
\btitle{Sparse Bayesian infinite factor models}.
\bjournal{Biometrika}
\bvolume{98}
\bpages{291--306}.
\end{barticle}
\endbibitem

\bibitem[\protect\citeauthoryear{Bien, Taylor and Tibshirani}{2013}]{bien2013lasso}
\begin{barticle}[author]
\bauthor{\bsnm{Bien},~\bfnm{Jacob}\binits{J.}}, \bauthor{\bsnm{Taylor},~\bfnm{Jonathan}\binits{J.}} \AND \bauthor{\bsnm{Tibshirani},~\bfnm{Robert}\binits{R.}}
(\byear{2013}).
\btitle{A lasso for hierarchical interactions}.
\bjournal{Annals of Statistics}
\bvolume{41}
\bpages{1111}.
\end{barticle}
\endbibitem

\bibitem[\protect\citeauthoryear{Bobb et~al.}{2015}]{bobb2015bayesian}
\begin{barticle}[author]
\bauthor{\bsnm{Bobb},~\bfnm{Jennifer~F}\binits{J.~F.}}, \bauthor{\bsnm{Valeri},~\bfnm{Linda}\binits{L.}}, \bauthor{\bsnm{Claus~Henn},~\bfnm{Birgit}\binits{B.}}, \bauthor{\bsnm{Christiani},~\bfnm{David~C}\binits{D.~C.}}, \bauthor{\bsnm{Wright},~\bfnm{Robert~O}\binits{R.~O.}}, \bauthor{\bsnm{Mazumdar},~\bfnm{Maitreyi}\binits{M.}}, \bauthor{\bsnm{Godleski},~\bfnm{John~J}\binits{J.~J.}} \AND \bauthor{\bsnm{Coull},~\bfnm{Brent~A}\binits{B.~A.}}
(\byear{2015}).
\btitle{Bayesian kernel machine regression for estimating the health effects of multi-pollutant mixtures}.
\bjournal{Biostatistics}
\bvolume{16}
\bpages{493--508}.
\end{barticle}
\endbibitem

\bibitem[\protect\citeauthoryear{Bobb et~al.}{2018}]{bobb2018statistical}
\begin{barticle}[author]
\bauthor{\bsnm{Bobb},~\bfnm{Jennifer~F}\binits{J.~F.}}, \bauthor{\bsnm{Claus~Henn},~\bfnm{Birgit}\binits{B.}}, \bauthor{\bsnm{Valeri},~\bfnm{Linda}\binits{L.}} \AND \bauthor{\bsnm{Coull},~\bfnm{Brent~A}\binits{B.~A.}}
(\byear{2018}).
\btitle{Statistical software for analyzing the health effects of multiple concurrent exposures via Bayesian kernel machine regression}.
\bjournal{Environmental Health}
\bvolume{17}
\bpages{1--10}.
\end{barticle}
\endbibitem

\bibitem[\protect\citeauthoryear{Bornehag et~al.}{2004}]{bornehag2004association}
\begin{barticle}[author]
\bauthor{\bsnm{Bornehag},~\bfnm{Carl-Gustaf}\binits{C.-G.}}, \bauthor{\bsnm{Sundell},~\bfnm{Jan}\binits{J.}}, \bauthor{\bsnm{Weschler},~\bfnm{Charles~J}\binits{C.~J.}}, \bauthor{\bsnm{Sigsgaard},~\bfnm{Torben}\binits{T.}}, \bauthor{\bsnm{Lundgren},~\bfnm{Bj{\"o}rn}\binits{B.}}, \bauthor{\bsnm{Hasselgren},~\bfnm{Mikael}\binits{M.}} \AND \bauthor{\bsnm{H{\"a}gerhed-Engman},~\bfnm{Linda}\binits{L.}}
(\byear{2004}).
\btitle{The association between asthma and allergic symptoms in children and phthalates in house dust: a nested case--control study}.
\bjournal{Environmental Health Perspectives}
\bvolume{112}
\bpages{1393--1397}.
\end{barticle}
\endbibitem

\bibitem[\protect\citeauthoryear{Braun, Sathyanarayana and Hauser}{2013}]{braun2013phthalate}
\begin{barticle}[author]
\bauthor{\bsnm{Braun},~\bfnm{Joseph~M}\binits{J.~M.}}, \bauthor{\bsnm{Sathyanarayana},~\bfnm{Sheela}\binits{S.}} \AND \bauthor{\bsnm{Hauser},~\bfnm{Russ}\binits{R.}}
(\byear{2013}).
\btitle{Phthalate exposure and children’s health}.
\bjournal{Current Opinion in Pediatrics}
\bvolume{25}
\bpages{247}.
\end{barticle}
\endbibitem

\bibitem[\protect\citeauthoryear{Buckley, Hamra and Braun}{2019}]{buckley2019statistical}
\begin{barticle}[author]
\bauthor{\bsnm{Buckley},~\bfnm{Jessie~P}\binits{J.~P.}}, \bauthor{\bsnm{Hamra},~\bfnm{Ghassan~B}\binits{G.~B.}} \AND \bauthor{\bsnm{Braun},~\bfnm{Joseph~M}\binits{J.~M.}}
(\byear{2019}).
\btitle{Statistical approaches for investigating periods of susceptibility in children’s environmental health research}.
\bjournal{Current Environmental Health Reports}
\bvolume{6}
\bpages{1--7}.
\end{barticle}
\endbibitem

\bibitem[\protect\citeauthoryear{Carpenter et~al.}{2017}]{carpenter2017stan}
\begin{barticle}[author]
\bauthor{\bsnm{Carpenter},~\bfnm{Bob}\binits{B.}}, \bauthor{\bsnm{Gelman},~\bfnm{Andrew}\binits{A.}}, \bauthor{\bsnm{Hoffman},~\bfnm{Matthew~D}\binits{M.~D.}}, \bauthor{\bsnm{Lee},~\bfnm{Daniel}\binits{D.}}, \bauthor{\bsnm{Goodrich},~\bfnm{Ben}\binits{B.}}, \bauthor{\bsnm{Betancourt},~\bfnm{Michael}\binits{M.}}, \bauthor{\bsnm{Brubaker},~\bfnm{Marcus}\binits{M.}}, \bauthor{\bsnm{Guo},~\bfnm{Jiqiang}\binits{J.}}, \bauthor{\bsnm{Li},~\bfnm{Peter}\binits{P.}} \AND \bauthor{\bsnm{Riddell},~\bfnm{Allen}\binits{A.}}
(\byear{2017}).
\btitle{Stan: A probabilistic programming language}.
\bjournal{Journal of Statistical Software}
\bvolume{76}.
\end{barticle}
\endbibitem

\bibitem[\protect\citeauthoryear{Carrico et~al.}{2015}]{carrico2015characterization}
\begin{barticle}[author]
\bauthor{\bsnm{Carrico},~\bfnm{Caroline}\binits{C.}}, \bauthor{\bsnm{Gennings},~\bfnm{Chris}\binits{C.}}, \bauthor{\bsnm{Wheeler},~\bfnm{David~C}\binits{D.~C.}} \AND \bauthor{\bsnm{Factor-Litvak},~\bfnm{Pam}\binits{P.}}
(\byear{2015}).
\btitle{Characterization of weighted quantile sum regression for highly correlated data in a risk analysis setting}.
\bjournal{Journal of Agricultural, Biological, and Environmental Statistics}
\bvolume{20}
\bpages{100--120}.
\end{barticle}
\endbibitem

\bibitem[\protect\citeauthoryear{Carvalho, Polson and Scott}{2009}]{carvalho2009handling}
\begin{binproceedings}[author]
\bauthor{\bsnm{Carvalho},~\bfnm{Carlos~M}\binits{C.~M.}}, \bauthor{\bsnm{Polson},~\bfnm{Nicholas~G}\binits{N.~G.}} \AND \bauthor{\bsnm{Scott},~\bfnm{James~G}\binits{J.~G.}}
(\byear{2009}).
\btitle{Handling sparsity via the horseshoe}.
In \bbooktitle{Artificial Intelligence and Statistics}
\bpages{73--80}.
\bpublisher{PMLR}.
\end{binproceedings}
\endbibitem

\bibitem[\protect\citeauthoryear{Chang et~al.}{2015}]{chang2015assessment}
\begin{barticle}[author]
\bauthor{\bsnm{Chang},~\bfnm{Howard~H}\binits{H.~H.}}, \bauthor{\bsnm{Warren},~\bfnm{Joshua~L}\binits{J.~L.}}, \bauthor{\bsnm{Darrow},~\bfnm{Lnydsey~A}\binits{L.~A.}}, \bauthor{\bsnm{Reich},~\bfnm{Brian~J}\binits{B.~J.}} \AND \bauthor{\bsnm{Waller},~\bfnm{Lance~A}\binits{L.~A.}}
(\byear{2015}).
\btitle{Assessment of critical exposure and outcome windows in time-to-event analysis with application to air pollution and preterm birth study}.
\bjournal{Biostatistics}
\bvolume{16}
\bpages{509--521}.
\end{barticle}
\endbibitem

\bibitem[\protect\citeauthoryear{Chen, Mukherjee and Berrocal}{2019}]{chen2019distributed}
\begin{barticle}[author]
\bauthor{\bsnm{Chen},~\bfnm{Yin-Hsiu}\binits{Y.-H.}}, \bauthor{\bsnm{Mukherjee},~\bfnm{Bhramar}\binits{B.}} \AND \bauthor{\bsnm{Berrocal},~\bfnm{Veronica~J}\binits{V.~J.}}
(\byear{2019}).
\btitle{Distributed lag interaction models with two pollutants}.
\bjournal{Journal of the Royal Statistical Society: Series C}
\bvolume{68}
\bpages{79}.
\end{barticle}
\endbibitem

\bibitem[\protect\citeauthoryear{Colicino et~al.}{2020}]{colicino2020per}
\begin{barticle}[author]
\bauthor{\bsnm{Colicino},~\bfnm{Elena}\binits{E.}}, \bauthor{\bsnm{Pedretti},~\bfnm{Nicolo~Foppa}\binits{N.~F.}}, \bauthor{\bsnm{Busgang},~\bfnm{Stefanie~A}\binits{S.~A.}} \AND \bauthor{\bsnm{Gennings},~\bfnm{Chris}\binits{C.}}
(\byear{2020}).
\btitle{Per-and poly-fluoroalkyl substances and bone mineral density: results from the Bayesian weighted quantile sum regression}.
\bjournal{Environmental Epidemiology}
\bvolume{4}.
\end{barticle}
\endbibitem

\bibitem[\protect\citeauthoryear{Curtin et~al.}{2021}]{curtin2021random}
\begin{barticle}[author]
\bauthor{\bsnm{Curtin},~\bfnm{Paul}\binits{P.}}, \bauthor{\bsnm{Kellogg},~\bfnm{Joshua}\binits{J.}}, \bauthor{\bsnm{Cech},~\bfnm{Nadja}\binits{N.}} \AND \bauthor{\bsnm{Gennings},~\bfnm{Chris}\binits{C.}}
(\byear{2021}).
\btitle{A random subset implementation of weighted quantile sum (WQSRS) regression for analysis of high-dimensional mixtures}.
\bjournal{Communications in Statistics-Simulation and Computation}
\bvolume{50}
\bpages{1119--1134}.
\end{barticle}
\endbibitem

\bibitem[\protect\citeauthoryear{Czarnota, Gennings and Wheeler}{2015}]{czarnota2015assessment}
\begin{barticle}[author]
\bauthor{\bsnm{Czarnota},~\bfnm{Jenna}\binits{J.}}, \bauthor{\bsnm{Gennings},~\bfnm{Chris}\binits{C.}} \AND \bauthor{\bsnm{Wheeler},~\bfnm{David~C}\binits{D.~C.}}
(\byear{2015}).
\btitle{Assessment of weighted quantile sum regression for modeling chemical mixtures and cancer risk}.
\bjournal{Cancer Informatics}
\bvolume{14}
\bpages{CIN--S17295}.
\end{barticle}
\endbibitem

\bibitem[\protect\citeauthoryear{de~Micheaux}{2017}]{de2017package}
\begin{barticle}[author]
\bauthor{\bparticle{de} \bsnm{Micheaux},~\bfnm{P~Lafaye}\binits{P.~L.}}
(\byear{2017}).
\btitle{CompQuadForm: Distribution function of quadratic forms in normal variables}.
\bjournal{CRAN Repository}.
\bnote{R package version 1.4.3}.
\end{barticle}
\endbibitem

\bibitem[\protect\citeauthoryear{De~Waal}{1985}]{de1985matrix}
\begin{barticle}[author]
\bauthor{\bsnm{De~Waal},~\bfnm{DJ}\binits{D.}}
(\byear{1985}).
\btitle{Matrix-Valued Distributions}.
\bjournal{Encyclopedia of Statistical Sciences}.
\end{barticle}
\endbibitem

\bibitem[\protect\citeauthoryear{Desvergne, Feige and Casals-Casas}{2009}]{desvergne2009ppar}
\begin{barticle}[author]
\bauthor{\bsnm{Desvergne},~\bfnm{B{\'e}atrice}\binits{B.}}, \bauthor{\bsnm{Feige},~\bfnm{J{\'e}r{\^o}me~N}\binits{J.~N.}} \AND \bauthor{\bsnm{Casals-Casas},~\bfnm{Cristina}\binits{C.}}
(\byear{2009}).
\btitle{PPAR-mediated activity of phthalates: a link to the obesity epidemic?}
\bjournal{Molecular and Cellular Endocrinology}
\bvolume{304}
\bpages{43--48}.
\end{barticle}
\endbibitem

\bibitem[\protect\citeauthoryear{Dominici et~al.}{2010}]{dominici2010opinion}
\begin{barticle}[author]
\bauthor{\bsnm{Dominici},~\bfnm{Francesca}\binits{F.}}, \bauthor{\bsnm{Peng},~\bfnm{Roger~D}\binits{R.~D.}}, \bauthor{\bsnm{Barr},~\bfnm{Christopher~D}\binits{C.~D.}} \AND \bauthor{\bsnm{Bell},~\bfnm{Michelle~L}\binits{M.~L.}}
(\byear{2010}).
\btitle{Opinion: Protecting human health from air pollution: shifting from a single-pollutant to a multipollutant approach}.
\bjournal{Epidemiology}
\bpages{187--194}.
\end{barticle}
\endbibitem

\bibitem[\protect\citeauthoryear{Eng et~al.}{2013}]{eng2013bisphenol}
\begin{barticle}[author]
\bauthor{\bsnm{Eng},~\bfnm{Donna~S}\binits{D.~S.}}, \bauthor{\bsnm{Lee},~\bfnm{Joyce~M}\binits{J.~M.}}, \bauthor{\bsnm{Gebremariam},~\bfnm{Achamyeleh}\binits{A.}}, \bauthor{\bsnm{Meeker},~\bfnm{John~D}\binits{J.~D.}}, \bauthor{\bsnm{Peterson},~\bfnm{Karen}\binits{K.}} \AND \bauthor{\bsnm{Padmanabhan},~\bfnm{Vasantha}\binits{V.}}
(\byear{2013}).
\btitle{Bisphenol A and chronic disease risk factors in US children}.
\bjournal{Pediatrics}
\bvolume{132}
\bpages{e637--e645}.
\end{barticle}
\endbibitem

\bibitem[\protect\citeauthoryear{Ferrari and Dunson}{2021}]{ferrari2021bayesian}
\begin{barticle}[author]
\bauthor{\bsnm{Ferrari},~\bfnm{Federico}\binits{F.}} \AND \bauthor{\bsnm{Dunson},~\bfnm{David~B}\binits{D.~B.}}
(\byear{2021}).
\btitle{Bayesian factor analysis for inference on interactions}.
\bjournal{Journal of the American Statistical Association}
\bvolume{116}
\bpages{1521--1532}.
\end{barticle}
\endbibitem

\bibitem[\protect\citeauthoryear{Fosdick and Hoff}{2014}]{fosdick2014separable}
\begin{barticle}[author]
\bauthor{\bsnm{Fosdick},~\bfnm{Bailey~K}\binits{B.~K.}} \AND \bauthor{\bsnm{Hoff},~\bfnm{Peter~D}\binits{P.~D.}}
(\byear{2014}).
\btitle{Separable factor analysis with applications to mortality data}.
\bjournal{The Annals of Applied Statistics}
\bvolume{8}
\bpages{120}.
\end{barticle}
\endbibitem

\bibitem[\protect\citeauthoryear{Frederiksen et~al.}{2012}]{frederiksen2012high}
\begin{barticle}[author]
\bauthor{\bsnm{Frederiksen},~\bfnm{H}\binits{H.}}, \bauthor{\bsnm{S{\o}rensen},~\bfnm{K}\binits{K.}}, \bauthor{\bsnm{Mouritsen},~\bfnm{A}\binits{A.}}, \bauthor{\bsnm{Aksglaede},~\bfnm{L}\binits{L.}}, \bauthor{\bsnm{Hagen},~\bfnm{CP}\binits{C.}}, \bauthor{\bsnm{Petersen},~\bfnm{JH}\binits{J.}}, \bauthor{\bsnm{Skakkebaek},~\bfnm{NE}\binits{N.}}, \bauthor{\bsnm{Andersson},~\bfnm{A-M}\binits{A.-M.}} \AND \bauthor{\bsnm{Juul},~\bfnm{A}\binits{A.}}
(\byear{2012}).
\btitle{High urinary phthalate concentration associated with delayed pubarche in girls}.
\bjournal{International Journal of Andrology}
\bvolume{35}
\bpages{216--226}.
\end{barticle}
\endbibitem

\bibitem[\protect\citeauthoryear{Gasparrini, Armstrong and Kenward}{2010}]{gasparrini2010distributed}
\begin{barticle}[author]
\bauthor{\bsnm{Gasparrini},~\bfnm{Antonio}\binits{A.}}, \bauthor{\bsnm{Armstrong},~\bfnm{Ben}\binits{B.}} \AND \bauthor{\bsnm{Kenward},~\bfnm{Mike~G}\binits{M.~G.}}
(\byear{2010}).
\btitle{Distributed lag non-linear models}.
\bjournal{Statistics in Medicine}
\bvolume{29}
\bpages{2224--2234}.
\end{barticle}
\endbibitem

\bibitem[\protect\citeauthoryear{Gasparrini et~al.}{2017}]{gasparrini2017penalized}
\begin{barticle}[author]
\bauthor{\bsnm{Gasparrini},~\bfnm{Antonio}\binits{A.}}, \bauthor{\bsnm{Scheipl},~\bfnm{Fabian}\binits{F.}}, \bauthor{\bsnm{Armstrong},~\bfnm{Ben}\binits{B.}} \AND \bauthor{\bsnm{Kenward},~\bfnm{Michael~G}\binits{M.~G.}}
(\byear{2017}).
\btitle{A penalized framework for distributed lag non-linear models}.
\bjournal{Biometrics}
\bvolume{73}
\bpages{938--948}.
\end{barticle}
\endbibitem

\bibitem[\protect\citeauthoryear{Gennings et~al.}{2020}]{gennings2020lagged}
\begin{barticle}[author]
\bauthor{\bsnm{Gennings},~\bfnm{Chris}\binits{C.}}, \bauthor{\bsnm{Curtin},~\bfnm{Paul}\binits{P.}}, \bauthor{\bsnm{Bello},~\bfnm{Ghalib}\binits{G.}}, \bauthor{\bsnm{Wright},~\bfnm{Robert}\binits{R.}}, \bauthor{\bsnm{Arora},~\bfnm{Manish}\binits{M.}} \AND \bauthor{\bsnm{Austin},~\bfnm{Christine}\binits{C.}}
(\byear{2020}).
\btitle{Lagged WQS regression for mixtures with many components}.
\bjournal{Environmental Research}
\bvolume{186}
\bpages{109529}.
\end{barticle}
\endbibitem

\bibitem[\protect\citeauthoryear{Goodrich et~al.}{2022a}]{goodrich2022trimester}
\begin{barticle}[author]
\bauthor{\bsnm{Goodrich},~\bfnm{Jaclyn~M}\binits{J.~M.}}, \bauthor{\bsnm{Tang},~\bfnm{Lu}\binits{L.}}, \bauthor{\bsnm{Carmona},~\bfnm{Yanelli~R}\binits{Y.~R.}}, \bauthor{\bsnm{Meijer},~\bfnm{Jennifer~L}\binits{J.~L.}}, \bauthor{\bsnm{Perng},~\bfnm{Wei}\binits{W.}}, \bauthor{\bsnm{Watkins},~\bfnm{Deborah~J}\binits{D.~J.}}, \bauthor{\bsnm{Meeker},~\bfnm{John~D}\binits{J.~D.}}, \bauthor{\bsnm{Mercado-Garc{\'\i}a},~\bfnm{Adriana}\binits{A.}}, \bauthor{\bsnm{Cantoral},~\bfnm{Alejandra}\binits{A.}}, \bauthor{\bsnm{Song},~\bfnm{Peter~X}\binits{P.~X.}} \betal{et~al.}
(\byear{2022}a).
\btitle{Trimester-specific phthalate exposures in pregnancy are associated with circulating metabolites in children}.
\bjournal{Plos one}
\bvolume{17}
\bpages{e0272794}.
\end{barticle}
\endbibitem

\bibitem[\protect\citeauthoryear{Goodrich et~al.}{2022b}]{goodrich2022data}
\begin{bmisc}[author]
\bauthor{\bsnm{Goodrich},~\bfnm{Jaclyn~M}\binits{J.~M.}}, \bauthor{\bsnm{Tang},~\bfnm{Lu}\binits{L.}}, \bauthor{\bsnm{Carmona},~\bfnm{Yanelli~R}\binits{Y.~R.}}, \bauthor{\bsnm{Meijer},~\bfnm{Jennifer~L}\binits{J.~L.}}, \bauthor{\bsnm{Perng},~\bfnm{Wei}\binits{W.}}, \bauthor{\bsnm{Watkins},~\bfnm{Deborah~J}\binits{D.~J.}}, \bauthor{\bsnm{Meeker},~\bfnm{John~D}\binits{J.~D.}}, \bauthor{\bsnm{Mercado-Garc{\'\i}a},~\bfnm{Adriana}\binits{A.}}, \bauthor{\bsnm{Cantoral},~\bfnm{Alejandra}\binits{A.}}, \bauthor{\bsnm{Song},~\bfnm{Peter~X}\binits{P.~X.}} \betal{et~al.}
(\byear{2022}b).
\btitle{Trimester-specific phthalate exposures in pregnancy are associated with circulating metabolites in children [Data set]}.
\bdoi{10.7302/pehh-r785}
\end{bmisc}
\endbibitem

\bibitem[\protect\citeauthoryear{Grimaldi et~al.}{2015}]{grimaldi2015reporter}
\begin{barticle}[author]
\bauthor{\bsnm{Grimaldi},~\bfnm{Marina}\binits{M.}}, \bauthor{\bsnm{Boulahtouf},~\bfnm{Abdelhay}\binits{A.}}, \bauthor{\bsnm{Delfosse},~\bfnm{Vanessa}\binits{V.}}, \bauthor{\bsnm{Thouennon},~\bfnm{Erwan}\binits{E.}}, \bauthor{\bsnm{Bourguet},~\bfnm{William}\binits{W.}} \AND \bauthor{\bsnm{Balaguer},~\bfnm{Patrick}\binits{P.}}
(\byear{2015}).
\btitle{Reporter cell lines for the characterization of the interactions between human nuclear receptors and endocrine disruptors}.
\bjournal{Frontiers in Endocrinology}
\bvolume{6}
\bpages{62}.
\end{barticle}
\endbibitem

\bibitem[\protect\citeauthoryear{Gr{\"u}n and Blumberg}{2007}]{grun2007perturbed}
\begin{barticle}[author]
\bauthor{\bsnm{Gr{\"u}n},~\bfnm{Felix}\binits{F.}} \AND \bauthor{\bsnm{Blumberg},~\bfnm{Bruce}\binits{B.}}
(\byear{2007}).
\btitle{Perturbed nuclear receptor signaling by environmental obesogens as emerging factors in the obesity crisis}.
\bjournal{Reviews in Endocrine and Metabolic Disorders}
\bvolume{8}
\bpages{161--171}.
\end{barticle}
\endbibitem

\bibitem[\protect\citeauthoryear{G{\"u}ven et~al.}{2016}]{guven2016low}
\begin{barticle}[author]
\bauthor{\bsnm{G{\"u}ven},~\bfnm{Celal}\binits{C.}}, \bauthor{\bsnm{Dal},~\bfnm{Fulya}\binits{F.}}, \bauthor{\bsnm{Ahbab},~\bfnm{M{\"u}fide~Aydo{\u{g}}an}\binits{M.~A.}}, \bauthor{\bsnm{Taskin},~\bfnm{Eylem}\binits{E.}}, \bauthor{\bsnm{Ahbab},~\bfnm{S{\"u}leyman}\binits{S.}}, \bauthor{\bsnm{Cinar},~\bfnm{Suzan~Adin}\binits{S.~A.}}, \bauthor{\bsnm{Ekmek{\c{c}}i},~\bfnm{Sema~S{\i}rma}\binits{S.~S.}}, \bauthor{\bsnm{G{\"u}le{\c{c}}},~\bfnm{{\c{C}}a{\u{g}}r{\i}}\binits{{\c{C}}.}}, \bauthor{\bsnm{Abac{\i}},~\bfnm{Neslihan}\binits{N.}} \AND \bauthor{\bsnm{Ak{\c{c}}akaya},~\bfnm{Handan}\binits{H.}}
(\byear{2016}).
\btitle{Low dose monoethyl phthalate (MEP) exposure triggers proliferation by activating PDX-1 at 1.1 B4 human pancreatic beta cells}.
\bjournal{Food and Chemical Toxicology}
\bvolume{93}
\bpages{41--50}.
\end{barticle}
\endbibitem

\bibitem[\protect\citeauthoryear{Hahn, Carvalho and Mukherjee}{2013}]{hahn2013partial}
\begin{barticle}[author]
\bauthor{\bsnm{Hahn},~\bfnm{P~Richard}\binits{P.~R.}}, \bauthor{\bsnm{Carvalho},~\bfnm{Carlos~M}\binits{C.~M.}} \AND \bauthor{\bsnm{Mukherjee},~\bfnm{Sayan}\binits{S.}}
(\byear{2013}).
\btitle{Partial factor modeling: Predictor-dependent shrinkage for linear regression}.
\bjournal{Journal of the American Statistical Association}
\bvolume{108}
\bpages{999--1008}.
\end{barticle}
\endbibitem

\bibitem[\protect\citeauthoryear{Hamra and Buckley}{2018}]{hamra2018environmental}
\begin{barticle}[author]
\bauthor{\bsnm{Hamra},~\bfnm{Ghassan~B}\binits{G.~B.}} \AND \bauthor{\bsnm{Buckley},~\bfnm{Jessie~P}\binits{J.~P.}}
(\byear{2018}).
\btitle{Environmental exposure mixtures: questions and methods to address them}.
\bjournal{Current Epidemiology Reports}
\bvolume{5}
\bpages{160--165}.
\end{barticle}
\endbibitem

\bibitem[\protect\citeauthoryear{Hao, Feng and Zhang}{2018}]{hao2018model}
\begin{barticle}[author]
\bauthor{\bsnm{Hao},~\bfnm{Ning}\binits{N.}}, \bauthor{\bsnm{Feng},~\bfnm{Yang}\binits{Y.}} \AND \bauthor{\bsnm{Zhang},~\bfnm{Hao~Helen}\binits{H.~H.}}
(\byear{2018}).
\btitle{Model selection for high-dimensional quadratic regression via regularization}.
\bjournal{Journal of the American Statistical Association}
\bvolume{113}
\bpages{615--625}.
\end{barticle}
\endbibitem

\bibitem[\protect\citeauthoryear{Hoerl and Kennard}{1970}]{hoerl1970ridge}
\begin{barticle}[author]
\bauthor{\bsnm{Hoerl},~\bfnm{Arthur~E}\binits{A.~E.}} \AND \bauthor{\bsnm{Kennard},~\bfnm{Robert~W}\binits{R.~W.}}
(\byear{1970}).
\btitle{Ridge regression: Biased estimation for nonorthogonal problems}.
\bjournal{Technometrics}
\bvolume{12}
\bpages{55--67}.
\end{barticle}
\endbibitem

\bibitem[\protect\citeauthoryear{Hoff}{2007}]{hoff2007extending}
\begin{barticle}[author]
\bauthor{\bsnm{Hoff},~\bfnm{Peter~D}\binits{P.~D.}}
(\byear{2007}).
\btitle{Extending the Rank Likelihood for Semiparametric Copula Estimation}.
\bjournal{The Annals of Applied Statistics}
\bpages{265--283}.
\end{barticle}
\endbibitem

\bibitem[\protect\citeauthoryear{Hoff and Niu}{2012}]{hoff2012covariance}
\begin{barticle}[author]
\bauthor{\bsnm{Hoff},~\bfnm{Peter~D}\binits{P.~D.}} \AND \bauthor{\bsnm{Niu},~\bfnm{Xiaoyue}\binits{X.}}
(\byear{2012}).
\btitle{A covariance regression model}.
\bjournal{Statistica Sinica}
\bpages{729--753}.
\end{barticle}
\endbibitem

\bibitem[\protect\citeauthoryear{Hung and Wang}{2013}]{hung2013matrix}
\begin{barticle}[author]
\bauthor{\bsnm{Hung},~\bfnm{Hung}\binits{H.}} \AND \bauthor{\bsnm{Wang},~\bfnm{Chen-Chien}\binits{C.-C.}}
(\byear{2013}).
\btitle{Matrix variate logistic regression model with application to EEG data}.
\bjournal{Biostatistics}
\bvolume{14}
\bpages{189--202}.
\end{barticle}
\endbibitem

\bibitem[\protect\citeauthoryear{Hung et~al.}{2012}]{hung2012multilinear}
\begin{barticle}[author]
\bauthor{\bsnm{Hung},~\bfnm{Hung}\binits{H.}}, \bauthor{\bsnm{Wu},~\bfnm{Peishien}\binits{P.}}, \bauthor{\bsnm{Tu},~\bfnm{Iping}\binits{I.}} \AND \bauthor{\bsnm{Huang},~\bfnm{Suyun}\binits{S.}}
(\byear{2012}).
\btitle{On multilinear principal component analysis of order-two tensors}.
\bjournal{Biometrika}
\bvolume{99}
\bpages{569--583}.
\end{barticle}
\endbibitem

\bibitem[\protect\citeauthoryear{Hurst and Waxman}{2003}]{hurst2003activation}
\begin{barticle}[author]
\bauthor{\bsnm{Hurst},~\bfnm{Christopher~H}\binits{C.~H.}} \AND \bauthor{\bsnm{Waxman},~\bfnm{David~J}\binits{D.~J.}}
(\byear{2003}).
\btitle{Activation of PPAR$\alpha$ and PPAR$\gamma$ by environmental phthalate monoesters}.
\bjournal{Toxicological Sciences}
\bvolume{74}
\bpages{297--308}.
\end{barticle}
\endbibitem

\bibitem[\protect\citeauthoryear{James-Todd et~al.}{2012}]{james2012urinary}
\begin{barticle}[author]
\bauthor{\bsnm{James-Todd},~\bfnm{Tamarra}\binits{T.}}, \bauthor{\bsnm{Stahlhut},~\bfnm{Richard}\binits{R.}}, \bauthor{\bsnm{Meeker},~\bfnm{John~D}\binits{J.~D.}}, \bauthor{\bsnm{Powell},~\bfnm{Sheena-Gail}\binits{S.-G.}}, \bauthor{\bsnm{Hauser},~\bfnm{Russ}\binits{R.}}, \bauthor{\bsnm{Huang},~\bfnm{Tianyi}\binits{T.}} \AND \bauthor{\bsnm{Rich-Edwards},~\bfnm{Janet}\binits{J.}}
(\byear{2012}).
\btitle{Urinary phthalate metabolite concentrations and diabetes among women in the National Health and Nutrition Examination Survey (NHANES) 2001--2008}.
\bjournal{Environmental Health Perspectives}
\bvolume{120}
\bpages{1307--1313}.
\end{barticle}
\endbibitem

\bibitem[\protect\citeauthoryear{Jiang et~al.}{2020}]{jiang2020bayesian}
\begin{barticle}[author]
\bauthor{\bsnm{Jiang},~\bfnm{Bei}\binits{B.}}, \bauthor{\bsnm{Petkova},~\bfnm{Eva}\binits{E.}}, \bauthor{\bsnm{Tarpey},~\bfnm{Thaddeus}\binits{T.}} \AND \bauthor{\bsnm{Ogden},~\bfnm{R~Todd}\binits{R.~T.}}
(\byear{2020}).
\btitle{A Bayesian approach to joint modeling of matrix-valued imaging data and treatment outcome with applications to depression studies}.
\bjournal{Biometrics}
\bvolume{76}
\bpages{87--97}.
\end{barticle}
\endbibitem

\bibitem[\protect\citeauthoryear{Joubert et~al.}{2022}]{joubert2022powering}
\begin{barticle}[author]
\bauthor{\bsnm{Joubert},~\bfnm{Bonnie~R}\binits{B.~R.}}, \bauthor{\bsnm{Kioumourtzoglou},~\bfnm{Marianthi-Anna}\binits{M.-A.}}, \bauthor{\bsnm{Chamberlain},~\bfnm{Toccara}\binits{T.}}, \bauthor{\bsnm{Chen},~\bfnm{Hua~Yun}\binits{H.~Y.}}, \bauthor{\bsnm{Gennings},~\bfnm{Chris}\binits{C.}}, \bauthor{\bsnm{Turyk},~\bfnm{Mary~E}\binits{M.~E.}}, \bauthor{\bsnm{Miranda},~\bfnm{Marie~Lynn}\binits{M.~L.}}, \bauthor{\bsnm{Webster},~\bfnm{Thomas~F}\binits{T.~F.}}, \bauthor{\bsnm{Ensor},~\bfnm{Katherine~B}\binits{K.~B.}}, \bauthor{\bsnm{Dunson},~\bfnm{David~B}\binits{D.~B.}} \betal{et~al.}
(\byear{2022}).
\btitle{Powering research through innovative methods for mixtures in epidemiology (PRIME) program: novel and expanded statistical methods}.
\bjournal{International Journal of Environmental Research and Public Health}
\bvolume{19}
\bpages{1378}.
\end{barticle}
\endbibitem

\bibitem[\protect\citeauthoryear{Just et~al.}{2012}]{just2012prenatal}
\begin{barticle}[author]
\bauthor{\bsnm{Just},~\bfnm{Allan~C}\binits{A.~C.}}, \bauthor{\bsnm{Whyatt},~\bfnm{Robin~M}\binits{R.~M.}}, \bauthor{\bsnm{Perzanowski},~\bfnm{Matthew~S}\binits{M.~S.}}, \bauthor{\bsnm{Calafat},~\bfnm{Antonia~M}\binits{A.~M.}}, \bauthor{\bsnm{Perera},~\bfnm{Frederica~P}\binits{F.~P.}}, \bauthor{\bsnm{Goldstein},~\bfnm{Inge~F}\binits{I.~F.}}, \bauthor{\bsnm{Chen},~\bfnm{Qixuan}\binits{Q.}}, \bauthor{\bsnm{Rundle},~\bfnm{Andrew~G}\binits{A.~G.}} \AND \bauthor{\bsnm{Miller},~\bfnm{Rachel~L}\binits{R.~L.}}
(\byear{2012}).
\btitle{Prenatal exposure to butylbenzyl phthalate and early eczema in an urban cohort}.
\bjournal{Environmental Health Perspectives}
\bvolume{120}
\bpages{1475--1480}.
\end{barticle}
\endbibitem

\bibitem[\protect\citeauthoryear{Kahn et~al.}{2020}]{kahn2020endocrine}
\begin{barticle}[author]
\bauthor{\bsnm{Kahn},~\bfnm{Linda~G}\binits{L.~G.}}, \bauthor{\bsnm{Philippat},~\bfnm{Claire}\binits{C.}}, \bauthor{\bsnm{Nakayama},~\bfnm{Shoji~F}\binits{S.~F.}}, \bauthor{\bsnm{Slama},~\bfnm{R{\'e}my}\binits{R.}} \AND \bauthor{\bsnm{Trasande},~\bfnm{Leonardo}\binits{L.}}
(\byear{2020}).
\btitle{Endocrine-disrupting chemicals: implications for human health}.
\bjournal{The lancet Diabetes \& Endocrinology}
\bvolume{8}
\bpages{703--718}.
\end{barticle}
\endbibitem

\bibitem[\protect\citeauthoryear{Kim et~al.}{2011}]{kim2011prenatal}
\begin{barticle}[author]
\bauthor{\bsnm{Kim},~\bfnm{Yeni}\binits{Y.}}, \bauthor{\bsnm{Ha},~\bfnm{Eun-Hee}\binits{E.-H.}}, \bauthor{\bsnm{Kim},~\bfnm{Eui-Jung}\binits{E.-J.}}, \bauthor{\bsnm{Park},~\bfnm{Hyesook}\binits{H.}}, \bauthor{\bsnm{Ha},~\bfnm{Mina}\binits{M.}}, \bauthor{\bsnm{Kim},~\bfnm{Ja-Hyeong}\binits{J.-H.}}, \bauthor{\bsnm{Hong},~\bfnm{Yun-Chul}\binits{Y.-C.}}, \bauthor{\bsnm{Chang},~\bfnm{Namsoo}\binits{N.}} \AND \bauthor{\bsnm{Kim},~\bfnm{Bung-Nyun}\binits{B.-N.}}
(\byear{2011}).
\btitle{Prenatal exposure to phthalates and infant development at 6 months: prospective Mothers and Children’s Environmental Health (MOCEH) study}.
\bjournal{Environmental Health Perspectives}
\bvolume{119}
\bpages{1495--1500}.
\end{barticle}
\endbibitem

\bibitem[\protect\citeauthoryear{Koch et~al.}{2012}]{koch2012di}
\begin{barticle}[author]
\bauthor{\bsnm{Koch},~\bfnm{HM}\binits{H.}}, \bauthor{\bsnm{Christensen},~\bfnm{KLY}\binits{K.}}, \bauthor{\bsnm{Harth},~\bfnm{V}\binits{V.}}, \bauthor{\bsnm{Lorber},~\bfnm{M}\binits{M.}} \AND \bauthor{\bsnm{Br{\"u}ning},~\bfnm{TH}\binits{T.}}
(\byear{2012}).
\btitle{Di-n-butyl phthalate (DnBP) and diisobutyl phthalate (DiBP) metabolism in a human volunteer after single oral doses}.
\bjournal{Archives of Toxicology}
\bvolume{86}
\bpages{1829--1839}.
\end{barticle}
\endbibitem

\bibitem[\protect\citeauthoryear{Lang et~al.}{2008}]{lang2008association}
\begin{barticle}[author]
\bauthor{\bsnm{Lang},~\bfnm{Iain~A}\binits{I.~A.}}, \bauthor{\bsnm{Galloway},~\bfnm{Tamara~S}\binits{T.~S.}}, \bauthor{\bsnm{Scarlett},~\bfnm{Alan}\binits{A.}}, \bauthor{\bsnm{Henley},~\bfnm{William~E}\binits{W.~E.}}, \bauthor{\bsnm{Depledge},~\bfnm{Michael}\binits{M.}}, \bauthor{\bsnm{Wallace},~\bfnm{Robert~B}\binits{R.~B.}} \AND \bauthor{\bsnm{Melzer},~\bfnm{David}\binits{D.}}
(\byear{2008}).
\btitle{Association of urinary bisphenol A concentration with medical disorders and laboratory abnormalities in adults}.
\bjournal{JAMA}
\bvolume{300}
\bpages{1303--1310}.
\end{barticle}
\endbibitem

\bibitem[\protect\citeauthoryear{Legramanti, Durante and Dunson}{2020}]{legramanti2020bayesian}
\begin{barticle}[author]
\bauthor{\bsnm{Legramanti},~\bfnm{Sirio}\binits{S.}}, \bauthor{\bsnm{Durante},~\bfnm{Daniele}\binits{D.}} \AND \bauthor{\bsnm{Dunson},~\bfnm{David~B}\binits{D.~B.}}
(\byear{2020}).
\btitle{Bayesian cumulative shrinkage for infinite factorizations}.
\bjournal{Biometrika}
\bvolume{107}
\bpages{745--752}.
\end{barticle}
\endbibitem

\bibitem[\protect\citeauthoryear{Liu and Wu}{1999}]{liu1999parameter}
\begin{barticle}[author]
\bauthor{\bsnm{Liu},~\bfnm{Jun~S}\binits{J.~S.}} \AND \bauthor{\bsnm{Wu},~\bfnm{Ying~Nian}\binits{Y.~N.}}
(\byear{1999}).
\btitle{Parameter expansion for data augmentation}.
\bjournal{Journal of the American Statistical Association}
\bvolume{94}
\bpages{1264--1274}.
\end{barticle}
\endbibitem

\bibitem[\protect\citeauthoryear{Liu et~al.}{2018}]{liu2018lagged}
\begin{barticle}[author]
\bauthor{\bsnm{Liu},~\bfnm{Shelley~H}\binits{S.~H.}}, \bauthor{\bsnm{Bobb},~\bfnm{Jennifer~F}\binits{J.~F.}}, \bauthor{\bsnm{Lee},~\bfnm{Kyu~Ha}\binits{K.~H.}}, \bauthor{\bsnm{Gennings},~\bfnm{Chris}\binits{C.}}, \bauthor{\bsnm{Claus~Henn},~\bfnm{Birgit}\binits{B.}}, \bauthor{\bsnm{Bellinger},~\bfnm{David}\binits{D.}}, \bauthor{\bsnm{Austin},~\bfnm{Christine}\binits{C.}}, \bauthor{\bsnm{Schnaas},~\bfnm{Lourdes}\binits{L.}}, \bauthor{\bsnm{Tellez-Rojo},~\bfnm{Martha~M}\binits{M.~M.}}, \bauthor{\bsnm{Hu},~\bfnm{Howard}\binits{H.}} \betal{et~al.}
(\byear{2018}).
\btitle{Lagged kernel machine regression for identifying time windows of susceptibility to exposures of complex mixtures}.
\bjournal{Biostatistics}
\bvolume{19}
\bpages{325--341}.
\end{barticle}
\endbibitem

\bibitem[\protect\citeauthoryear{Liu et~al.}{2022}]{liu2022cross}
\begin{barticle}[author]
\bauthor{\bsnm{Liu},~\bfnm{Jeremiah~Zhe}\binits{J.~Z.}}, \bauthor{\bsnm{Deng},~\bfnm{Wenying}\binits{W.}}, \bauthor{\bsnm{Lee},~\bfnm{Jane}\binits{J.}}, \bauthor{\bsnm{Lin},~\bfnm{Pi-i~Debby}\binits{P.-i.~D.}}, \bauthor{\bsnm{Valeri},~\bfnm{Linda}\binits{L.}}, \bauthor{\bsnm{Christiani},~\bfnm{David~C}\binits{D.~C.}}, \bauthor{\bsnm{Bellinger},~\bfnm{David~C}\binits{D.~C.}}, \bauthor{\bsnm{Wright},~\bfnm{Robert~O}\binits{R.~O.}}, \bauthor{\bsnm{Mazumdar},~\bfnm{Maitreyi~M}\binits{M.~M.}} \AND \bauthor{\bsnm{Coull},~\bfnm{Brent~A}\binits{B.~A.}}
(\byear{2022}).
\btitle{A cross-validated ensemble approach to robust hypothesis testing of continuous nonlinear interactions: application to nutrition-environment studies}.
\bjournal{Journal of the American Statistical Association}
\bvolume{117}
\bpages{561--573}.
\end{barticle}
\endbibitem

\bibitem[\protect\citeauthoryear{Massy}{1965}]{massy1965principal}
\begin{barticle}[author]
\bauthor{\bsnm{Massy},~\bfnm{William~F}\binits{W.~F.}}
(\byear{1965}).
\btitle{Principal components regression in exploratory statistical research}.
\bjournal{Journal of the American Statistical Association}
\bvolume{60}
\bpages{234--256}.
\end{barticle}
\endbibitem

\bibitem[\protect\citeauthoryear{Mork and Wilson}{2022}]{mork2022treed}
\begin{barticle}[author]
\bauthor{\bsnm{Mork},~\bfnm{Daniel}\binits{D.}} \AND \bauthor{\bsnm{Wilson},~\bfnm{Ander}\binits{A.}}
(\byear{2022}).
\btitle{Treed distributed lag nonlinear models}.
\bjournal{Biostatistics}
\bvolume{23}
\bpages{754--771}.
\end{barticle}
\endbibitem

\bibitem[\protect\citeauthoryear{Murray et~al.}{2013}]{murray2013bayesian}
\begin{barticle}[author]
\bauthor{\bsnm{Murray},~\bfnm{Jared~S}\binits{J.~S.}}, \bauthor{\bsnm{Dunson},~\bfnm{David~B}\binits{D.~B.}}, \bauthor{\bsnm{Carin},~\bfnm{Lawrence}\binits{L.}} \AND \bauthor{\bsnm{Lucas},~\bfnm{Joseph~E}\binits{J.~E.}}
(\byear{2013}).
\btitle{Bayesian Gaussian copula factor models for mixed data}.
\bjournal{Journal of the American Statistical Association}
\bvolume{108}
\bpages{656--665}.
\end{barticle}
\endbibitem

\bibitem[\protect\citeauthoryear{Nguyen, Herring and Engel}{2023}]{nguyen2023power}
\begin{barticle}[author]
\bauthor{\bsnm{Nguyen},~\bfnm{Phuc~H}\binits{P.~H.}}, \bauthor{\bsnm{Herring},~\bfnm{Amy~H}\binits{A.~H.}} \AND \bauthor{\bsnm{Engel},~\bfnm{Stephanie~M}\binits{S.~M.}}
(\byear{2023}).
\btitle{Power analysis of exposure mixture studies via monte carlo simulations}.
\bjournal{Statistics in Biosciences}
\bpages{1--26}.
\end{barticle}
\endbibitem

\bibitem[\protect\citeauthoryear{Ovaskainen et~al.}{2017}]{ovaskainen2017make}
\begin{barticle}[author]
\bauthor{\bsnm{Ovaskainen},~\bfnm{Otso}\binits{O.}}, \bauthor{\bsnm{Tikhonov},~\bfnm{Gleb}\binits{G.}}, \bauthor{\bsnm{Norberg},~\bfnm{Anna}\binits{A.}}, \bauthor{\bsnm{Guillaume~Blanchet},~\bfnm{F}\binits{F.}}, \bauthor{\bsnm{Duan},~\bfnm{Leo}\binits{L.}}, \bauthor{\bsnm{Dunson},~\bfnm{David}\binits{D.}}, \bauthor{\bsnm{Roslin},~\bfnm{Tomas}\binits{T.}} \AND \bauthor{\bsnm{Abrego},~\bfnm{Nerea}\binits{N.}}
(\byear{2017}).
\btitle{How to make more out of community data? A conceptual framework and its implementation as models and software}.
\bjournal{Ecology letters}
\bvolume{20}
\bpages{561--576}.
\end{barticle}
\endbibitem

\bibitem[\protect\citeauthoryear{Poworoznek}{2020}]{poworoznek2020package}
\begin{barticle}[author]
\bauthor{\bsnm{Poworoznek},~\bfnm{Evan}\binits{E.}}
(\byear{2020}).
\btitle{infinitefactor: Bayesian infinite factor models}.
\bjournal{CRAN Repository}.
\bnote{R package version 1.0}.
\end{barticle}
\endbibitem

\bibitem[\protect\citeauthoryear{Shanik et~al.}{2008}]{shanik2008insulin}
\begin{barticle}[author]
\bauthor{\bsnm{Shanik},~\bfnm{Michael~H}\binits{M.~H.}}, \bauthor{\bsnm{Xu},~\bfnm{Yuping}\binits{Y.}}, \bauthor{\bsnm{Skrha},~\bfnm{Jan}\binits{J.}}, \bauthor{\bsnm{Dankner},~\bfnm{Rachel}\binits{R.}}, \bauthor{\bsnm{Zick},~\bfnm{Yehiel}\binits{Y.}} \AND \bauthor{\bsnm{Roth},~\bfnm{Jesse}\binits{J.}}
(\byear{2008}).
\btitle{Insulin resistance and hyperinsulinemia: Is hyperinsulinemia the cart or the horse?}
\bjournal{Diabetes Care}
\bvolume{31}
\bpages{S262--S268}.
\end{barticle}
\endbibitem

\bibitem[\protect\citeauthoryear{Silva et~al.}{2004}]{silva2004urinary}
\begin{barticle}[author]
\bauthor{\bsnm{Silva},~\bfnm{Manori~J}\binits{M.~J.}}, \bauthor{\bsnm{Barr},~\bfnm{Dana~B}\binits{D.~B.}}, \bauthor{\bsnm{Reidy},~\bfnm{John~A}\binits{J.~A.}}, \bauthor{\bsnm{Malek},~\bfnm{Nicole~A}\binits{N.~A.}}, \bauthor{\bsnm{Hodge},~\bfnm{Carolyn~C}\binits{C.~C.}}, \bauthor{\bsnm{Caudill},~\bfnm{Samuel~P}\binits{S.~P.}}, \bauthor{\bsnm{Brock},~\bfnm{John~W}\binits{J.~W.}}, \bauthor{\bsnm{Needham},~\bfnm{Larry~L}\binits{L.~L.}} \AND \bauthor{\bsnm{Calafat},~\bfnm{Antonia~M}\binits{A.~M.}}
(\byear{2004}).
\btitle{Urinary levels of seven phthalate metabolites in the US population from the National Health and Nutrition Examination Survey (NHANES) 1999-2000.}
\bjournal{Environmental Health Perspectives}
\bvolume{112}
\bpages{331--338}.
\end{barticle}
\endbibitem

\bibitem[\protect\citeauthoryear{Spearman}{1904}]{spearman1904general}
\begin{barticle}[author]
\bauthor{\bsnm{Spearman},~\bfnm{Charles}\binits{C.}}
(\byear{1904}).
\btitle{General intelligence, objectively determined and measured}.
\bjournal{American Journal of Psychology}
\bvolume{15}
\bpages{201--293}.
\end{barticle}
\endbibitem

\bibitem[\protect\citeauthoryear{Stafoggia et~al.}{2017}]{stafoggia2017statistical}
\begin{barticle}[author]
\bauthor{\bsnm{Stafoggia},~\bfnm{Massimo}\binits{M.}}, \bauthor{\bsnm{Breitner},~\bfnm{Susanne}\binits{S.}}, \bauthor{\bsnm{Hampel},~\bfnm{Regina}\binits{R.}} \AND \bauthor{\bsnm{Basaga{\~n}a},~\bfnm{Xavier}\binits{X.}}
(\byear{2017}).
\btitle{Statistical approaches to address multi-pollutant mixtures and multiple exposures: the state of the science}.
\bjournal{Current Environmental Health Reports}
\bvolume{4}
\bpages{481--490}.
\end{barticle}
\endbibitem

\bibitem[\protect\citeauthoryear{Stahlhut et~al.}{2007}]{stahlhut2007concentrations}
\begin{barticle}[author]
\bauthor{\bsnm{Stahlhut},~\bfnm{Richard~W}\binits{R.~W.}}, \bauthor{\bparticle{van} \bsnm{Wijngaarden},~\bfnm{Edwin}\binits{E.}}, \bauthor{\bsnm{Dye},~\bfnm{Timothy~D}\binits{T.~D.}}, \bauthor{\bsnm{Cook},~\bfnm{Stephen}\binits{S.}} \AND \bauthor{\bsnm{Swan},~\bfnm{Shanna~H}\binits{S.~H.}}
(\byear{2007}).
\btitle{Concentrations of urinary phthalate metabolites are associated with increased waist circumference and insulin resistance in adult US males}.
\bjournal{Environmental Health Perspectives}
\bvolume{115}
\bpages{876--882}.
\end{barticle}
\endbibitem

\bibitem[\protect\citeauthoryear{Stojanoska et~al.}{2017}]{stojanoska2017influence}
\begin{barticle}[author]
\bauthor{\bsnm{Stojanoska},~\bfnm{Milica~Medic}\binits{M.~M.}}, \bauthor{\bsnm{Milosevic},~\bfnm{Natasa}\binits{N.}}, \bauthor{\bsnm{Milic},~\bfnm{Natasa}\binits{N.}} \AND \bauthor{\bsnm{Abenavoli},~\bfnm{Ludovico}\binits{L.}}
(\byear{2017}).
\btitle{The influence of phthalates and bisphenol A on the obesity development and glucose metabolism disorders}.
\bjournal{Endocrine}
\bvolume{55}
\bpages{666--681}.
\end{barticle}
\endbibitem

\bibitem[\protect\citeauthoryear{Stuart}{2010}]{stuart2010matching}
\begin{barticle}[author]
\bauthor{\bsnm{Stuart},~\bfnm{Elizabeth~A}\binits{E.~A.}}
(\byear{2010}).
\btitle{Matching methods for causal inference: A review and a look forward}.
\bjournal{Statistical Science}
\bvolume{25}
\bpages{1--21}.
\end{barticle}
\endbibitem

\bibitem[\protect\citeauthoryear{Sun et~al.}{2014}]{sun2014association}
\begin{barticle}[author]
\bauthor{\bsnm{Sun},~\bfnm{Qi}\binits{Q.}}, \bauthor{\bsnm{Cornelis},~\bfnm{Marilyn~C}\binits{M.~C.}}, \bauthor{\bsnm{Townsend},~\bfnm{Mary~K}\binits{M.~K.}}, \bauthor{\bsnm{Tobias},~\bfnm{Deirdre~K}\binits{D.~K.}}, \bauthor{\bsnm{Eliassen},~\bfnm{A~Heather}\binits{A.~H.}}, \bauthor{\bsnm{Franke},~\bfnm{Adrian~A}\binits{A.~A.}}, \bauthor{\bsnm{Hauser},~\bfnm{Russ}\binits{R.}} \AND \bauthor{\bsnm{Hu},~\bfnm{Frank~B}\binits{F.~B.}}
(\byear{2014}).
\btitle{Association of urinary concentrations of bisphenol A and phthalate metabolites with risk of type 2 diabetes: a prospective investigation in the Nurses’ Health Study (NHS) and NHSII cohorts}.
\bjournal{Environmental Health Perspectives}
\bvolume{122}
\bpages{616--623}.
\end{barticle}
\endbibitem

\bibitem[\protect\citeauthoryear{Tanner, Bornehag and Gennings}{2019}]{tanner2019repeated}
\begin{barticle}[author]
\bauthor{\bsnm{Tanner},~\bfnm{Eva~M}\binits{E.~M.}}, \bauthor{\bsnm{Bornehag},~\bfnm{Carl-Gustaf}\binits{C.-G.}} \AND \bauthor{\bsnm{Gennings},~\bfnm{Chris}\binits{C.}}
(\byear{2019}).
\btitle{Repeated holdout validation for weighted quantile sum regression}.
\bjournal{MethodsX}
\bvolume{6}
\bpages{2855--2860}.
\end{barticle}
\endbibitem

\bibitem[\protect\citeauthoryear{Tibshirani}{1996}]{tibshirani1996regression}
\begin{barticle}[author]
\bauthor{\bsnm{Tibshirani},~\bfnm{Robert}\binits{R.}}
(\byear{1996}).
\btitle{Regression shrinkage and selection via the lasso}.
\bjournal{Journal of the Royal Statistical Society: Series B}
\bvolume{58}
\bpages{267--288}.
\end{barticle}
\endbibitem

\bibitem[\protect\citeauthoryear{Tibshirani et~al.}{2005}]{tibshirani2005sparsity}
\begin{barticle}[author]
\bauthor{\bsnm{Tibshirani},~\bfnm{Robert}\binits{R.}}, \bauthor{\bsnm{Saunders},~\bfnm{Michael}\binits{M.}}, \bauthor{\bsnm{Rosset},~\bfnm{Saharon}\binits{S.}}, \bauthor{\bsnm{Zhu},~\bfnm{Ji}\binits{J.}} \AND \bauthor{\bsnm{Knight},~\bfnm{Keith}\binits{K.}}
(\byear{2005}).
\btitle{Sparsity and smoothness via the fused lasso}.
\bjournal{Journal of the Royal Statistical Society: Series B}
\bvolume{67}
\bpages{91--108}.
\end{barticle}
\endbibitem

\bibitem[\protect\citeauthoryear{Van~Buuren and Groothuis-Oudshoorn}{2011}]{van2011mice}
\begin{barticle}[author]
\bauthor{\bsnm{Van~Buuren},~\bfnm{Stef}\binits{S.}} \AND \bauthor{\bsnm{Groothuis-Oudshoorn},~\bfnm{Karin}\binits{K.}}
(\byear{2011}).
\btitle{mice: Multivariate imputation by chained equations in R}.
\bjournal{Journal of Statistical Software}
\bvolume{45}
\bpages{1--67}.
\end{barticle}
\endbibitem

\bibitem[\protect\citeauthoryear{van~der Pas et~al.}{2016}]{van2016horseshoe}
\begin{barticle}[author]
\bauthor{\bparticle{van~der} \bsnm{Pas},~\bfnm{Stephanie}\binits{S.}}, \bauthor{\bsnm{Scott},~\bfnm{James}\binits{J.}}, \bauthor{\bsnm{Chakraborty},~\bfnm{Antik}\binits{A.}} \AND \bauthor{\bsnm{Bhattacharya},~\bfnm{Anirban}\binits{A.}}
(\byear{2016}).
\btitle{horseshoe: Implementation of the horseshoe prior}.
\bjournal{CRAN Repository}.
\bnote{R package version 0.2.0}.
\end{barticle}
\endbibitem

\bibitem[\protect\citeauthoryear{Wang, Jiang and Zhu}{2021}]{wang2021penalized}
\begin{barticle}[author]
\bauthor{\bsnm{Wang},~\bfnm{Cheng}\binits{C.}}, \bauthor{\bsnm{Jiang},~\bfnm{Binyan}\binits{B.}} \AND \bauthor{\bsnm{Zhu},~\bfnm{Liping}\binits{L.}}
(\byear{2021}).
\btitle{Penalized interaction estimation for ultrahigh dimensional quadratic regression}.
\bjournal{Statistica Sinica}
\bvolume{31}
\bpages{1549--1570}.
\end{barticle}
\endbibitem

\bibitem[\protect\citeauthoryear{Wang et~al.}{2022}]{wang2022semiparametric}
\begin{barticle}[author]
\bauthor{\bsnm{Wang},~\bfnm{Yuyan}\binits{Y.}}, \bauthor{\bsnm{Ghassabian},~\bfnm{Akhgar}\binits{A.}}, \bauthor{\bsnm{Gu},~\bfnm{Bo}\binits{B.}}, \bauthor{\bsnm{Afanasyeva},~\bfnm{Yelena}\binits{Y.}}, \bauthor{\bsnm{Li},~\bfnm{Yiwei}\binits{Y.}}, \bauthor{\bsnm{Trasande},~\bfnm{Leonardo}\binits{L.}} \AND \bauthor{\bsnm{Liu},~\bfnm{Mengling}\binits{M.}}
(\byear{2022}).
\btitle{Semiparametric distributed lag quantile regression for modeling time-dependent exposure mixtures}.
\bjournal{Biometrics}.
\end{barticle}
\endbibitem

\bibitem[\protect\citeauthoryear{Warren et~al.}{2012}]{warren2012spatial}
\begin{barticle}[author]
\bauthor{\bsnm{Warren},~\bfnm{Joshua}\binits{J.}}, \bauthor{\bsnm{Fuentes},~\bfnm{Montserrat}\binits{M.}}, \bauthor{\bsnm{Herring},~\bfnm{Amy}\binits{A.}} \AND \bauthor{\bsnm{Langlois},~\bfnm{Peter}\binits{P.}}
(\byear{2012}).
\btitle{Spatial-temporal modeling of the association between air pollution exposure and preterm birth: identifying critical windows of exposure}.
\bjournal{Biometrics}
\bvolume{68}
\bpages{1157--1167}.
\end{barticle}
\endbibitem

\bibitem[\protect\citeauthoryear{Warren et~al.}{2016}]{warren2016bayesian}
\begin{barticle}[author]
\bauthor{\bsnm{Warren},~\bfnm{Joshua~L}\binits{J.~L.}}, \bauthor{\bsnm{Stingone},~\bfnm{Jeanette~A}\binits{J.~A.}}, \bauthor{\bsnm{Herring},~\bfnm{Amy~H}\binits{A.~H.}}, \bauthor{\bsnm{Luben},~\bfnm{Thomas~J}\binits{T.~J.}}, \bauthor{\bsnm{Fuentes},~\bfnm{Montserrat}\binits{M.}}, \bauthor{\bsnm{Aylsworth},~\bfnm{Arthur~S}\binits{A.~S.}}, \bauthor{\bsnm{Langlois},~\bfnm{Peter~H}\binits{P.~H.}}, \bauthor{\bsnm{Botto},~\bfnm{Lorenzo~D}\binits{L.~D.}}, \bauthor{\bsnm{Correa},~\bfnm{Adolfo}\binits{A.}}, \bauthor{\bsnm{Olshan},~\bfnm{Andrew~F}\binits{A.~F.}} \betal{et~al.}
(\byear{2016}).
\btitle{Bayesian multinomial probit modeling of daily windows of susceptibility for maternal PM2. 5 exposure and congenital heart defects}.
\bjournal{Statistics in Medicine}
\bvolume{35}
\bpages{2786--2801}.
\end{barticle}
\endbibitem

\bibitem[\protect\citeauthoryear{Warren et~al.}{2020}]{warren2020critical}
\begin{barticle}[author]
\bauthor{\bsnm{Warren},~\bfnm{Joshua~L}\binits{J.~L.}}, \bauthor{\bsnm{Kong},~\bfnm{Wenjing}\binits{W.}}, \bauthor{\bsnm{Luben},~\bfnm{Thomas~J}\binits{T.~J.}} \AND \bauthor{\bsnm{Chang},~\bfnm{Howard~H}\binits{H.~H.}}
(\byear{2020}).
\btitle{Critical window variable selection: estimating the impact of air pollution on very preterm birth}.
\bjournal{Biostatistics}
\bvolume{21}
\bpages{790--806}.
\end{barticle}
\endbibitem

\bibitem[\protect\citeauthoryear{Warren et~al.}{2022}]{warren2022critical}
\begin{barticle}[author]
\bauthor{\bsnm{Warren},~\bfnm{Joshua~L}\binits{J.~L.}}, \bauthor{\bsnm{Chang},~\bfnm{Howard~H}\binits{H.~H.}}, \bauthor{\bsnm{Warren},~\bfnm{Lauren~K}\binits{L.~K.}}, \bauthor{\bsnm{Strickland},~\bfnm{Matthew~J}\binits{M.~J.}}, \bauthor{\bsnm{Darrow},~\bfnm{Lyndsey~A}\binits{L.~A.}} \AND \bauthor{\bsnm{Mulholland},~\bfnm{James~A}\binits{J.~A.}}
(\byear{2022}).
\btitle{Critical window variable selection for mixtures: estimating the impact of multiple air pollutants on stillbirth}.
\bjournal{The Annals of Applied Statistics}
\bvolume{16}
\bpages{1633--1652}.
\end{barticle}
\endbibitem

\bibitem[\protect\citeauthoryear{Welty et~al.}{2009}]{welty2009bayesian}
\begin{barticle}[author]
\bauthor{\bsnm{Welty},~\bfnm{Leah~J}\binits{L.~J.}}, \bauthor{\bsnm{Peng},~\bfnm{Roger~D}\binits{R.~D.}}, \bauthor{\bsnm{Zeger},~\bfnm{Scott~L}\binits{S.~L.}} \AND \bauthor{\bsnm{Dominici},~\bfnm{Francesca}\binits{F.}}
(\byear{2009}).
\btitle{Bayesian distributed lag models: estimating effects of particulate matter air pollution on daily mortality}.
\bjournal{Biometrics}
\bvolume{65}
\bpages{282--291}.
\end{barticle}
\endbibitem

\bibitem[\protect\citeauthoryear{Wenzel et~al.}{2018}]{wenzel2018prevalence}
\begin{barticle}[author]
\bauthor{\bsnm{Wenzel},~\bfnm{Abby~G}\binits{A.~G.}}, \bauthor{\bsnm{Brock},~\bfnm{John~W}\binits{J.~W.}}, \bauthor{\bsnm{Cruze},~\bfnm{Lori}\binits{L.}}, \bauthor{\bsnm{Newman},~\bfnm{Roger~B}\binits{R.~B.}}, \bauthor{\bsnm{Unal},~\bfnm{Elizabeth~R}\binits{E.~R.}}, \bauthor{\bsnm{Wolf},~\bfnm{Bethany~J}\binits{B.~J.}}, \bauthor{\bsnm{Somerville},~\bfnm{Stephen~E}\binits{S.~E.}} \AND \bauthor{\bsnm{Kucklick},~\bfnm{John~R}\binits{J.~R.}}
(\byear{2018}).
\btitle{Prevalence and predictors of phthalate exposure in pregnant women in Charleston, SC}.
\bjournal{Chemosphere}
\bvolume{193}
\bpages{394--402}.
\end{barticle}
\endbibitem

\bibitem[\protect\citeauthoryear{West}{2003}]{west2003bayesian}
\begin{barticle}[author]
\bauthor{\bsnm{West},~\bfnm{Mike}\binits{M.}}
(\byear{2003}).
\btitle{Bayesian factor regression models in the “large p, small n” paradigm}.
\bjournal{Bayesian Statistics}.
\end{barticle}
\endbibitem

\bibitem[\protect\citeauthoryear{Whyatt et~al.}{2012}]{whyatt2012maternal}
\begin{barticle}[author]
\bauthor{\bsnm{Whyatt},~\bfnm{Robin~M}\binits{R.~M.}}, \bauthor{\bsnm{Liu},~\bfnm{Xinhua}\binits{X.}}, \bauthor{\bsnm{Rauh},~\bfnm{Virginia~A}\binits{V.~A.}}, \bauthor{\bsnm{Calafat},~\bfnm{Antonia~M}\binits{A.~M.}}, \bauthor{\bsnm{Just},~\bfnm{Allan~C}\binits{A.~C.}}, \bauthor{\bsnm{Hoepner},~\bfnm{Lori}\binits{L.}}, \bauthor{\bsnm{Diaz},~\bfnm{Diurka}\binits{D.}}, \bauthor{\bsnm{Quinn},~\bfnm{James}\binits{J.}}, \bauthor{\bsnm{Adibi},~\bfnm{Jennifer}\binits{J.}}, \bauthor{\bsnm{Perera},~\bfnm{Frederica~P}\binits{F.~P.}} \betal{et~al.}
(\byear{2012}).
\btitle{Maternal prenatal urinary phthalate metabolite concentrations and child mental, psychomotor, and behavioral development at 3 years of age}.
\bjournal{Environmental Health Perspectives}
\bvolume{120}
\bpages{290--295}.
\end{barticle}
\endbibitem

\bibitem[\protect\citeauthoryear{Wilson et~al.}{2022}]{wilson2022kernel}
\begin{barticle}[author]
\bauthor{\bsnm{Wilson},~\bfnm{Ander}\binits{A.}}, \bauthor{\bsnm{Hsu},~\bfnm{Hsiao-Hsien~Leon}\binits{H.-H.~L.}}, \bauthor{\bsnm{Chiu},~\bfnm{Yueh-Hsiu~Mathilda}\binits{Y.-H.~M.}}, \bauthor{\bsnm{Wright},~\bfnm{Robert~O}\binits{R.~O.}}, \bauthor{\bsnm{Wright},~\bfnm{Rosalind~J}\binits{R.~J.}} \AND \bauthor{\bsnm{Coull},~\bfnm{Brent~A}\binits{B.~A.}}
(\byear{2022}).
\btitle{Kernel machine and distributed lag models for assessing windows of susceptibility to environmental mixtures in children’s health studies}.
\bjournal{The Annals of Applied Statistics}
\bvolume{16}
\bpages{1090--1110}.
\end{barticle}
\endbibitem

\bibitem[\protect\citeauthoryear{Yuan and Lin}{2006}]{yuan2006model}
\begin{barticle}[author]
\bauthor{\bsnm{Yuan},~\bfnm{Ming}\binits{M.}} \AND \bauthor{\bsnm{Lin},~\bfnm{Yi}\binits{Y.}}
(\byear{2006}).
\btitle{Model selection and estimation in regression with grouped variables}.
\bjournal{Journal of the Royal Statistical Society: Series B}
\bvolume{68}
\bpages{49--67}.
\end{barticle}
\endbibitem

\bibitem[\protect\citeauthoryear{Zhou, Li and Zhu}{2013}]{zhou2013tensor}
\begin{barticle}[author]
\bauthor{\bsnm{Zhou},~\bfnm{Hua}\binits{H.}}, \bauthor{\bsnm{Li},~\bfnm{Lexin}\binits{L.}} \AND \bauthor{\bsnm{Zhu},~\bfnm{Hongtu}\binits{H.}}
(\byear{2013}).
\btitle{Tensor regression with applications in neuroimaging data analysis}.
\bjournal{Journal of the American Statistical Association}
\bvolume{108}
\bpages{540--552}.
\end{barticle}
\endbibitem

\end{thebibliography}

%% or include bibliography directly:
% \begin{thebibliography}{}
% \bibitem[\protect\citeauthoryear{???}{???}]{b1}
% \end{thebibliography}

\end{document}

\maketitle

\section{Data generation for Section 1 simulations}

For both the rank-1 simulation in Figure 1 and the rank-2 simulation in Figure 2, we generated $x_i$ for $i=1,...,200$ as
$$x_i \sim N_{15}(0, I_{5} \otimes \Psi)$$
where $x_i = (x_{i11},x_{i12},x_{i13},...,x_{i51},x_{i52}, x_{i53})^T$ and $\Psi \in \mathbb{R}^{3 \times 3}$ is a compound-symmetric correlation matrix with correlation $0.7$. Then, we generated $y_i$ for $i=1,...,200$ as
$$y_i \sim N(\theta^T x_i, 5)$$
where for the rank-1 simulations, $\theta = vec(\omega \beta^T)$ with $\omega = (0.25,0.25,0.5)^T$ and $\beta = (1,3,2,-1,-2)^T$, and for the rank-2 simulations, $\theta = vec(\omega_1 \beta_1^T + \omega_2 \beta_2^T)$ with $\omega_1=(0.25,0.25,0.5)^T$, $\beta_1=(1,3,2,0,0)^T$, $\omega_2=(0.8,0.1,0.1)^T$, and $\beta_2=(0,0,0,-1,-2)^T$.

\section{Full LowFR model description including prior distributions}\label{app_model}
Let $y_i \in \mathbb{R}$, $x_i = (x_{i11},...,x_{i1T},...,x_{ip1},...,x_{ipT})^T \in \mathbb{R}^{pT}$, and $\eta_i = (\eta_{i11},...,\eta_{i1T}, \\ ...,\eta_{ik1},...,\eta_{ikT})^T \in \mathbb{R}^{kT}$. Then we define the following factor model for $x_i$:
$$x_{i} = (\Lambda \otimes I_T) \eta_{i} + \varepsilon_{i}$$
$$\eta_{i} \sim N_{kT}(0, I_k \otimes \Phi), \,\,\,\,\,\,\,\,\, \varepsilon_{i} \sim N_{pT}(0, \Sigma \otimes \Phi)$$
$$\Sigma = \text{diag}(\sigma_1^2,...,\sigma_p^2)$$
$$\Phi = \begin{pmatrix}
    1 & \phi & \hdots & \phi \\
    \phi & 1 & \hdots & \phi \\
    \vdots & \ddots & & \vdots \\
    \phi & \hdots & & 1 
\end{pmatrix}$$
$$\lambda_{jh} \sim N(0,10) \,\,\,\,\,\, \text{ for } j=1,...,p, \,\,\, h=1,...,k$$
$$\phi \sim Unif(0,1), \,\,\,\,\, \sigma_1^2,...,\sigma_p^2 \sim IG(1,1) \,\,\,\,\,\, \text{ for } j =1,...,p$$
Given the above, we model $y_i$ for $i=1,...,n$ as
$$y_i \sim N(\mu + \theta^T \eta_i + \eta_i^T \Omega \eta_i, \sigma^2)$$
$$\sigma^2 \sim IG(1,1)$$
where the main effects $\theta$ are modeled as
$$\theta = vec(\sum_{l=1}^{H} \omega_l \beta_l^T)$$
$$\beta_{lj} \sim N(0,\xi_{\beta,l,j}^{-1} \tau_l^{-1}), \,\,\,\, \text{ for } l = 1,...,H, \,\,\,\, j=1,...,k$$
$$\omega_{lt} \sim N(0,\xi_{\omega,l,t}^{-1} \tau_l^{-1}), \,\,\,\, \text{ for } l = 1,...,H, \,\,\,\, t=1,...,T$$
$$\xi_{\beta,l,j} \sim Gamma(3/2, 3/2) \,\,\,\,\, \text{ for } l=1,...,H, \,\,\,\, j = 1,..., k$$
$$\xi_{\omega,l,t} \sim Gamma(3/2, 3/2) \,\,\,\,\, \text{ for } l=1,...,H, \,\,\,\, t = 1,..., T$$
$$\tau_l = \prod_{m=1}^l \delta_l, \,\,\,\,\,\, \delta_1 \sim Gamma(a_1, 1), \,\,\,\,\,\, \delta_2,...,\delta_H \sim Gamma(a_2, 1)$$
$$a_1, a_2 \sim Gamma(2,1)$$
and the interaction matrix $\Omega$ is modeled as
$$\Omega = B \otimes W$$
$$b_{qr} \sim N(0,\xi_{B,q,r}^{-1} \tau_{int}^{-1}) \,\,\,\,\,\, \text{ for } q=1,...,k, \,\,\,\, r=1,...,k$$
$$w_{st} \sim N(0,\xi_{W,s,r}^{-1} \tau_{int}^{-1}) \,\,\,\,\,\, \text{ for } s=1,...,T, \,\,\,\, t=1,...,T.$$

\section{Alternative Theorem 3.1 for general $Cov(\eta_i)$ and $Cov(\varepsilon_i)$}\label{app_alt_thm1}
Consider the same setup as in (5), but with no assumptions on the form of $Cov(\eta_i)$ or $Cov(\varepsilon_i)$:
$$y_i = \mu + \theta^T \eta_i + \eta_i^T \Omega \eta_i + \varepsilon_i^{(y)}, \,\,\,\,\, \varepsilon_i^{(y)} \sim N(0, \sigma^2),$$
$$x_i = (\Lambda \otimes I_T) \eta_i + \varepsilon_i, \,\,\,\,\, \varepsilon_i \sim N(0, \Psi^{(\varepsilon)}),$$
$$\eta_i \sim N(0, \Psi^{(\eta)})$$
where $\Psi^{(\varepsilon)}$ and $\Psi^{(\eta)}$ are generic covariance matrices. Everything about the calculation in Appendix A goes through in this case, except we end up with a more general form for $\eta_i | x_i$:
$$\eta_i | x_i \sim N(Ax_i, V)$$
where
$$V = ((\Lambda^T \otimes I_T) \Psi^{(\varepsilon) -1} (\Lambda \otimes I_T) + \Psi^{(\eta) -1})^{-1}$$
and
$$A = V(\Lambda^T \otimes I_T) \Psi^{(\varepsilon) -1}.$$
Thus, we again get that
\begin{align*}
    \mathbb{E}[y_i | x_i] &= \mu + tr(\Omega V) + (\theta^T A) x_i + x_i^T (A^T \Omega A) x_i
\end{align*}
where $A$ and $V$ are specified above.

\section{CorrQuadReg model description}
We model $y_i | x_i$ as normally distributed such that
$$y_i = \mu + \sum_{j=1}^p \sum_{t=1}^T \theta_{jt} x_{ijt} + \sum_{j_1=1}^p \sum_{j_2=1}^{j_2} \sum_{t_1=1}^T \sum_{t_2=1}^T \gamma_{j_1,j_2,t_1,t_2} x_{i j_1 t_1} x_{i j_2 t_2} + \varepsilon_i$$
with $\varepsilon_i \sim N(0, \sigma^2)$. We give $\mu$ a $N(0,10)$ prior. Each of $\theta$ and $\Gamma$ is given a mean-zero multivariate normal prior such that $Var(\theta_{jt})=\nu_{main}$, $Var(\gamma_{j_1,j_2,t_1,t_2}) = \nu_{int}$, with
$$Cor(\theta_{jt}, \theta_{j't'}) = \begin{cases}
  1 & \text{if } j=j', t=t' \\
  \psi_{main}  & \text{if } j=j', t \neq t' \\
  0 & \text{otherwise}
\end{cases}$$
and
$$Cor(\gamma_{j_1,j_2,t_1,t_2}, \gamma_{j_1',j_2',t_1',t_2'}) = \begin{cases}
  1 & \text{if } j_1=j_1', j_2=j_2', t_1=t_1', t_2=t_2' \\
  \psi_{main}  & \text{if } j_1=j_1', j_2=j_2', (t_1,t_2) \neq (t_1', t_2') \\
  0 & \text{otherwise}
\end{cases}$$
with $\nu_{main} \sim IG(1,1)$, $\nu_{int} \sim IG(1,1)$, $\sigma \sim IG(1,1)$, $\psi_{main} \sim Unif(0,1)$, and $\psi_{int} \sim Unif(0,1)$.

\section{Marginal 2-DG posterior predictive plots}

\begin{figure}[]
\includegraphics[width=0.8\textwidth]{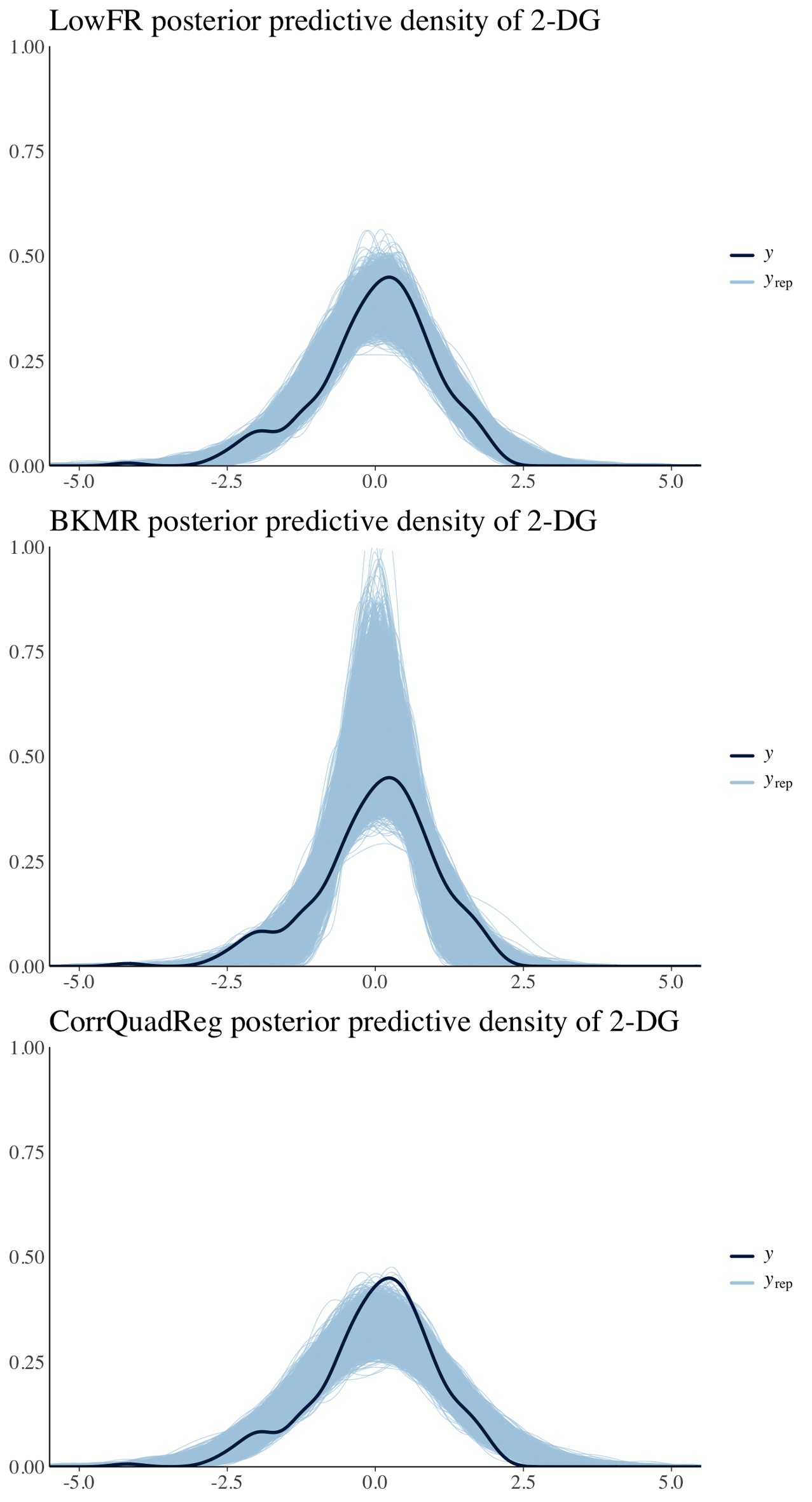}
\caption{Posterior predictive density of 2-DG for LowFR (top), BKMR (center), and CorrQuadReg (bottom).}
\label{fig_ppcheck_marginal}
\end{figure}

\newpage

\section{Alternate Theorem 3.1 for the model of Hung et al. (2012) and Jiang et al. (2020)}

In both Jiang et al. (2020) and Hung et al. (2012), the model for $x_i$, if we assume centered data and express $x_i = (x_{i11},...,x_{i1T}, ..., x_{ip1},..., x_{ipT})^T \in \mathbb{R}^{pT}$ as we do in our work, is
$$x_i = vec(B \otimes A) \eta_i + \varepsilon_i$$ 
$$\eta_i \sim N_{k T_0}(0, I), \,\,\,\,\, \varepsilon_i \sim N_{pT}(0, \phi^{-1} I)$$
where $B \in \mathbb{R}^{p \times k}$, $A \in \mathbb{R}^{T \times T_0}$, $k < p$, and $T_0 < T$. This induces the covariance model
$$x_i \sim N_{pT}(0, (B B^T) \otimes (A A^T) + \phi^{-1} I).$$
Given this setup, if we again use the outcome model
$$y_i = \mu + \theta^T \eta_i + \eta_i^T \Omega \eta_i + \varepsilon_i^{(y)}, \,\,\,\,\, \varepsilon_i^{(y)} \sim N(0, \sigma^2),$$
we can go through a very similar calculation to the one in the proof of Theorem 3.1 to yield
$$\eta_i | x_i \sim N_{kT_0}(M x_i, V)$$
where
$$V = (\phi (B^T B) \otimes (A^T A) + I_{kT_0})^{-1}$$
and
$$M = \phi V (B^T \otimes A^T).$$
And thus, we get
\begin{align*}
    \mathbb{E}[y_i | x_i] &= \mu + tr(\Omega V) + (\theta^T M) x_i + x_i^T (M^T \Omega M) x_i
\end{align*}
where $M$ and $V$ are defined above. Note that $M$ does not have a Kronecker structure here, as it does in our initial Theorem 3.1. Thus, the appealing properties noted in Corollary 3.2 do not apply in this case.

\newpage

\section{Simulation results for $T=p=10$}

To evaluate whether LowFR can find low-dimensional structure for larger values of $T$, we simulated data using the same generation mechanism as our Scenario 2 described in the paper, but with both $T$ and $p$ set to $10$. As can be seen from the results below, LowFR dramatically outperforms all competitors in this scenario, suggesting that it is able to take advantage of the rank-2 structure of the linear coefficients, as well as being well-specified for the Kronecker structure of the interactions, thus making estimation of the large number of interaction effects more tractable than for competitors.

\begin{table}[h]
\resizebox{\textwidth}{!}{
\begin{tabular}{llllllll}
\hline
                     &               & LowFR                        & BKMR                         & BKMR\_group                  & FIN                          & Horseshoe                    & CorrQuadReg                  \\ \hline
\multirow{10}{*}{S4} & Main MSE      & 5.4 * 10\textasciicircum{}-4 & 2.4 * 10\textasciicircum{}-3 & 1.6 * 10\textasciicircum{}-2 & 4.7 * 10\textasciicircum{}-2 & 6.6 * 10\textasciicircum{}-3 & 5.6 * 10\textasciicircum{}-3 \\
                     & Int MSE       & 2.0 * 10\textasciicircum{}-6 & 5.8 * 10\textasciicircum{}-1 & 9.2 * 10\textasciicircum{}-2 & 4.2 * 10\textasciicircum{}-5 & 9.9 * 10\textasciicircum{}-6 & 1.3 * 10\textasciicircum{}-3 \\
                     & Main Coverage & 0.99                         & 1.00                         & 0.17                         & 1.00                         & 0.042                        & 0.98                         \\
                     & Int Coverage  & 1.00                         & 0.83                         & 0.14                         & 1.00                         & 0.42                         & 0.97                         \\
                     & Main TP Rate  & 0.11                         & 0.00                         & 0.024                        & 0.00                         & 0.030                        & 0.17                         \\
                     & Main TN Rate  & N/A                          & N/A                          & N/A                          & N/A                          & N/A                          & N/A                          \\ \cline{2-8} 
                     & CE MSE        & 0.021                        & 0.33                         & 0.28                         & 3.06                         & 0.29                         & 0.14                         \\
                     & CE Coverage   & 0.99                         & 1.00                         & 0.83                         & 1.00                         & 0.22                         & 0.96                         \\
                     & CE TP Rate    & 0.61                         & 0.00                         & 0.37                         & 0.00                         & 0.26                         & 0.29                         \\
                     & CE TN Rate    & N/A                          & N/A                          & N/A                          & N/A                          & N/A                          & N/A                         
\end{tabular}}
\caption{Comparison of six models for simulated data with $n=200$, $p=10$, and $T=10$. Longitudinal exposures were generated from a factor model as in the LowFR setup. The quadratic regression coefficients were generated using the same generation scheme as from Scenario 2 described in the paper. In particular, the linear coefficients are rank-2, and the interaction matrix is generated as a Kronecker product of $p \times p$ and $T \times T$ matrices. CE indicates cumulative effects, defined as the expected change in outcome if a single exposure increases from $-1$ to $1$ at all measurement times, with all other exposures held constant at zero. Results for LowFR, BKMR, BKMR\_group, Horseshoe, and CorrQuadReg are averaged over 100 replicates. Due to computational time, results for FIN are averaged over 64 replicates.}
\end{table}

\newpage

\section{ELEMENT results with transformed data}

\begin{figure}[h]
\includegraphics[width=\textwidth]{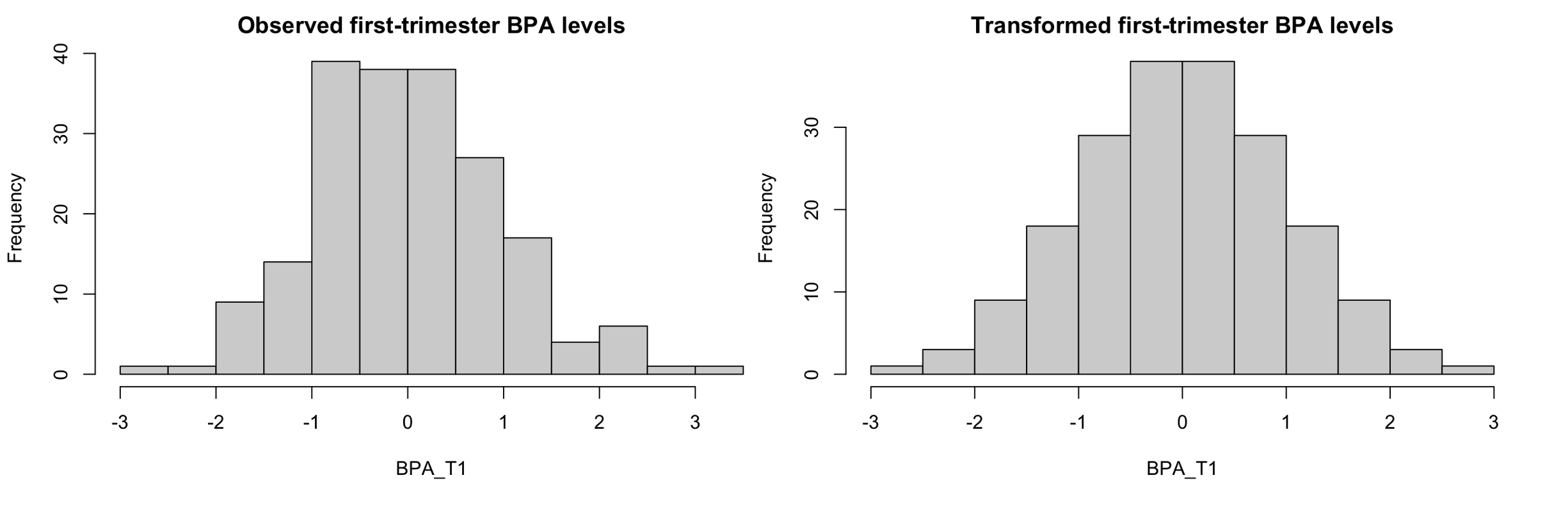}
\caption{Distribution of standardized and log-transformed first-trimester BPA levels (left), and the distribution of the same exposures after applying the inverse normal CDF transformation (right).}
\label{fig_transform_hist}
\end{figure}

To check sensitivity of our modeling approach to the assumption of multivariate normality of all the chemical exposures, we re-ran our ELEMENT analysis after applying an inverse CDF transform. This ensures that the marginal distribution of each exposure is well-modeled as Gaussian. For reference, histograms of first-trimester BPA values before and after transforming are shown in Figure \ref{fig_transform_hist}.

\begin{figure}[h]
\includegraphics[width=\textwidth]{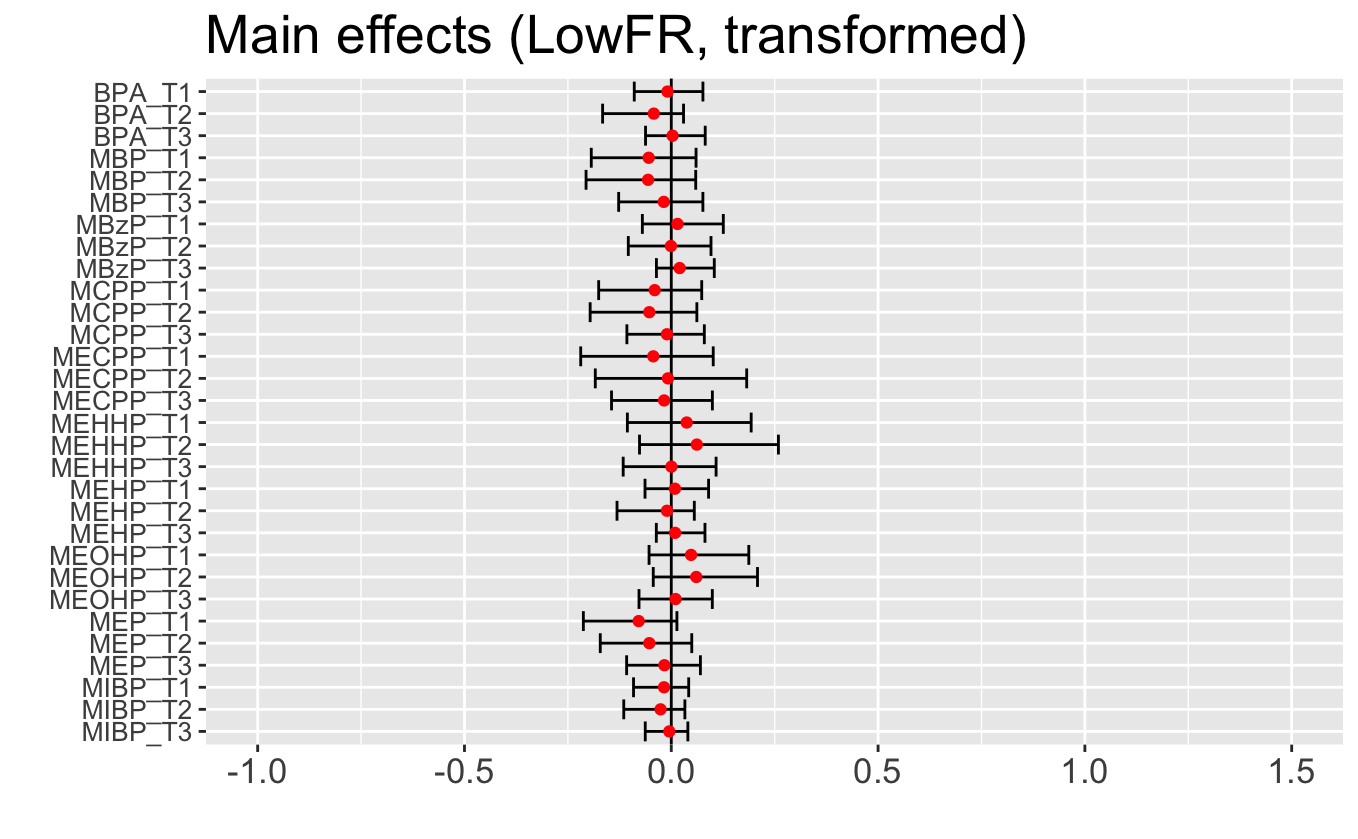}
\caption{Main effect coefficients for all measured exposures. (Transformed data.)}
\label{fig8a_transformed}
\end{figure}

\begin{figure}[h]
\includegraphics[width=\textwidth]{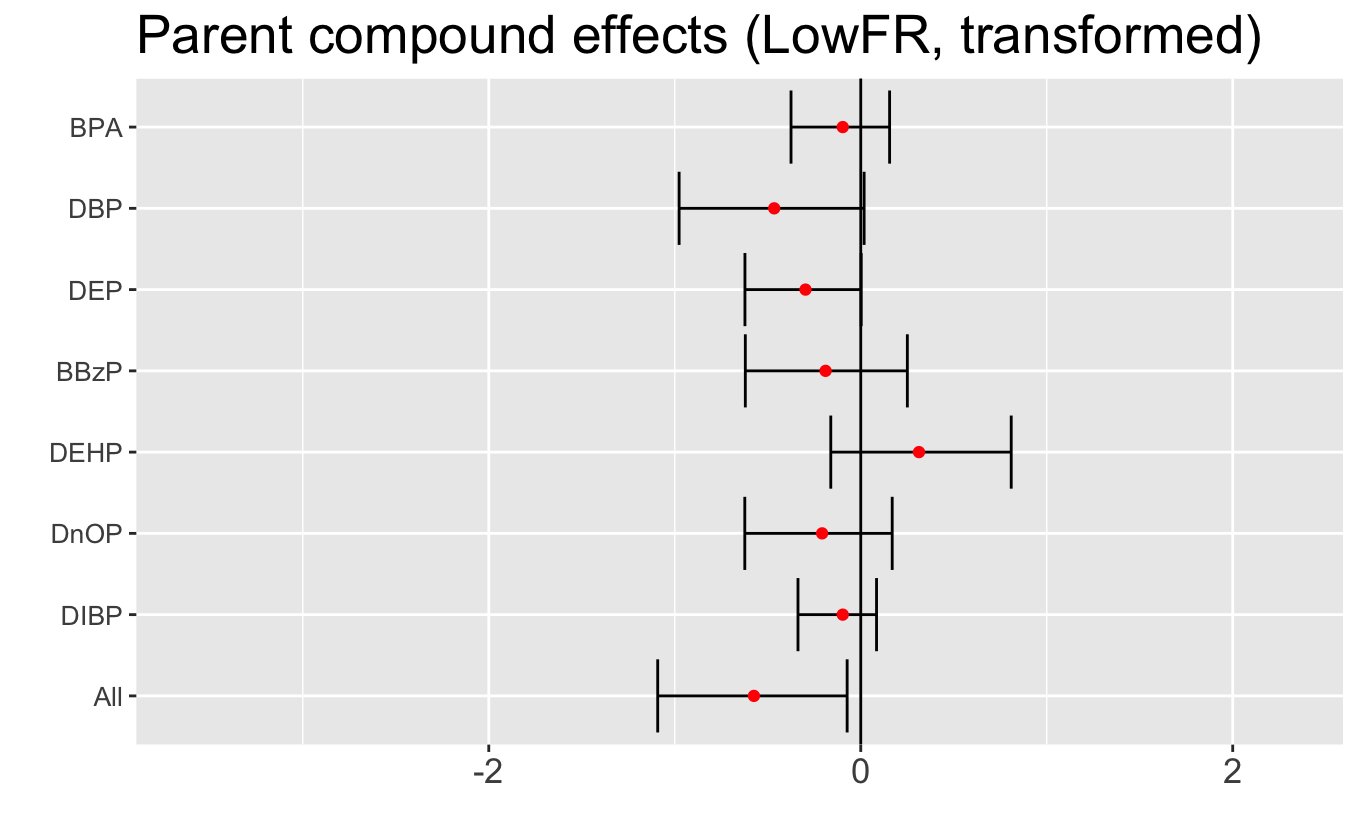}
\caption{Expected change in 2-DG when all metabolites of the given phthalates are increased from $-1$ to $1$ at all three trimesters. (Transformed data.)}
\label{fig8b_transformed}
\end{figure}

\begin{figure}[h]
\includegraphics[width=\textwidth]{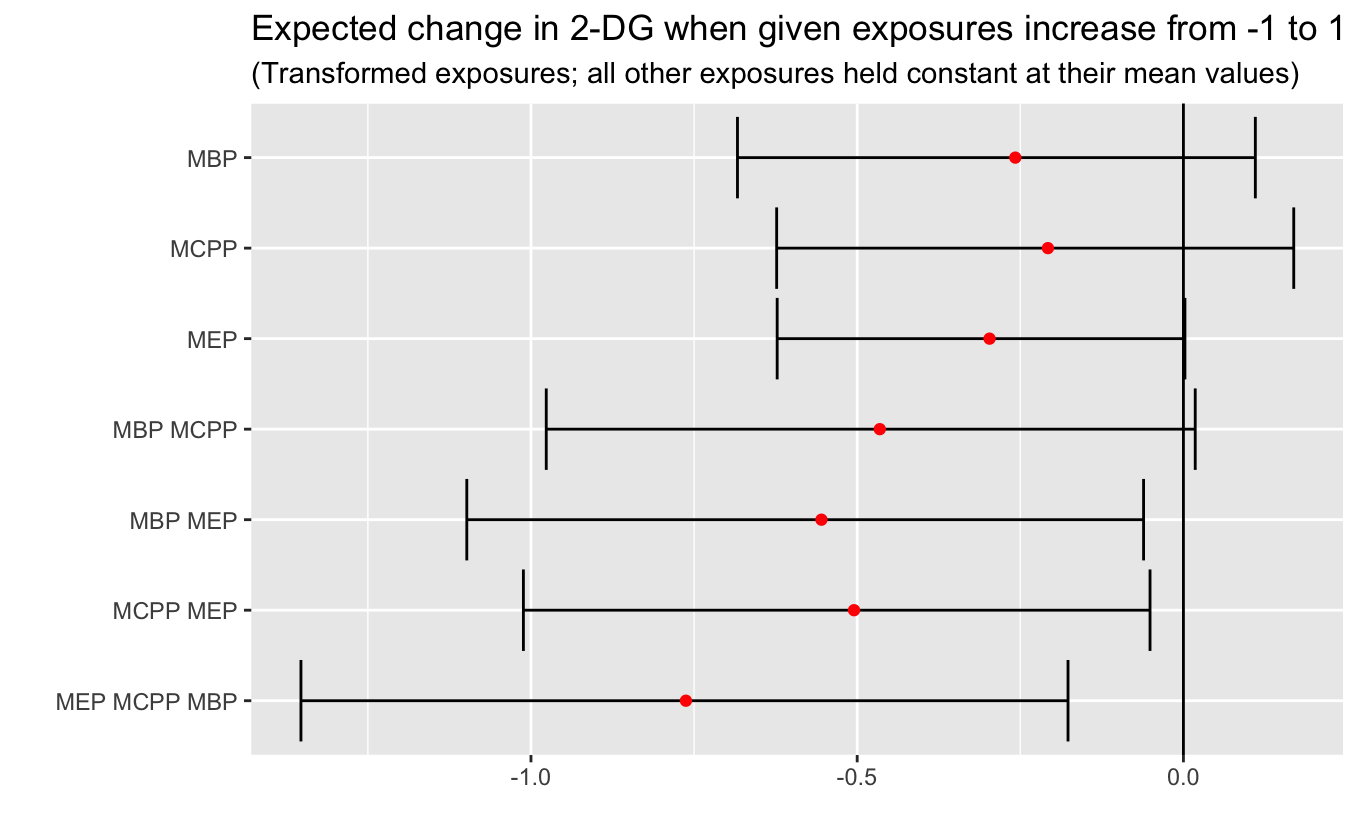}
\caption{Expected change in 2-DG level when the noted combinations of exposures are increased from $-1$ to $1$ at all three trimesters. (Transformed data.)}
\label{fig9_transformed}
\end{figure}

Figures \ref{fig8a_transformed}-\ref{fig9_transformed} show the results for LowFR corresponding to Figures 8 and 9 in the paper, but for this transformed data set. Observe that although some intervals shift slightly, the results are qualitatively quite similar. In particular, Figure \ref{fig9_transformed} shows that given an increase in MEP, MBP, and MCPP levels, expected followup 2-DG is significantly lower. Thus, we still conclude that the metabolites of DEP and DBP may be associated with decreased 2-DG, as we did in the paper.

%% figure of main effects

% parent compound effects

%% figure of effects for different combos of mbp, mcpp, and mep

\newpage